\documentclass{aa}  

\usepackage{graphicx}
\usepackage{txfonts}
\usepackage[]{natbib}
\usepackage{graphicx}   
\usepackage{amsmath}    
\usepackage[flushleft]{threeparttable}
\usepackage{lscape}
\usepackage{ulem}
\usepackage[usenames,dvipsnames]{xcolor}
\usepackage{dsfont}
\usepackage{soul} 
\usepackage{placeins}

\newcommand{\teff}{$T_{\rm eff}$}
\newcommand{\logg}{$\log{g}$}

\newcommand{\fluxunit}{erg/cm$^{2}$/s/\AA}

\usepackage[colorlinks=true,     linkcolor=blue, citecolor=blue, filecolor=blue, urlcolor=blue]{hyperref}
\begin{document}


\title{Photometric segregation of dwarf and giant FGK stars using the SVO Filter Profile Service and photometric tools}\titlerunning{Photometric FGK dwarf / giant segregation using SVO services}

   \author{Carlos Rodrigo\inst{1}\thanks{{\sl In Memoriam}: This work is dedicated to the memory of our dearest friend and colleague Carlos Rodrigo. The Spanish Virtual Observatory will certainly miss one of its best ambassadors.}
          \and
          Patricia Cruz\inst{1}
          \and
          John F. Aguilar\inst{2,3}
          \and
          Alba Aller\inst{4}
          \and
          Enrique Solano\inst{1}
          \and
          Maria Cruz G{\'a}lvez-Ortiz\inst{1}
          \and
          Francisco Jim{\'e}nez-Esteban\inst{1}
          \and
          Pedro Mas-Buitrago\inst{1}
          \and
          Amelia Bayo\inst{5}
          \and
          Miriam Cort{\'e}s-Contreras\inst{6}
          \and
          Raquel Murillo-Ojeda\inst{1}
          \and
          Silvia Bonoli\inst{7,8}
          \and
          Javier Cenarro\inst{8}
          \and
          Renato Dupke\inst{9}
          \and
          Carlos López-Sanjuan\inst{8}
          \and
          Antonio Marín-Franch\inst{8}
          \and
          Claudia {Mendes de Oliveira}\inst{10}
          \and
          Mariano Moles\inst{8}
          \and
          Keith Taylor\inst{11}
          \and
          Jesús Varela\inst{8}
          \and
          Héctor {Vázquez Ramió}\inst{8}
          }

   \institute{Centro de Astrobiolog\'{\i}a (CAB), CSIC-INTA, Camino Bajo del Castillo s/n, E-28692, Villanueva de la Ca\~{n}ada, Madrid, Spain.\\              \email{pcruz@cab.inta-csic.es/esm@cab.inta-csic.es}
         \and
         Departamento de Matem\'aticas, Universidad Militar Nueva Granada, kil\'ometro 2 v\'ia Cajic\'a - Zipaquir\'a, c\'odigo postal 110111, Colombia.
         \and
         PhD Programme in Astrophysics, Doctoral School, Universidad Aut\'onoma de Madrid, Ciudad Universitaria de Cantoblanco, 28049 Madrid, Spain.
         \and
         Observatorio Astronómico Nacional (OAN), Alfonso XII 3, 28014, Madrid, Spain.
         \and
         European Southern Observatory, Karl-Schwarzschild-Strasse 2, 85748 Garching bei M\"unchen, Germany.
         \and
         Departamento de F\'isica de la Tierra y Astrof\'isica, Facultad de Ciencias F\'isicas, e IPARCOS-UCM (Instituto de Física de Partículas y del Cosmos de la UCM), Universidad Complutense de Madrid, 28040 Madrid, Spain.
         \and
         Donostia International Physics Center (DIPC), Manuel Lardizabal Ibilbidea, 4, San Sebastián, Spain.
         \and
         Centro de Estudios de Física del Cosmos de Aragón (CEFCA), Plaza San Juan, 1, E-44001, Teruel, Spain.
         \and
         Observatório Nacional, Rua General José Cristino, 77, São Cristóvão, 20921-400, Rio de Janeiro, RJ, Brazil.
         \and
         Departamento de Astronomia, Instituto de Astronomia, Geofísica e Ciências Atmosféricas, Universidade de São Paulo, São Paulo, Brazil.
         \and
         Instruments4, 4121 Pembury Place, La Canada Flintridge, CA 91011, U.S.A.
         }

   \date{Received 15 March 2024 / Accepted 20 May 2024}

 
  \abstract
   {}
   {This paper is focused on the segregation of FGK dwarf and giant stars through narrow-band photometric data using the Spanish Virtual Observatory (SVO) Filter Profile Service and associated photometric tools.}
   {We selected spectra from the MILES, STELIB, and ELODIE stellar libraries, and used SVO photometric tools to derive the synthetic photometry in 15 J-PAS narrow filters, which were especially selected to cover spectral features sensitive to gravity changes. Using machine-learning techniques as the Gaussian mixture model and the support vector machine, we defined several criteria based on J-PAS colours to discriminate between dwarf and giant stars.}
   {We selected five colour-colour diagrams that presented the most promising separation between both samples. Our results show an overall accuracy in the studied sample of $\sim$0.97 for FGK stars, although a dependence on the luminosity type and the stellar effective temperature was found. 
   We also defined a colour-temperature relation for dwarf stars with effective temperatures between 4\,000 and 7\,000\,K, which allows one to estimate the stellar effective temperature from four J-PAS filters ($J0450$, $J0510$, $J0550$, and $J0620$). Additionally, we extended the study to M-type giant and dwarf stars, achieving a similar accuracy to that for FGK stars. }
   {}


   \maketitle
%

\section{Introduction}

Since  the  beginning  of  the  21st  century, large-area, multi-filter surveys have provided photometric information for millions of astronomical objects. For instance, some widely used surveys are the {\sl Gaia} mission \citep[{\sl Gaia};][]{Gaia16}, the Sloan Digital Sky Survey \citep[SDSS;][]{York00}, and the Panoramic Survey Telescope and Rapid Response System survey \citep[Pan-STARRS;][]{Chambers16} 
at visible wavelengths, or the Visible and Infrared Survey Telescope for Astronomy \citep[VISTA;][]{Sutherland15}, the UKIRT Infrared Deep Sky Survey \citep[UKIDSS;][]{Lawrence07}, the Two Micron All Sky Survey \citep[2MASS;][]{Skrutskie06}, and the Wide-Field Infrared Survey Explorer \citep[WISE;][]{Wright10} at infrared wavelengths, to name a few. 

Astronomical photometry refers to measuring the apparent brightness of an astrophysical object. A photometric system can be defined as a set of filters at different wavelengths with a well-characterised response 
to the incident radiation. They are typically designed to be sensitive to 
regions of the electromagnetic spectrum in which the variations in a given parameter (for instance, stellar effective temperature, surface gravity, and metallicity) are more prominent. 
A compilation of photometric systems is of great importance for many astrophysical purposes. Although some compilations were already available in the last decades of the 20th century, the first big work in this field was the Asiago Database on Photometric Systems \citep[ADPS;][]{Moro00}, which provided information on 218 optical, ultraviolet, and infrared photometric systems in its latest updates \citep{Fiorucci03} and played an important role in the design of the {\sl Gaia} photometric system \citep{Jordi06}. 

The combination of photometric data coming from different resources requires these measurements to be described and characterised in sufficient detail to enable the conversion into compatible flux density and spectral energy units. Nevertheless, information on the properties of filters is not always easily available; sometimes it is only specified in manuals (or in the literature) and often it is not in a digital form.

To unambiguously understand the (typically) heterogeneous information provided by the different photometric systems and to allow comparisons to be made between them, it is therefore mandatory to establish a common set of parameters for all photometric systems. 
The International Virtual Observatory Alliance (IVOA\footnote{\url{http://www.ivoa.net}}) in 2013 developed the IVOA Photometry Data Model \citep{Salgado13} to describe in a homogeneous way the essential elements of flux density measurements made across the full electromagnetic spectrum. 

By comparing information at different wavelengths using colours, it is possible to estimate physical parameters that can be used in very diverse research projects. Of particularly interest to many astrophysical studies is the separation between giant and dwarf stars. For instance, studies of the structure, kinematics, and chemical distribution of our Galaxy, in particular the outer disc and halo, need a clean sample of red giants \citep{Thomas2018}. Also, studies on the star formation history in the solar neighbourhood or on the occurrence rate of exoplanets as a function of the mass of the host stars require a clean sample of giants \citep[see, for instance][]{Wolthoff2022}. In both cases, contamination from members of the non-desired class may lead to incorrect conclusions. Furthermore, working with contaminated samples may result in telescope time being wasted when researchers attempt, for instance, to spectroscopically characterise these objects.

The above-mentioned cases are only a few recent examples; the question of how to discriminate between giant and dwarf stars has been of interest for a long time. 
\citet{Lindblad1919} found that stars on the list of \cite{Adams1917} could be arranged in two distinct series: one representing giants, the other dwarf stars. 
\citet{Neff1966} showed that late-type (G5-M0) dwarfs could be distinguished from giants by using filters in the blue part of the electromagnetic spectrum. Later, \citet{Geisler_1984} used the strong surface gravity (\logg) sensitivity of the Mg\,I triplet and MgH bands to define a good discrimination colour index in the optical for G and K stars also. 
More recently, \citet{Bilir_2006} used near-infrared 2MASS $J$, $H$, and $K$, and optical $V$ photometry to determine a linear separation between dwarfs and giants. They applied their results to member stars of the open cluster NGC 752 \citep{Bilir_2006b}.

In recent years, machine-learning techniques have joined the cause. \citet{Klement2011} used a support vector machine algorithm, reduced proper motions, and different apparent magnitudes in the optical and the infrared (USNO-B $B2R2$, \citealt{Monet_2003}; DENIS $IJK$, \citealt{Epchteun_1999}; and 2MASS $JHK$, \citealt{Skrutskie06}) to separate giants and dwarfs in a sample of bright stars with \logg\  estimates from the RAVE DR2 survey \citep{Steinmetz_2006}, achieving a contamination rate (giants classified as dwarfs) that remained 
above 20\%. \citet{Thomas2019} present a random forest algorithm to discriminate between dwarf and giant stars using optical photometry from the Canada–France Imaging Survey, Pan-STARRS, and {\sl Gaia}, reaching a dwarf contamination fraction below 30\% for the metallicity regime of [Fe/H]$<$-1.2\,dex. More recently, \citet{Yuan2023a} used a deep-learning technique to address the selection of red giants. Their method used data from the SkyMapper Southern Survey \citep[SMSS;][]{Wolf2018} and the Large Sky Area Multi-Object Fiber Spectroscopic Telescope \citep[LAMOST;][]{Wang2020}, obtaining over 20 thousand potential red giant candidates out of more than 304 thousand observed stars. Their model has a precision rate of 0.92 and an estimated contamination of 6.4\%.

One approach to reduce contamination rates is to use narrow-band filters, since the features that show temperature, gravity, or metallicity dependence can be better constrained, and thus provide better physical parameter estimations. Apart from the work mentioned by \citet{Geisler_1984}, \citet{Casey18} used a combination of wide and narrow filters ($g$ and $i$ bands and the DDO$_{51}$ filter from the David Dunlap Observatory; \citealt{McClure_1668}) to obtain a sample of giant stars to be used as tracers of the disc, bulge, and halo. They got a contamination rate (dwarfs classified as giants) of less that 1\% and a completeness of 80\% for giants with an effective temperature (\teff) of less than 5\,250\,K. 

More recently, \citet{Yuan23b} explored the discrimination of dwarfs and giants in the miniJPAS sample \citep[][]{Bonoli21}, using a extreme gradient boosting classification model and the collection of 56 J-PAS narrow-band filters, from the Javalambre Physics of the Accelerating universe Astrophysical Survey \citep[J-PAS;][]{Benitez14}, plus 4 SDSS-like filters. They ranked the J-PAS filters by their ability to separate giant and dwarf stars, finding that two filters, $J0510$ and $J0520$, may play an important role in the classification. 

The segregation of FGK dwarf and giant stars may seem evident and easy after 100 years of papers presenting different colour-magnitude, colour-colour, colour-index, or index-index discriminating diagrams. However, this separation is not so straightforward. 
From a compromise between filter width and gravity-dependent features, we further explore this topic of research and present new results using J-PAS narrow filters and the photometric services from the Spanish Virtual Observatory (SVO).

The paper is structured as follows. In Sect.~\ref{FTP}, we describe the Filter Profile Service (FPS), a key resource with which to efficiently obtain photometric filters information. In Sect.~\ref{DvsG}, we present the adopted dataset. The separation between giant and dwarf FGK stars using narrow-band filters is shown in Sect.~\ref{sec_selec}. Our discussions are presented in Sect.~\ref{discus}, in which a comparison to the classical approach using broad-band colours is given, followed by the application of our results to the miniJPAS sample. Additionally, in Sect.~\ref{discus}, we present a colour-temperature relation for FGK dwarfs and the case of M stars. Finally, the conclusions are summarised in Sect.~\ref{conclusions}.

\section{The Filter Profile Service}\label{FTP}

The FPS is a web service built and maintained by the SVO project.\footnote{\url{https://svo.cab.inta-csic.es}} It was originally conceived as a mechanism to efficiently manage the photometric information required by the VO Sed Analyzer\footnote{\url{http://svo2.cab.inta-csic.es/theory/vosa/}} \citep[\texttt{VOSA};][]{Bayo08}, a Virtual Observatory (VO) tool also built and maintained by the SVO project. \texttt{VOSA} is designed for the analysis of the spectral energy distribution (SED) of stellar objects based on the comparison of photometric observations and theoretical models. In particular, \texttt{VOSA} needs to transform observed magnitudes, either provided by the user or from multiple photometric databases available through VO archives, into fluxes and calculate synthetic photometry from theoretical spectra in order to compare them. 
In this sense, the FPS is not only a repository of information but also a central resource around which other services and applications can be built. 

To accomplish the previous tasks, it is mandatory to have well-controlled, well-structured information on the photometric filters. In this sense, the information provided by the FPS is in compliance with the IVOA Photometry Data Model. The FPS is regularly updated with new photometric systems found in the literature or requested by the astronomical community. At the time of writing (March 2024), it contains the largest public collection of filters available to the community. It provides information on more than 10\,000 filters, including not only astronomical but also Solar System and Earth-observation collections of filters. 
 
Most of the filter properties are calculated, by default, directly from the transmission curve. Nevertheless, in those cases in which there is a quantity provided by the filter's owners or developers (for instance, a zero point), the service uses that value instead. The information is accessible not only through a web interface\footnote{\url{http://svo2.cab.inta-csic.es/theory/fps/}} but also through VO protocols.\footnote{\url{http://svo2.cab.inta-csic.es/theory/fps/index.php?mode=voservice}} 

The service has been heavily used by the community, as is demonstrated by the fact that, during the last four years, it has received an average of around 720\,000 requests a month. Likewise, 115 refereed papers making use of the FPS for very different purposes\footnote{The comprehensive list of publications can be found here: \url{https://sdc.cab.inta-csic.es/vopubs/jsp/result.jsp?order=pub_id&bib=&com_id=26&com=&m_in=01&y_in=2020&m_en=12&y_en=2022&submit=Submit}.} have been published in the period 2020$-$2022. Moreover, archives like the Canadian Astronomy Data Center\footnote{\url{https://www.cadc-ccda.hia-iha.nrc-cnrc.gc.ca/en/search}} or Vizier\footnote{\url{http://vizier.unistra.fr/vizier/sed/}} make heavy use of the FPS. Special mention goes to \texttt{GaiaXPy},\footnote{\url{https://gaia-dpci.github.io/GaiaXPy-website/}} which, used in combination with the FPS, allows one to generate photometric information in a plethora of photometric systems using spectra from {\sl Gaia} Data Release 3 \citep[{\sl Gaia} DR3; ][]{Gaia22}.

A detailed description of the FPS can be found at Appendix \ref{appenA}. In addition to the FPS, other photometric services and tools developed by the SVO have been used in this paper (see Sect.~\ref{libraries}). They are all described in Appendix \ref{appenB}.

\section{Observational spectral data and synthetic photometry}\label{DvsG}

To explore the photometric segregation of dwarfs and giants among FGK stars, we made use of spectroscopic data available at SVO archives, as is described below, as well as several SVO photometric services to obtain from these spectra the synthetic photometry used later in the analysis.

\subsection{Adopted stellar libraries}\label{libraries}

We selected several spectra from libraries that contain observational data from both dwarf and giants stars, with spectra covering the visible wavelength range. We used three libraries, which are briefly described below:

\begin{itemize}

\item The Medium-resolution Isaac Newton telescope Library of Empirical Spectra \citep[MILES;][]{Sanchez06,Cenarro07} contains spectra with a wavelength coverage from 3\,525 to 7\,500\,\AA\, at a (full width at half maximum, FWHM) spectral resolution of 2.5\,\AA\ . The library covers a large range of atmospheric parameters, with \teff\  up to 9\,950\,K, \logg\  from $\sim$0 to 5.5\,dex, and [Fe/H] from $-$2.93 to $+$1.65\,dex. The SVO service contains spectra of 985 stars.\footnote{\url{http://svocats.cab.inta-csic.es/miles}}\\

\item The STELIB Stellar Library \citep{LeBorgne03} contains spectra from 3\,200 to 9\,500\,\AA\ at a (FWHM) spectral resolution of 3.0\,\AA\ . The covered atmospheric parameters go from 3\,252 to 31\,500\,K in \teff, from 0.2 to 4.8\,dex in \logg, and from $-$2.64 to $+$0.67\,dex in [Fe/H], and thus include stars of all spectral types. The SVO service contains spectra of 254 stars.\footnote{\url{http://svocats.cab.inta-csic.es/stelib/}}\\ 

\item The ELODIE.3.1 library \citep[][and references therein]{Prugniel07}\footnote{\url{http://atlas.obs-hp.fr/elodie/intro.html}} includes 1\,962 spectra of 1\,388 stars in the wavelength range from 3\,900 to 6\,800\,\AA, with a nominal resolution of $R=42\,000$. 
The library provides a large coverage of \teff, ranging from 3\,100 to 50\,000\,K, \logg from $-$0.25 to 4.9\,dex, and [Fe/H] from $-$3.0 to $+$1.0\,dex. 

\end{itemize}

\begin{figure}[!t]
        \includegraphics[width=\columnwidth]{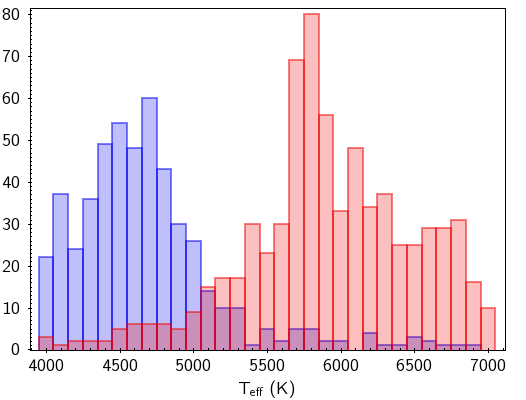}
    \caption{Distribution of effective temperatures of the selected spectra from the adopted sample from MILES, STELIB, and ELODIE espectral libraries together, from 4\,000 to 7\,000\,K, where the blue and the red histograms show the distributions for giant and dwarf stars, respectively.}
    \label{fig_histteff}
\end{figure}

In all cases, only stars with effective temperatures varying from 4\,000 to 7\,000\,K, roughly corresponding to FGK spectral types according to \citet{Pecaut13},\footnote{Available at \url{https://www.pas.rochester.edu/~emamajek/EEM_dwarf_UBVIJHK_colors_Teff.txt}.} were kept. Also, the samples of dwarf and giant stars were defined according to the available surface gravity values, assigning stars with $\log g \ge 4$ to our sample of dwarf stars and with $\log g \le 3$ to our sample of giant stars. Sources with $3<\log{g}<4$\,dex were not used to avoid the transitional subgiant phase with its fast evolution \citep[e.g.][]{Cayrel2001}, which may hamper the separation among both classes. 
In total, we selected 1\,200 spectra from the above-described libraries, consisting of 701 from dwarfs and 499 from giants. The distribution of their \teff\ is presented in Fig.~\ref{fig_histteff}, where the blue and the red histograms show the distributions of giant and dwarf stars, respectively.

\subsection{Selection of narrow-band J-PAS filters}\label{sec_filters}

\begin{table}
\centering
\caption{Fe\,II lines that are sensitive to gravity changes within the wavelength range covered by the libraries described in Sect.\,\ref{libraries}.} 
\begin{tabular}{lcccl}
\hline \hline
$\lambda_{\rm Fe\,II}$ & $\chi_{l}$ & $\log{\rm gf}$ & Ref. & J-PAS filter \\
(\AA) & (eV) &  &  & \\
\hline 
4508.28  & 2.86 & -2.403 & 1 & $J0450$ \\
4520.22 & 2.81 & -2.563 & 1 & $J0450$, $J0460$ \\
4534.17 & 2.86 & -3.203 & 1 & $J0450$, $J0460$ \\
4541.52 & 2.86 & -2.762 & 1 & $J0450$, $J0460$ \\
4576.34 & 2.84 & -2.947 & 1 & $J0450$, $J0460$ \\
4582.84 & 2.84 & -3.076 & 1 & $J0450$, $J0460$ \\
4620.51 & 2.83 & -3.234 & 1 & $J0460$, $J0470$ \\
4629.34 & 2.81 & -2.262 & 1 & $J0460$, $J0470$ \\
4656.98 & 2.89 & -3.676 & 1 & $J0460$, $J0470$ \\
4670.17 & 2.58 & -4.011 & 1 & $J0460$, $J0470$ \\
4923.93 & 2.89 & -1.541 & 1 & $J0490$ \\
4993.36 & 2.81 & -3.684 & 2 & $J0490$ \\
5132.67 & 2.81 & -4.094 & 2 & $J0510$, $J0520$ \\
5197.57 & 3.23 & -2.293 & 1 & $J0510$, $J0520$ \\
5234.63 & 3.22 & -2.235 & 1 & $J0520$, $J0530$ \\
5256.94 & 2.89 & -4.019 & 1 & $J0520$, $J0530$ \\
5264.81 & 3.23 & -3.091 & 1 & $J0520$, $J0530$ \\
5284.11 & 2.89 & -3.195 & 2 & $J0520$, $J0530$ \\
5414.07 & 3.22 & -3.580 & 2 & $J0540$, $J0550$ \\
5425.26 & 3.20 & -3.220 & 2 & $J0540$, $J0550$ \\
5427.81 & 6.72 & -1.481 & 1 & $J0540$, $J0550$ \\
5525.12 & 3.27 & -4.000 & 1 & $J0550$ \\
5534.85 & 3.24 & -2.865 & 2 & $J0550$ \\
5991.38 & 3.15 & -3.647 & 2 & $J0600$ \\
6084.11 & 3.20 & -3.881 & 2 & $J0600$, $J0610$ \\
6149.26 & 3.89 & -2.841 & 2 & $J0610$, $J0620$ \\
6238.39 & 3.89 & -2.600 & 2 & $J0620$, $J0630$ \\
6239.94 & 3.89 & -3.448 & 1 & $J0620$, $J0630$ \\
6247.56 & 3.89 & -2.347 & 1 & $J0620$, $J0630$ \\
6369.46 & 2.89 & -4.131 & 1 & $J0630$, $J0640$ \\
6407.29 & 3.89 & -2.899 & 1 & $J0630$, $J0640$ \\
6416.93 & 3.89 & -2.625 & 1 & $J0640$, $J0650$ \\
6432.69 & 2.89 & -3.572 & 1 & $J0640$, $J0650$ \\
6442.97 & 5.55 & -2.399 & 1 & $J0640$, $J0650$ \\
6456.39 & 3.90 & -2.110 & 1 & $J0640$, $J0650$ \\
6516.08 & 2.89 & -3.310 & 2 & $J0650$ \\
\hline \hline
\end{tabular}
\tablefoot{
The first column shows the line wavelength, and the second and third columns show the line's excitation potential and the oscillator strength, respectively, followed by its reference. In the last column, the J-PAS filters that contain the line mentioned are listed. (1) \citet{Sousa08}; (2) \citet{Sousa14}.}
\label{tab_FeIIlines}
\end{table}

\begin{table}
\centering
\caption{Selected J-PAS filters and their characteristics.} 
\begin{tabular}{lcccc}
\hline \hline
J-PAS filter & $\lambda_{\rm eff}$ & $\lambda_{\rm min}$ & $\lambda_{\rm max}$ & $W_{\rm eff}$ \\
 & (\AA) & (\AA) & (\AA) & (\AA) \\
\hline 

$J0450$ & 4514.42 & 4420.83 & 4613.85 & 143.74 \\
$J0460$ & 4608.52 & 4513.99 & 4706.85 & 143.95 \\
$J0470$ & 4704.40 & 4608.26 & 4803.61 & 140.72 \\
$J0490$ & 4911.15 & 4808.71 & 5009.24 & 148.71 \\
$J0510$ & 5101.53 & 5007.49 & 5199.79 & 146.66 \\
$J0520$ & 5207.86 & 5110.47 & 5308.04 & 149.27 \\
$J0530$ & 5303.52 & 5206.95 & 5403.67 & 150.59 \\
$J0540$ & 5396.85 & 5302.38 & 5496.61 & 149.54 \\
$J0550$ & 5501.64 & 5405.61 & 5600.81 & 146.03 \\
$J0600$ & 6009.76 & 5914.71 & 6108.40 & 148.38 \\
$J0610$ & 6117.19 & 6021.49 & 6215.13 & 146.63 \\
$J0620$ & 6208.79 & 6113.73 & 6307.07 & 147.18 \\
$J0630$ & 6312.31 & 6217.88 & 6409.61 & 147.53 \\
$J0640$ & 6410.21 & 6314.82 & 6508.32 & 146.51 \\
$J0650$ & 6503.43 & 6412.92 & 6604.47 & 146.86 \\
\hline \hline
\end{tabular}
\tablefoot{
$\lambda_{\rm eff}$ is the effective wavelength, $\lambda_{\rm min}$ and $\lambda_{\rm max}$ are the minimum and maximum wavelength, respectively, and $W_{\rm eff}$ is the effective width. Additional information is provided on the SVO FPS webpage.}
\label{tab_filters}
\end{table}

\begin{table*}
\centering
\caption{Additional once-ionised metal lines with a measured equivalent width (EW) equal to or greater than 30\,m\AA\ in the solar spectrum atlas by \citet{Moore66}.} 
\begin{tabular}{lccl|lccl|lccl}
\hline \hline
Wavelength & $\chi_{l}$ & EW & Identified & Wavelength & $\chi_{l}$ & EW & Identified & Wavelength & $\chi_{l}$ & EW & Identified \\
(\AA) & (eV) & (m\AA) & element & (\AA) & (eV) & (m\AA) & element & (\AA) & (eV) & (m\AA) & element \\
\hline 

4421.944 & 2.06 & 51 & Ti\,II & 4555.892 & 2.83 & 77 & Fe\,II & 5129.162 & 1.89 & 70 & Ti\,II \\
4429.906 & 0.23 & 30$^{a}$ & La\,II & 4558.650 & 4.07 & 66 & Cr\,II & 5146.119 & 2.83 & 37$^{b}$ & Fe\,II \\
4431.360 & 0.61 & 30 & Sc\,II & 4563.766 & 1.22 & 120 & Ti\,II & 5154.075 & 1.57 & 73 & Ti\,II \\
4440.482 & 1.21 & 46$^{b}$ & Zr\,II & 4571.982 & 1.57 & 126 & Ti\,II & 5169.050 & 2.89 & 154 & Fe\,II \\
4441.719 & 1.18 & 79$^{b}$ & Ti\,II & 4583.839 & 2.81 & 124$^{a}$ & Fe\,II & 5185.908 & 1.89 & 58 & Ti\,II \\
4443.812 & 1.08 & 124 & Ti\,II & 4588.204 & 4.07 & 66 & Cr\,II & 5188.698 & 1.58 & 202$^{b}$ & Ti\,II \\
4444.562 & 1.12 & 50 & Ti\,II & 4589.953 & 1.24 & 70 & Ti\,II & 5200.415 & 0.99 & 37 & Y\,II \\
4450.491 & 1.08 & 79 & Ti\,II & 4592.057 & 4.07 & 44 & Cr\,II & 5205.730 & 1.03 & 52 & Y\,II \\
4457.437 & 1.18 & 81$^{b}$ & Zr\,II & 4616.628 & 4.07 & 37 & Cr\,II & 5226.545 & 1.57 & 94 & Ti\,II \\
4460.225 & 0.48 & 100$^{b}$ & Ce\,II & 4618.792 & 4.07 & 78$^{b}$ & Cr\,II & 5237.325 & 4.07 & 49 & Cr\,II \\
4461.205 & 1.01 & 48$^{b}$ & Zr\,II & 4634.079 & 4.07 & 53$^{a}$ & Cr\,II & 5239.823 & 1.45 & 55 & Sc\,II \\
4461.429 & 2.58 & 55$^{b}$ & Fe\,II & 4657.204 & 1.24 & 38 & Ti\,II & 5262.150 & 1.58 & 128$^{b}$ & Ti\,II \\
4464.457 & 1.16 & 68 & Ti\,II & 4666.754 & 2.83 & 45 & Fe\,II & 5274.979 & 4.07 & 48$^{a}$ & Cr\,II \\
4468.500 & 1.13 & 120 & Ti\,II & 4670.413 & 1.36 & 55 & Sc\,II & 5276.002 & 3.20 & 152$^{a}$ & Fe\,II \\
4469.154 & 1.08 & 49 & Ti\,II & 4682.351 & 0.41 & 39$^{b}$ & Y\,II & 5313.585 & 4.07 & 35 & Cr\,II \\
4470.858 & 1.16 & 54 & Ti\,II & 4708.672 & 1.24 & 46 & Ti\,II & 5316.620 & 3.15 & 112 & Fe\,II \\
4472.930 & 2.84 & 39 & Fe\,II & 4731.473 & 2.89 & 79$^{a}$ & Fe\,II & 5316.780 & 3.22 & 97$^{a}$ & Fe\,II \\
4481.140 & 8.86 & 63 & Mg\,II & 4762.782 & 1.08 & 30 & Ti\,II & 5325.560 & 3.22 & 45 & Fe\,II \\
4481.338 & 8.86 & 97$^{b}$ & Mg\,II & 4779.984 & 2.05 & 76$^{a}$ & Ti\,II & 5332.665 & 2.27 & 45$^{b}$ & V\,II \\
4488.329 & 3.12 & 45 & Ti\,II & 4798.537 & 1.08 & 50 & Ti\,II & 5334.870 & 4.07 & 32$^{b}$ & Cr\,II \\
4489.184 & 2.83 & 61 & Fe\,II & 4805.099 & 2.06 & 128$^{a}$ & Ti\,II & 5336.794 & 1.58 & 71 & Ti\,II \\
4491.408 & 2.85 & 66 & Fe\,II & 4812.352 & 3.86 & 41 & Cr\,II & 5337.727 & 3.23 & 35$^{a}$ & Fe\,II \\
4501.278 & 1.12 & 111 & Ti\,II & 4824.143 & 3.87 & 94$^{a}$ & Cr\,II & 5337.760 & 4.07 & 35$^{b}$ & Cr\,II \\
4515.343 & 2.84 & 75$^{a}$ & Fe\,II & 4836.238 & 3.86 & 33$^{a}$ & Cr\,II & 5362.867 & 3.20 & 110$^{a}$ & Fe\,II \\
4522.638 & 2.84 & 101$^{a}$ & Fe\,II & 4848.252 & 3.86 & 52 & Cr\,II & 5381.028 & 1.57 & 56$^{a}$ & Ti\,II \\
4524.944 & 2.51 & 32 & Ba\,II & 4849.172 & 1.13 & 32$^{b}$ & Ti\,II & 5478.378 & 4.18 & 46$^{b}$ & Cr\,II \\
4529.492 & 1.57 & 99$^{a}$ & Ti\,II & 4854.873 & 0.99 & 41$^{a}$ & Y\,II & 5480.761 & 1.72 & 68$^{b}$ & Y\,II  \\
4533.970 & 1.24 & 109 & Ti\,II & 4864.323 & 3.86 & 50 & Cr\,II & 5497.356 & 1.75 & 128$^{b}$ & Y\,II \\
4539.777 & 0.33 & 37$^{b}$ & Ce\,II & 4865.618 & 1.12 & 34 & Ti\,II & 5526.821 & 1.77 & 76 & Sc\,II \\
4544.022 & 1.24 & 35 & Ti\,II & 4874.014 & 3.09 & 34 & Ti\,II & 6141.727 & 0.70 & 113$^{a}$ & Ba\,II \\
4545.143 & 1.13 & 42 & Ti\,II & 4876.401 & 3.85 & 41 & Cr\,II & 6147.742 & 3.89 & 76$^{b}$ & Fe\,II \\
4549.189 & 5.91 & 33$^{b}$ & Fe\,II & 4883.690 & 1.08 & 51 & Y\,II & 6245.620 & 1.51 & 30 & Sc\,II \\
4549.474 & 2.83 & 231$^{b}$ & Fe\,II & 4900.124 & 1.03 & 54 & Y\,II & 6347.095 & 8.12 & 54 & Si\,II \\
4549.638 & 1.58 & 231$^{a}$ & Ti\,II & 4911.199 & 3.12 & 50 & Ti\,II & 6371.355 & 8.12 & 35$^{a}$ & Si\,II \\
4549.820 & 1.18 & 53 & Ti\,II & 4921.785 & 0.24 & 40$^{b}$ & La\,II & 6491.582 & 2.06 & 45$^{a}$ & Ti\,II \\
4554.036 & 0.00 & 159 & Ba\,II & 4934.095 & 0.00 & 207$^{a}$ & Ba\,II & 6496.908 & 0.60 & 98 & Ba\,II \\
4554.992 & 4.07 & 39 & Cr\,II & & & & & & & &  \\
\hline \hline
\end{tabular}
\tablefoot{
The first column shows the line wavelength, and the second and third columns show the line's lower excitation potential and measured equivalent width, respectively, followed by the atomic species identified. 
\tablefoottext{a}{Main contributor to a blend.}
\tablefoottext{b}{Blended to a stronger line.}
In both cases, the EWs were measured considering the whole blend.
}
\label{tab_extraIIlines}
\end{table*}

To estimate the stellar luminosity class through photometry, we need to choose a collection of photometric filters that can offer both a large spectral coverage in the optical and enough resolution for our purpose. The J-PAS filter system\footnote{\url{http://svo2.cab.inta-csic.es/theory/fps/index.php?mode=browse&gname=OAJ&gname2=JPAS&asttype=}} \citep[][and references therein]{Bonoli21} is the largest collection of narrow-band filters available to date, with a total of 56 filters (54 narrow filters plus two broader ones) that cover the visible wavelength range from 3\,200 to 10\,800\,\AA, approximately. 
The narrow-band filters have a typical width of $\sim$145\,\AA, which all together can generate a photo spectrum with a resolution of R$\sim$60 \citep{Bonoli21}. 
Thus, the J-PAS filter system seems suitable for our purposes.

One way to address the problem is through colour-colour diagrams. Considering the 56 J-PAS narrow-band filters, it would be extremely costly computationally to explore all possible combinations among them, as they generate 1\,540 different colours, which in turn can produce 1\,185\,030 different diagrams. So, we needed to select some of the J-PAS filters and generate a manageable number of colours considering the differences seen in the stellar spectra.

Dwarf and giant stars present different spectral features. For instance, the intensities of ionised-metal lines are frequently used to derive the stellar surface gravity, as they are sensitive to such changes in FGK stars. According to \citet[][and references therein]{Gray09}, ionised lines such as Fe\,II and Ti\,II discriminate well between FGK dwarfs and giants, not only due to the electron density in the ionisation equilibrium but also due to micro-turbulence, which increases with luminosity in these stars. 

We gathered from the literature a set of 36 Fe\,II lines that are sensitive to gravity changes and that have been frequently used for luminosity classification in spectroscopic analyses \citep[][and references therein]{Sousa08,Sousa14}. These lines are listed in Table \ref{tab_FeIIlines}. 
We did not explore stellar features beyond 6\,600\,\AA\  to avoid contamination by telluric lines, 
which may appear in this spectral range as strong absorption features, 
as is shown in the atlas of the solar spectrum by \citet{Moore66}. 
Additionally, there are other spectral features important for the luminosity classification of FGK stars that are present in the region of interest; for instance, those involving magnesium, such as the Mg\,I triplet ($\lambda\lambda$ 5\,167, 72, 83\,\AA), especially for late-G to mid-K, and the MgH/TiO blend near 4\,770\,\AA, which is dominant in dwarf stars for mid-K or later types \citep[][and references therein]{Gray09}. 

After having selected the sample of Fe\,II lines, we selected 15 J-PAS narrow-band filters, which cover the electromagnetic spectrum range in which the gravity-sensitive Fe\,II lines mentioned in Table~\ref{tab_FeIIlines} are present. These filters are listed in Table~\ref{tab_filters}, along with their main characteristics.

In addition to the Fe\,II lines mentioned, we have identified other gravity-sensitive spectral features (once-ionised metal lines) covered 
by the selected 15 J-PAS filters, which may help to segregate dwarfs and giants. These lines were selected from the atlas of the solar spectrum by \citet{Moore66} and are listed in Table~\ref{tab_extraIIlines}. 
These spectral features are very weak and may appear blended in low- and intermediate-resolution spectra. However, since this study is focused on the integrated flux, the accumulative effect of several tenuous spectral features affected by luminosity changes may have an impact on the photometry.

For the set of 1\,200 spectra gathered from the libraries presented in Sect.~\ref{libraries}, the great majority of them (1\,182 spectra) have enough wavelength coverage to completely cover the 15 J-PAS narrow-band filters listed in Table\,\ref{tab_filters}. These 1\,182 spectra were adopted as our final set. Among them, 690 are dwarfs ($\log g \ge 4$) and 492 are giant stars ($\log g \le 3$), according to their libraries of origin. 

Once we had defined the final set of spectra, we used the SVO utilities described in Appendix\,\ref{appenB} to obtain their corresponding 105 J-PAS photometric colours, associated with the adopted 15 J-PAS filters. In particular, we used the tool \texttt{COLCA}.\footnote{\url{http://svo2.cab.inta-csic.es/theory/newov2/wcolors.php}} This calculates photometric colours using the synthetic photometry available at the SVO Synthetic Photometry Server.\footnote{\url{http://svo2.cab.inta-csic.es/theory//newov2/syph.php}}

\section{Photometric characterisation of giant and dwarf FGK stars}\label{sec_selec}

The full dataset of 105 colours can be grouped two by two into 5\,460 different colour-colour diagrams. This number of possibilities is too high to be visually explored, and therefore we adopted machine-learning techniques to select the best discriminators, as is described below.

\begin{figure}
        \includegraphics[width=\columnwidth,clip]{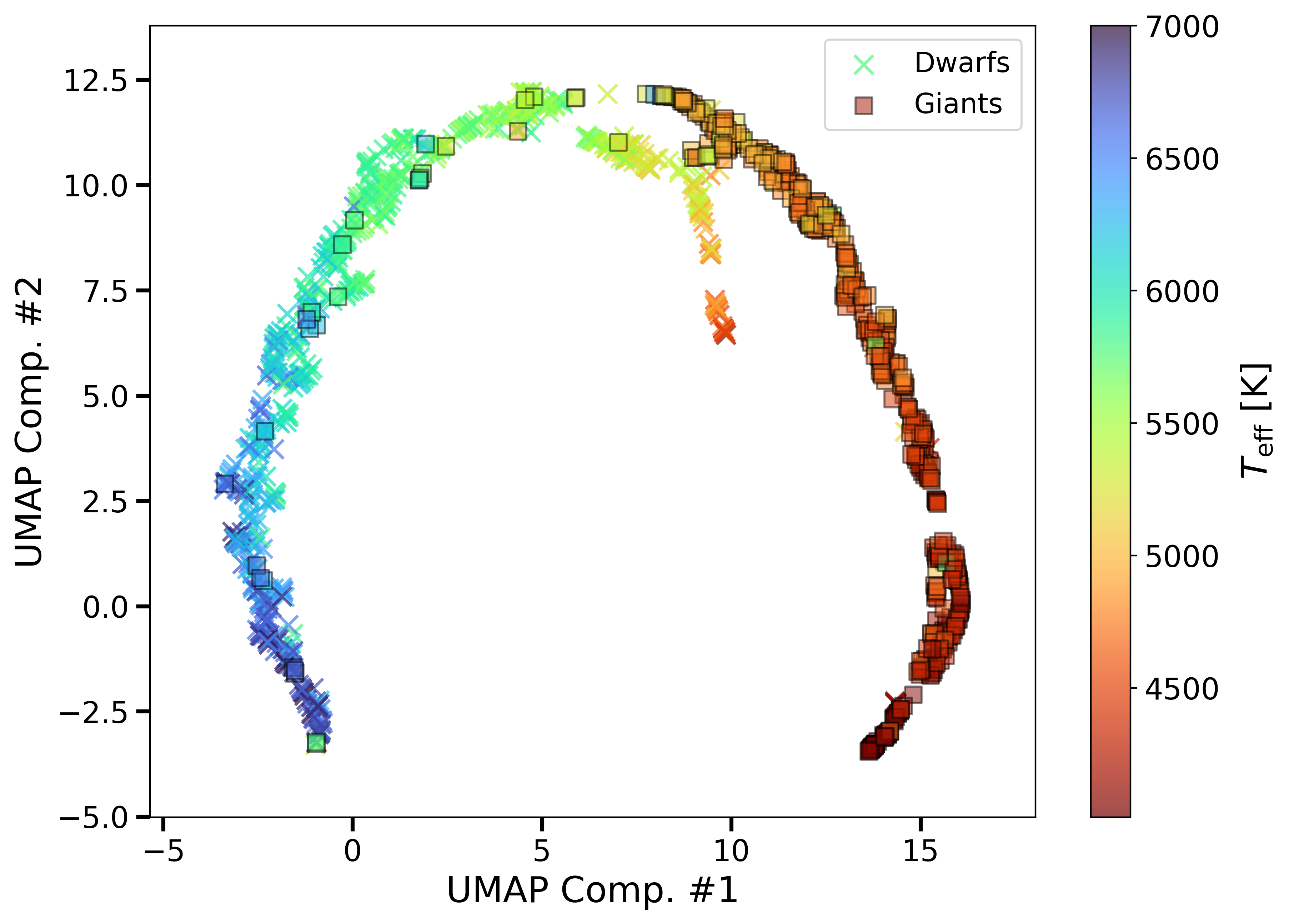}
    \caption{Two-dimensional UMAP projection of our sample of giants (filled squares with darker contours) and dwarfs (crosses), colour-coded by effective temperature.}
    \label{fig_umap}
\end{figure}

To estimate the extent to which the selected colours can be used to separate giants from dwarf stars, we used the uniform manifold approximation and projection \citep[UMAP;][]{mcinnes18} algorithm, which firstly constructs a graph representation of the high-dimensional data and then finds a low-dimensional representation that preserves as much of the topology as possible. 
Figure \ref{fig_umap} shows a two-dimensional (2D) UMAP projection for our sample of giants and dwarfs, represented by filled squares and crosses, respectively. Although there is an overlap for temperatures above approximately 5\,500\,K, both classes are clearly distinguishable for lower temperatures. This confirms the suitability of the selected colours for separating giants and dwarfs.

\begin{figure}
        \includegraphics[width=0.97\columnwidth,clip]{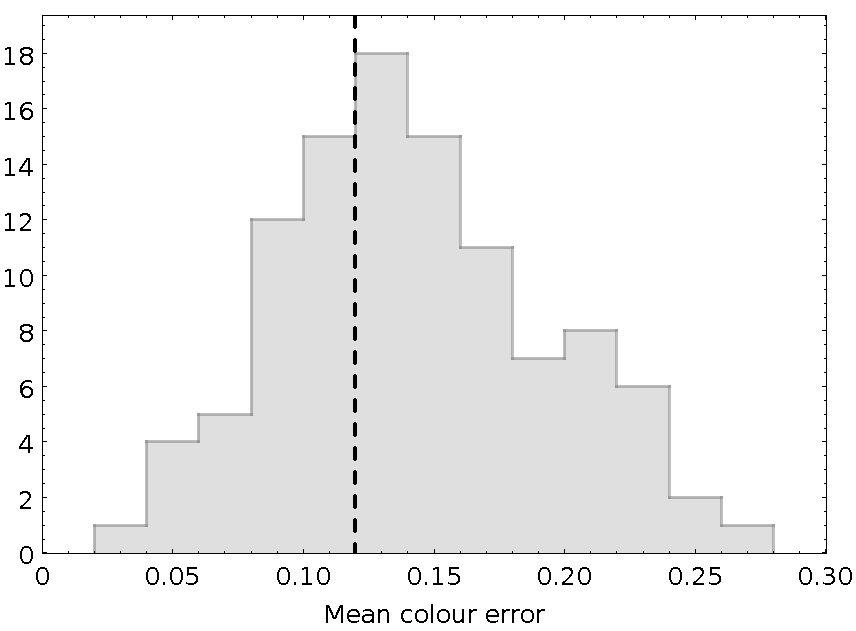}
    \caption{Mean colour error of the 105 J-PAS colours considered in this work, estimated from 2\,059 stars from the miniJPAS survey. The dashed vertical line depicts the threshold imposed in the mean colour error (see text).}
    \label{fig_color-error}
\end{figure}

To select the combination of colours that best discriminates both sets, we adopted the Gaussian mixture model \citep[GMM;][]{Scrucca16}, a clustering algorithm capable of assigning membership probability to each group (giants or dwarfs) based on a combination of Gaussian distributions. From the adopted 105 colours, we sought the pairs of colours that best discriminate dwarfs from giants. This means that the locus of each type of stars in the colour-colour space has the maximum possible separation. Once the sources in our sample were classified as dwarfs or giants, this was compared with their true classification from the spectral libraries. 
For that, we used the Rand index (RI), which is a metric that evaluates the quality of the results of classification algorithms. It is defined as

\begin{equation}
    {\rm RI = \frac{TP+TN}{TP+FP+FN+TN}},
\end{equation}

\noindent
where TP is the number of true positives, FP is the number of false positives, FN is the number of false negatives, and TN is the number of true negatives. The RI value is equal to unity when the classification given by the algorithm and the true classification are the same, and it tends to nullity when there is a high discrepancy. 

The highest RI value ($\sim$0.887) was found for the pair of colours $J0460-J0490$ versus $J0610-J0630$. 
However, $J0610-J0630$ presents a small dynamic range of colour values ($\sim$0.15\,mag), which seems to be good enough for the synthetic photometry obtained from the spectra collections used in this work (see Sect.\ref{libraries}). Nevertheless, when working with observed data from photometric surveys, like J-PAS, 
the observational errors will make the colour-colour diagram get blurred, hampering the segregation of luminosity classes. Thus, additional criteria were needed.

\begin{table*}
\centering
\caption{Five pairs of J-PAS colours selected to discriminate between giant and dwarf stars, according to the criteria described in Sect.~\ref{sec_selec}.} 
\begin{tabular}{lccccccc}
\hline \hline
Diagram & \multicolumn{2}{c}{J-PAS colour pair} & RI & \multicolumn{2}{c}{Dynamic range} &  \multicolumn{2}{c}{Mean error} \\
 & colour 1 & colour 2 & & colour 1 & colour 2 & colour 1 & colour 2 \\
 \hline 
1 &$J0450-J0510$ & $J0630-J0650$ & 0.823 & 1.395 & 0.785 & 0.103 & 0.054 \\
2 & $J0450-J0510$ & $J0620-J0650$ & 0.819 & 1.395 & 1.177 & 0.103 & 0.063 \\
3 & $J0450-J0620$ & $J0510-J0550$ & 0.773 & 2.971 & 0.929 & 0.093 & 0.110 \\
4 & $J0450-J0510$ & $J0610-J0650$ & 0.763 & 1.395 & 0.911 & 0.103 & 0.110 \\
5 & $J0450-J0510$ & $J0510-J0550$ & 0.756 & 1.395 & 0.929 & 0.103 & 0.110 \\
\hline \hline
\end{tabular}
\tablefoot{
The last four columns are given in magnitudes.
}
\label{tab_GMMcolours}
\end{table*}

The miniJPAS survey \citep{Bonoli21}, which observed around 1\,deg$^2$ of the sky with the J-PAS filter system, is the only data collection available to date made using such a filter system. We queried the miniJPAS database\footnote{\url{https://www.j-pas.org/datareleases/minijpas_public_data_release_pdr201912}} for stars adopting CLASS\_STAR $\ge$ 0.6, which is a morphological parameter provided by \texttt{SExtractor}\footnote{\texttt{SExtractor}  was used to detect objects in miniJPAS images. See \citet{Bonoli21} for details.} \citep{Bertin96} that indicates a point-like source. 
We considered the aperture-corrected magnitudes obtained from a 6-arcsec aperture in the dual-mode catalogue. 
We also adopted other criteria, as $r$ < 22\,mag, to avoid faint sources and quality flags equal to zero (both \texttt{SExtractor}  FLAGS and MASK\_FLAGS; see \citealt{Bonoli21} for details). Moreover, we selected those objects with magnitudes and error estimates available in all studied bands. This returned photometry data for 2\,059 stars. Using these data, we estimated for each of the 105 studied J-PAS colours its mean observational error and its dynamic range.

Figure\,\ref{fig_color-error} shows the distribution of the estimated mean colour errors. 
After different tests, we concluded that an observational error less than 0.12\,mag and a dynamic colour range six times larger is an adequate combination to distinguish between dwarf and giants in colour-colour diagrams. 
So, in addition to imposing a high RI $\ge 0.75$, two more conditions were defined from this diagram: 
i) a mean observational colour error of less than 0.12\,mag, shown with a dashed vertical line; ii) and a dynamic colour range greater than 0.75\,mag, to ensure a range of approximately six times the expected colour error.

By considering the three criteria mentioned above, we ended up with the five pairs of colours listed in Table\,\ref{tab_GMMcolours}, which are ranked by RI value.\footnote{The complete table with all solutions for the 5\,460 combinations is available online at \url{http://svocats.cab.inta-csic.es/fps_jpascolours/index.php}.} The first, second, and fourth pairs of colours have similar behaviour, as they have three of the four filters in common. The presence of the $J0510$ filter among the selected colours is not a surprise, as it contains the Mg\,I triplet). 
As was mentioned in Sect.~\ref{sec_filters}, this spectral feature is sensitive to luminosity changes from G- to K-type stars \citep[][and references therein]{Gray09}, which appear to be more intense in dwarf stars than in giants. \citet{Yuan23b} also stated that $J0510$ in one of the most important J-PAS filters for the discrimination of both luminosity classes, which is in agreement with our results. 

\begin{figure*}[!th]
\centering
        \includegraphics[width=0.99\columnwidth]{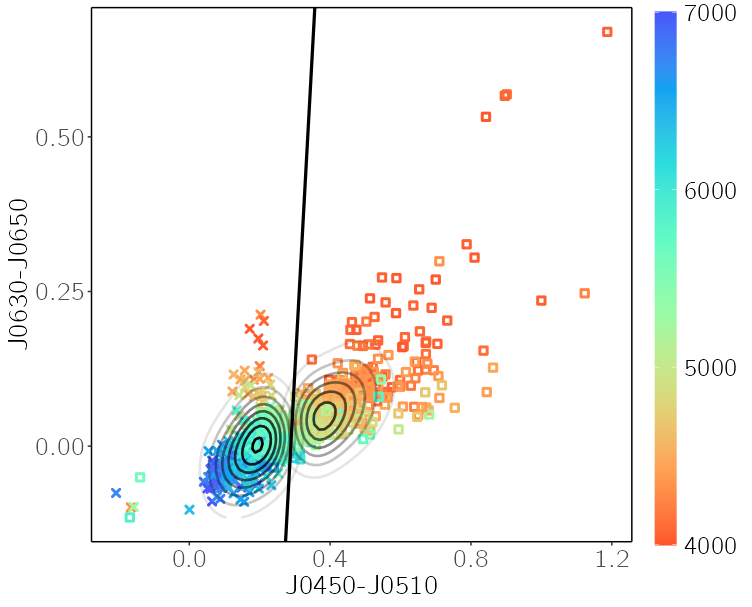}
        \includegraphics[width=0.99\columnwidth]{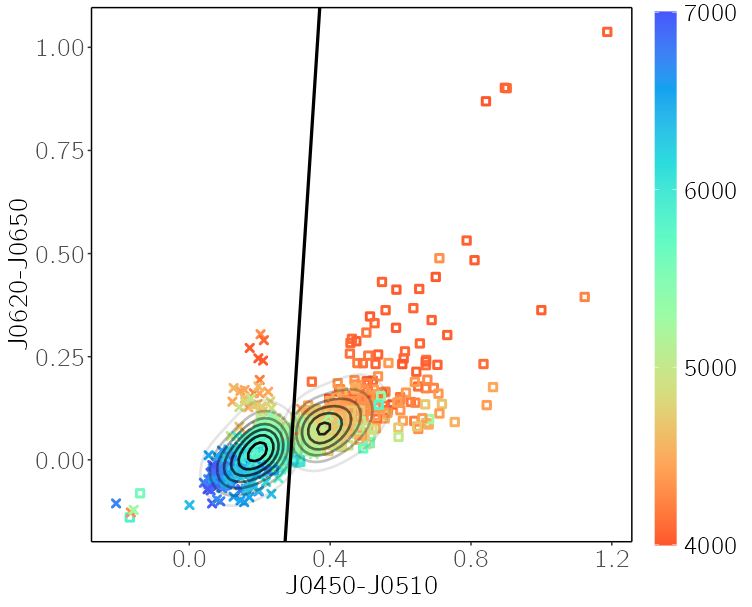}
        \includegraphics[width=0.98\columnwidth]{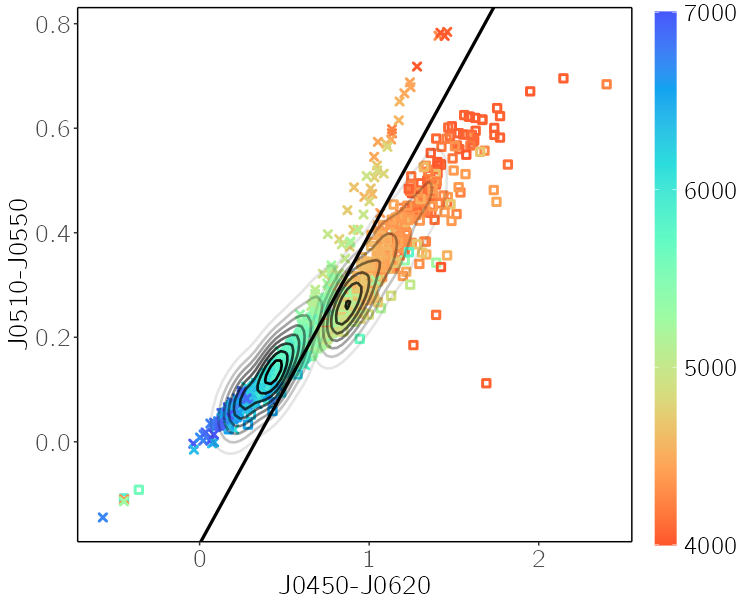}
        \includegraphics[width=0.99\columnwidth]{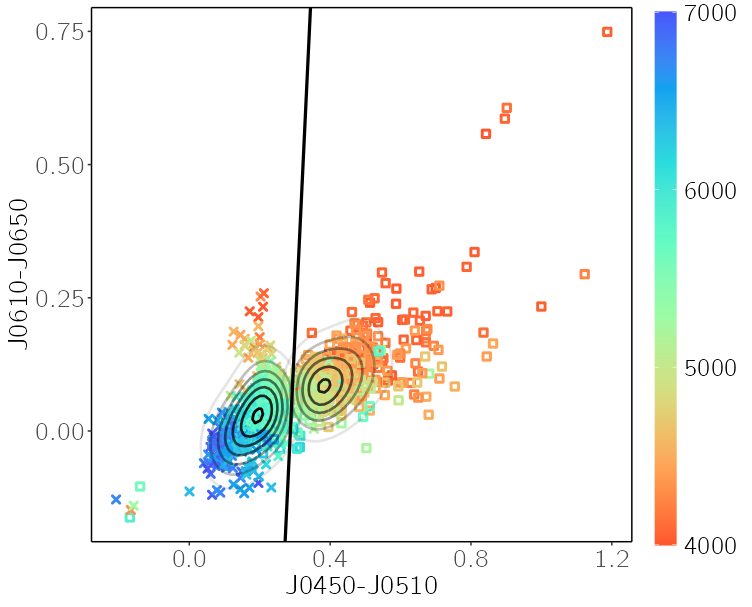}
        \includegraphics[width=0.99\columnwidth]{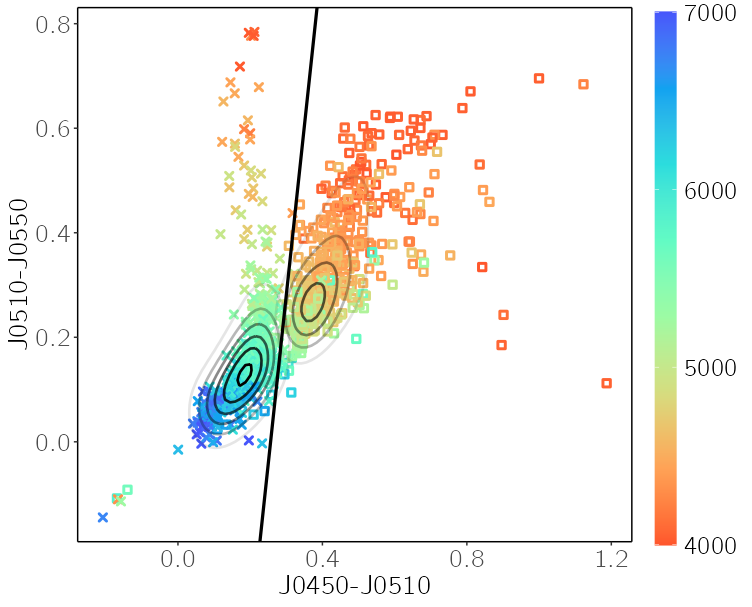}
    \caption{Colour-colour diagrams from J-PAS filters obtained with the GMM method, based on RI values, colour ranges, and expected colour errors. The dwarf stars from spectral libraries -- MILES, STELIB, and ELODIE -- are represented by crosses and giant stars by squares, both colour-coded according to their \teff\  values. The grey curves represent probability density contours across a 2D density plot. The darker the curve is, the higher the density is. The solid lines show the best separation between the two groups according to the SVM solution (see text).}
    \label{fig_coloursGMM}
\end{figure*}

The corresponding colour-colour diagrams are presented in Fig.\,\ref{fig_coloursGMM}, in which all dwarf and giant stars from the sample are presented as crosses and squares, respectively. They are colour-coded according to their \teff\  values to show that the separation between the classes is not caused by their differences in temperature.
In each plot, a 2D kernel density plot is also presented, which helps to visualise the distribution of stars across the 2D space, highlighting regions where there 
is a strong concentration of stars in each sample (zones with darker contours). This representation shows that the higher-probability-density zone of one type of stars is well differentiated from the other's.

Based on the GMM solutions, we were able to segregate a very high percentage of dwarfs and giants using the proposed J-PAS colour pairs. From the GMM results, we defined regions of the colour-colour plane where we expect to find more likely dwarf or giant stars. 
To establish these regions, we adopted the support vector machine \citep[SVM;][]{Cortes95} technique, a method that creates a separation hyperplane between the two samples by maximising the distance between the closest data points from different classes, also known as support vectors.

The SVM gives a hyperplane (or a kernel) as a result for each colour-colour diagram as the best possible separator between dwarfs and giants, according to the training set, the synthetic photometry, and the luminosity class 
from the used libraries. 
We adopted a linear kernel, since more complex functions did not 
achieve a better performance. 
The linear kernels obtained as SVM solutions for the five colour-colour diagrams are: 

\begin{subequations}
\begin{align}
    (J0630-J0650) =& -2.994 +10.388 \cdot (J0450-J0510) \tag{\theequation\,a}\label{eq_CCDa} \\
    (J0620-J0650) =& -3.781 +13.165 \cdot (J0450-J0510) \tag{\theequation\,b} \\
    (J0510-J0550) =& -0.195 +0.591 \cdot (J0450-J0620) \tag{\theequation\,c} \\
    (J0610-J0650) =& -3.995 +13.917 \cdot (J0450-J0510) \tag{\theequation\,d} \\
    (J0510-J0550) =& -1.657 +6.460 \cdot (J0450-J0510).\tag{\theequation\,e}\label{eq_CCDe}
\end{align}
\label{eq:svm_hyp}
\end{subequations}
\noindent
These boundary functions are shown in Fig.\,\ref{fig_coloursGMM} as solid lines.


Although SVM is a robust method in determining classes in different types among samples, the standard approach of dividing the sample into 70\% for training and 30\% for testing can be challenging for small datasets like ours. Thus, we first followed this approach, using the test set to assess that there was no overfitting, and then we trained the SVM model on the full set of 1\,182 spectra to obtain the final solutions. 

\begin{figure*}
\centering
        \includegraphics[width=0.67\columnwidth]{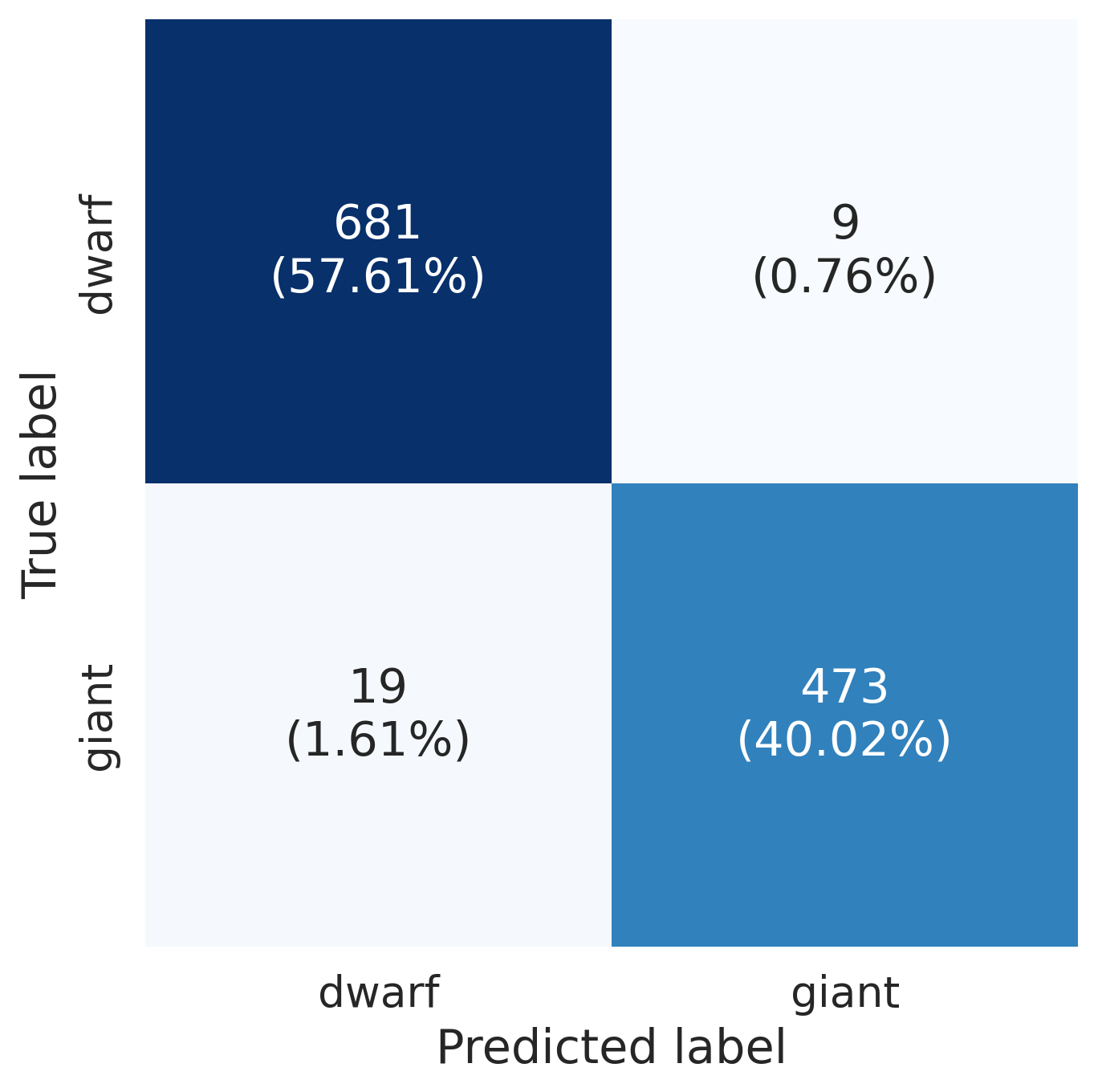}
        \includegraphics[width=0.67\columnwidth]{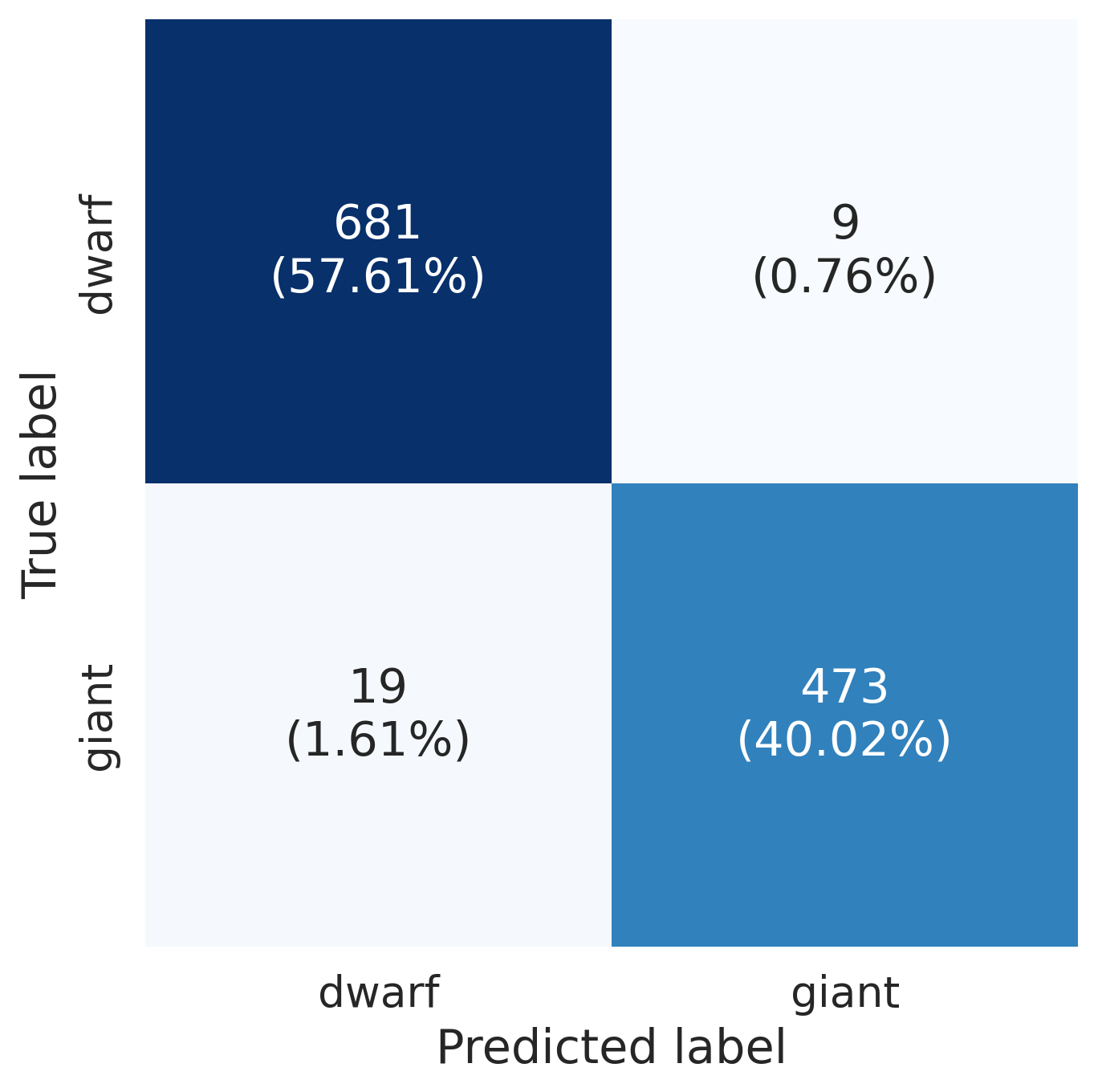}
        \includegraphics[width=0.67\columnwidth]{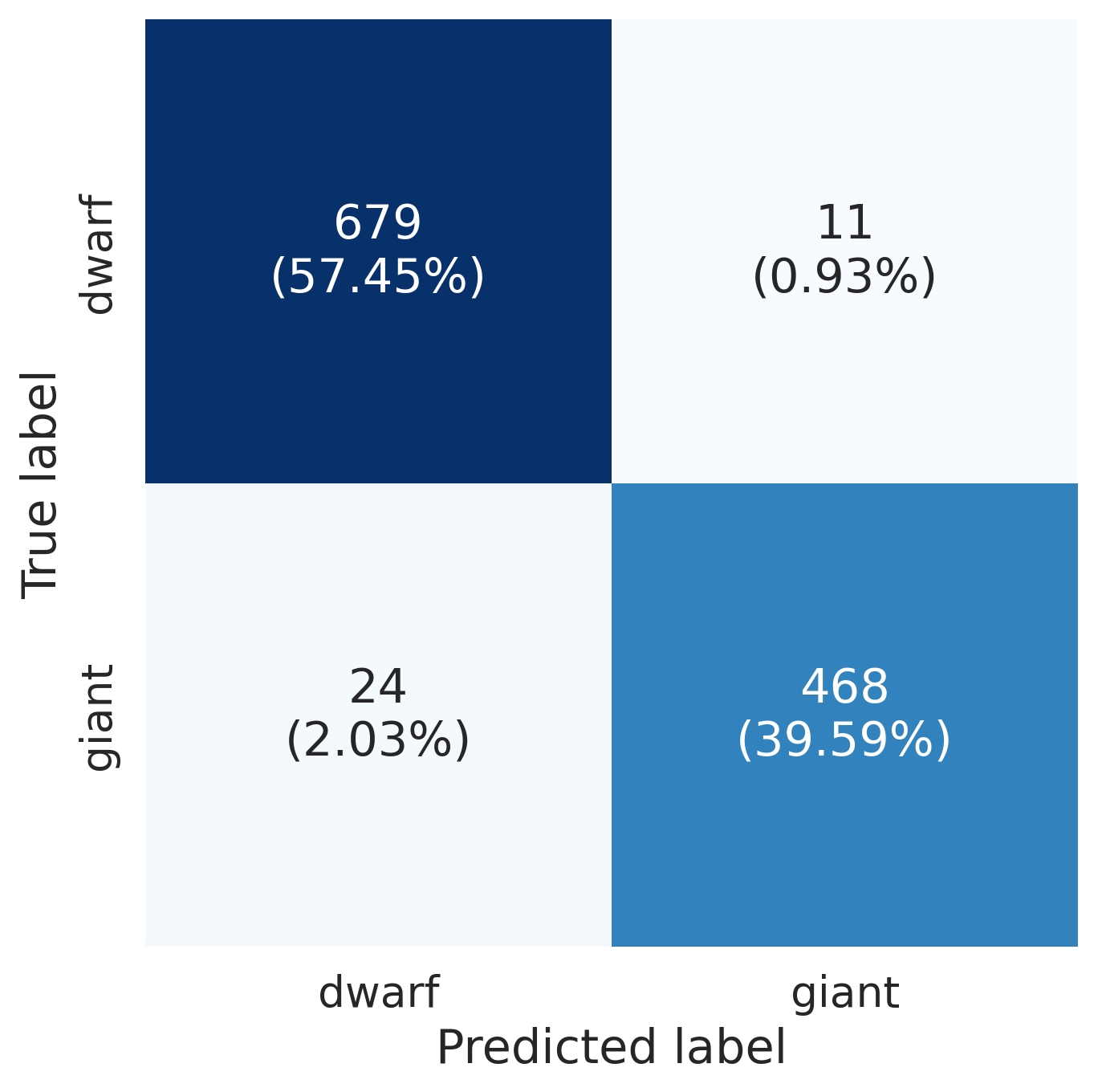}
        \includegraphics[width=0.67\columnwidth]{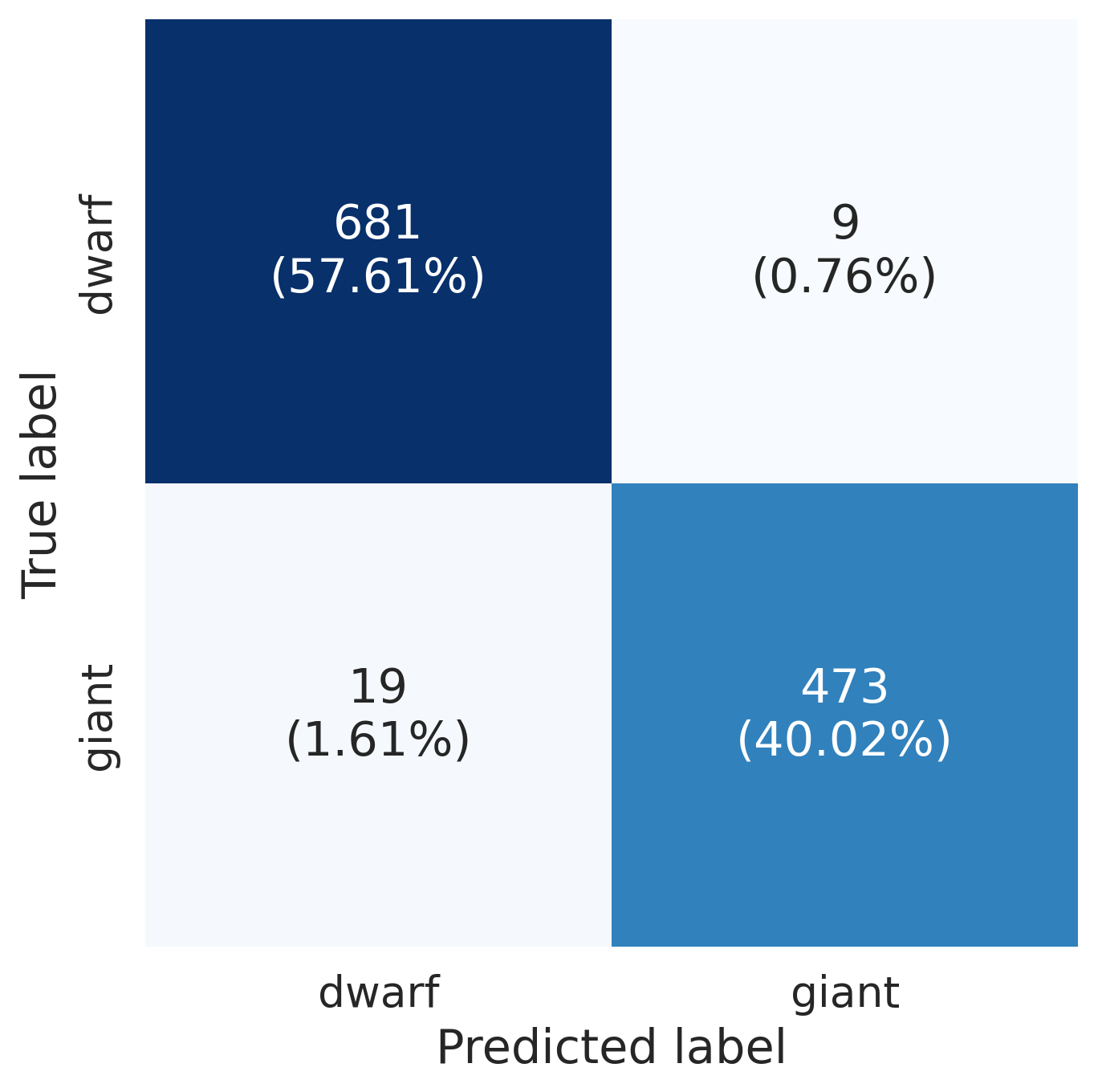}
        \includegraphics[width=0.8\columnwidth]{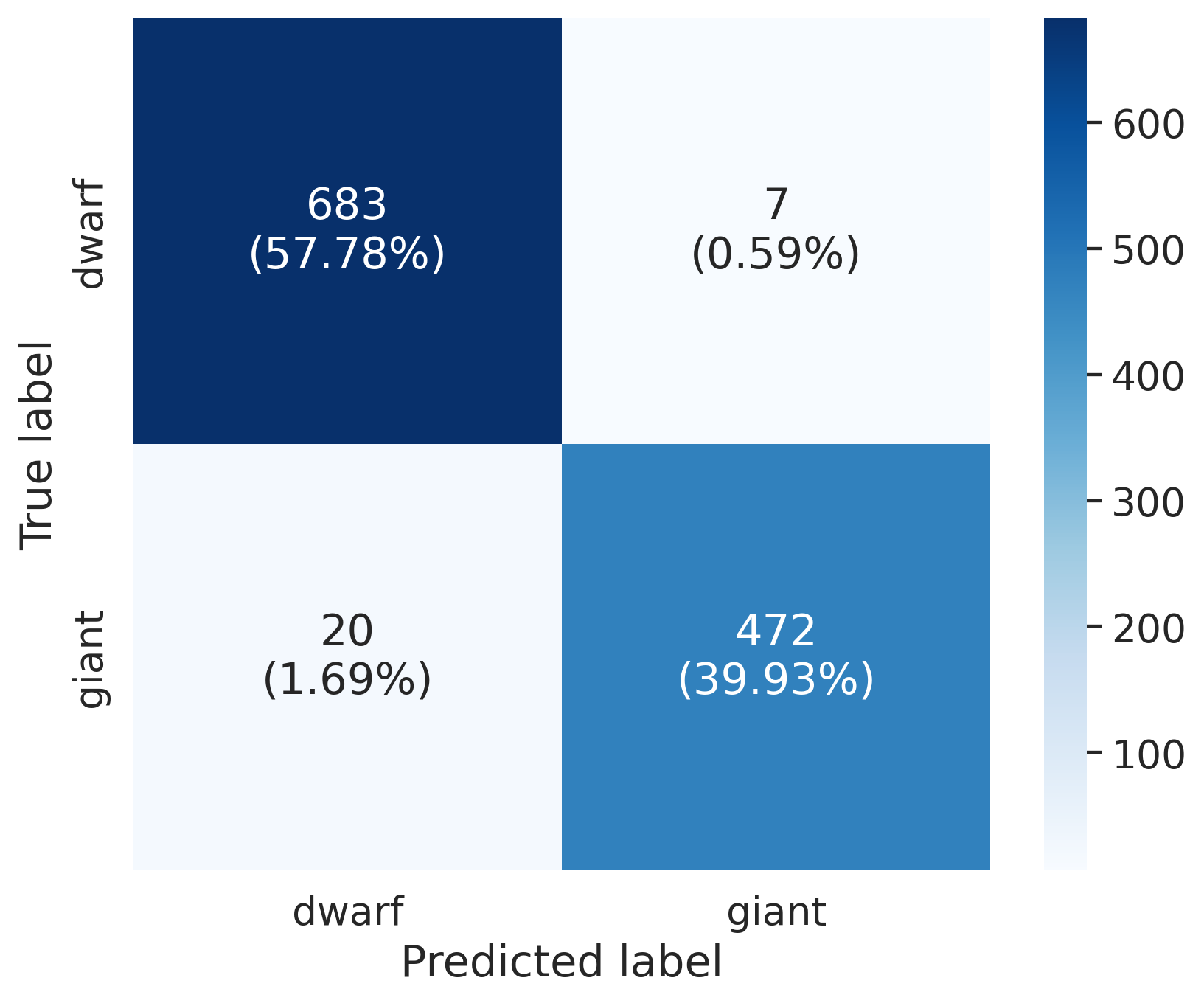}
    \caption{Confusion matrices from the dwarf-giant classification obtained from the SVM solutions for the five colour-colour diagrams presented in Table\,\ref{tab_GMMcolours}, and following the same order shown in the table, from left to right and from top to bottom. 
    The accuracy of our method for each pair of colours is 0.976, 0.976, 0.969, 0.976, and 0.977, respectively.}
    \label{fig_confusion}
\end{figure*}

In order to estimate the accuracy and the amount of contamination in each group, which means the false discovery rate or how many giants were classified as dwarfs and vice versa, we compared the label (dwarf or giant) predicted by the SVM solution to the original label (according to the \logg\  value) available in the stellar libraries. We obtained an accuracy of 0.976, 0.976, 0.969, 0.976, and 0.977 for the five diagrams, respectively, as is shown in the confusion matrices of Fig.\,\ref{fig_confusion}. It is worth mentioning that the accuracy may change for samples containing other luminosity classes than those considered in this work (e.g. subgiants with $3<\log{g}<4$). 
The contamination obtained on average for the five diagrams is only of $\sim$2.9\% of the giants classified as dwarfs, and only $\sim$1.9\% of the dwarfs classified as giants, considering the analysed sample.

As was anticipated by the UMAP in Fig.\,\ref{fig_umap}, it is not possible to find a kernel of complete separation between the two groups across the whole temperature range for FGK stars. 
So, we expect larger contamination at higher temperatures, as this is located in the lower left part of each plot in Fig.\,\ref{fig_coloursGMM}. 
Thus, to further explore this issue, we divided the studied sample into sources with \teff\ higher or lower than 5\,500\,K, and analysed both sub-samples separately. As was expected, the accuracy obtained was smaller for higher \teff\ and larger for lower \teff. Nevertheless, the accuracy was very high in both cases, 0.969 and 0.982 on average for the five J-PAS colour pairs for higher and lower \teff, respectively.

Furthermore, a closer look at the results shows that the rate of true classification depends on the class of star and the temperature range considered. In the case of dwarf stars, they are better classified in the higher temperature range, with just $\sim$0.3\% of misclassifications on average for the five pairs of colours, while this percent increases to $\sim$5.3\% on average for the lower temperature range. More extreme and in the opposite sense is the case of the giant stars. They are better classified in the lower temperature range, as was expected, with only $\sim$0.8\% of misclassifications, while the misclassification rate increases to as high as $\sim$60\% for the higher range of temperatures. It is worth highlighting that these rates were obtained based on the studied sample, which lacks a significant number of giants with higher temperatures and dwarfs with lower temperatures, as is illustrated in Fig.\,\ref{fig_histteff}.

\begin{figure*}
        \includegraphics[width=0.69\columnwidth,trim={0.3cm 0cm 0.1cm 0cm}]{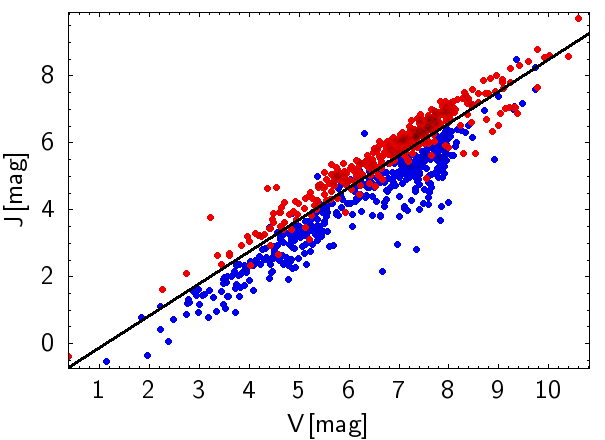}
        \includegraphics[width=0.69\columnwidth,trim={0.3cm 0cm 0.1cm 0cm}]{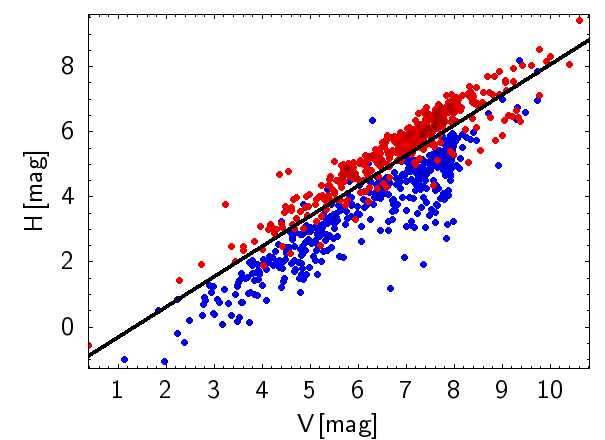} 
    \includegraphics[width=0.69\columnwidth,trim={0.3cm 0cm 0.1cm 0cm}]{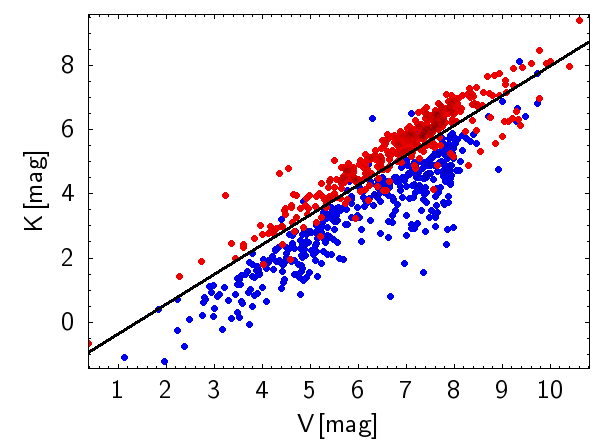}
    \caption{Panels representing the $J$, $H$, and $K$ magnitudes versus the $V$ magnitude for the sample of 437 giants and 637 dwarfs described in Sect.\,\ref{miniJPAS}, with the linear separation found by \citet{Bilir_2006} in each respective diagram. The dwarf  and giant stars are represented by red and blue dots, respectively.}
    \label{fig_comp1}
\end{figure*}

The spectra of a dwarf and a giant star present different features indeed. However, for F and early-G type stars, such discrepancies are more subtle because the majority of metal lines at these temperatures are shallow, and they grow in strength towards lower effective temperatures until mid-K types \citep{Gray09}. Therefore, when considering the integrated flux from a photometric filter (i.e. the synthetic photometry), the photometric colours in both luminosity classes, dwarfs and giants, are very similar, making discriminating between them extremely difficult at this higher \teff\ range. 
Similar results were found by \citet{Casey18}, who achieved a very low rate ($\sim$1\%) of dwarf contamination in giants, but only for temperatures under 5\,250\,K. They remark that gravity sensitivity is lost for giants at higher temperatures, for all metallicities, and cannot separate between classes with their photometry. 

The above results reflect, in fact, a limitation of photometric data, an issue that not even the excellent collection of $\sim$145\,\AA\  narrow-band filters from J-PAS can solve completely. Despite all that, we have been able to achieve a very high accuracy not achieved in previous works with other filter collections, as is further discussed in Sect.\,\ref{discus}.

\section{Discussion}\label{discus}

\subsection{The classical approach: A comparison with broad-band colours}\label{Class-appro}

At this point, it is important to quantify how much the separation between giant and dwarf stars is improved using the narrow J-PAS filters, instead of a collection of broader photometric filters.

In the literature, one can find several works related to this topic. For example, \cite{Huang2019} and \cite{Zhang2021} study different combination of SDSS colours, 
using parallaxes from {\sl Gaia} in the first case and estimating the dependence with metallicity in both cases. They find that the best separation is achieved when the $u$ or $v$ filters are included. Unfortunately, the vast majority of our targets have no accurate SDSS photometry, since they are much brighter than the saturation limit \citep[$r$ $\sim$ 14,][]{Stoughton2002}. As an alternative approach, we attempted to calculate the SDSS colours for the spectra in our sample using the tool \texttt{COLCA}. However, due to the gaps present in the spectra, this approach was unsuccessful. Thus, we were not able to compare our results with those obtained using SDSS colours.

Alternatively, and as we mentioned in the introduction, \cite{Bilir_2006} used 2MASS $J$, $H$, and $K$, and $V$ photometry to search for a methodology that discriminates between dwarf and giants. Since there is a degeneration between the infrared colours of dwarfs and giants, except for the latest types \citep{Majewski_2003}, they combined $J$, $H$, and $K$ with the $V$ band to discriminate between them. They used 196 field dwarfs and 156 field giants from the catalogues of \citet{Cayrel2001}, and defined linear fits to separate between the field dwarfs ($\log{g}>4$) and giants ($\log{g}<3$). Although Fig.\,1 of their paper shows good separation, no quantification of the procedure is given.

\begin{figure*}
\sidecaption
    \includegraphics[width=12cm]{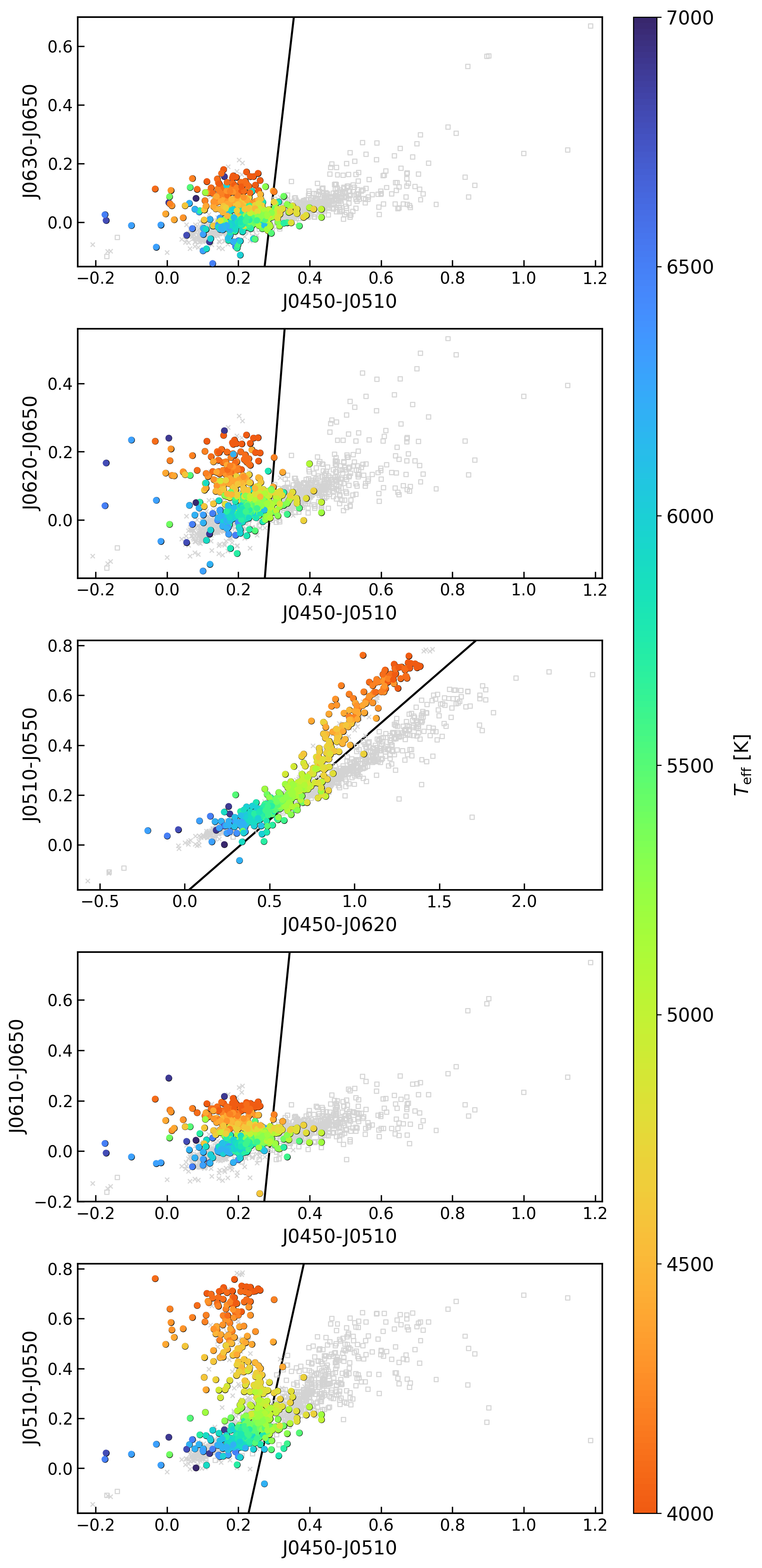}
    \caption{Colour-colour diagrams for 508 stars from the miniJPAS survey classified as FGK stars ($4\,000 \leq T_{\rm eff} \leq 7\,000$\,K), colour-coded according to their derived effective temperature. The comparison sample -- from the MILES, STELIB, and ELODIE libraries -- is shown in grey, and dwarfs and giants are represented by crosses and squares, respectively. The solid lines show the best separation between the two groups according to the SVM solutions.}
    \label{fig_miniJPAS_CMD}
\end{figure*}

In order to compare our results with those published in \cite{Bilir_2006}, we cross-matched our sample with the 2MASS catalogue to obtain the near infrared photometry and with the SIMBAD\footnote{\url{http://simbad.u-strasbg.fr/simbad/}} database \citep{Wenger00} to obtain the available $V$ photometry. In total, we ended up with a sample of 437 giants and 637 dwarfs that have both 2MASS and V photometry. 
Figure~\ref{fig_comp1} reproduce Fig.\,1 from \citet{Bilir_2006}, but for our samples of giants (blue dots) and dwarfs (red dots), 
including their functions (Eqs.\,1-3 of their paper) as a linear separation between dwarfs and giants (black line in each panel). With this method and criteria, we obtained a contamination rate of $\sim$17$-$18\% and $\sim$3$-$5\% for dwarfs and giants, respectively, which is very similar in the three panels and significantly higher than the degree of contamination estimated in Sect.\,\ref{sec_selec} ($\sim$2.9\% for dwarfs and $\sim$1.9\% for giants). 

Therefore, and according to these results, we can certainly conclude that the segregation of dwarf and giant stars is substantially improved when using our methodology based on J-PAS narrow-band filters.

\subsection{Application to miniJPAS stars}\label{miniJPAS}

As was mentioned in Sect.\,\ref{sec_selec}, the miniJPAS survey was the only data collection available with the J-PAS photometric system, and therefore it is the only one that we can use (at the moment) to test our methodology of separating giants and dwarfs using real data (for further details on the miniJPAS survey, see \citealt{Bonoli21}). 

Aiming to have a well-defined sample and minimise any odd behaviour in the measured magnitudes of the miniJPAS stars, we applied some additional criteria to the miniJPAS data, more restrictive than those adopted in Sect.\,\ref{sec_selec}. 
Firstly, we chose those objects that presented a CLASS\_STAR parameter greater than 0.8, to ensure the stellar nature of the miniJPAS sources. We also considered the TOTAL\_PROB\_STAR parameter, which is another morphological classifier designed to distinguish between point-like and extended sources through a Bayesian method \citep[see][for details]{LopezSanjuan19}, to be greater than 0.8, as a second indicator of a point-like source. 
In order to have a good determination of the \teff\  from their SEDs, a good spectral coverage was required. Thus, we selected only those objects with magnitudes and error estimates available in all 56 J-PAS bands. Lastly, we kept those that presented photometric errors smaller than 0.1\,mag in the seven J-PAS filters needed to generate the colour-colour diagrams listed in Table~\ref{tab_GMMcolours}. We ended up with a total sample of 594 miniJPAS stars.

We used \texttt{VOSA} to estimate the effective temperatures of the objects in our sample. For this, we adopted the same steps described in \citet{Yuan23b}, which are outlined as follows. 
For the SED fitting, we used the BT-Settl stellar atmospheric model \citep{Allard11}, which covers \teff\  from 400 to 70\,000\,K, \logg\ from 2.0 to 6.0\,dex, and metallicities from $-$4.0 to $+$0.5\,dex. It is worth mentioning that SEDs usually have a low sensitivity to changes in surface gravity and metallicity, 
which were allowed to vary with the aim of getting better-fitting solutions. This had a minimum effect on the determination of \teff. 
\texttt{VOSA} also corrects the photometric data from interstellar extinction using a known value or a range of values. 
We adopted the same range suggested for the miniJPAS region, $A_v=0$ to $0.1$, according to \citet{Bonoli21}. 
The obtained \teff\  values span from 3\,100 to 9\,000\,K from the analysis of 592 miniJPAS stars. The remaining objects were excluded for presenting unsatisfactory SED fittings; that is, they presented a visual goodness-of-fit, Vgf$_b$,\footnote{The description of the Vgf$_b$ parameter can be found at VOSA's help page at \url{http://svo2.cab.inta-csic.es/theory/vosa/helpw4.php?otype=star&action=help}.} greater than three \citep[for more details, see][]{Yuan23b}. 

We then plotted the stars from the miniJPAS sample, with extinction-corrected magnitudes, on the five colour-colour diagrams explored in Sect.\,\ref{sec_selec} and presented in Table~\ref{tab_GMMcolours}. These diagrams are presented in Fig.\,\ref{fig_miniJPAS_CMD}, 
in which the objects from miniJPAS classified as FGK stars (508 objects; $4\,000 \leq T_{\rm eff} \leq 7\,000$\,K) are colour-coded according to their estimated effective temperature. The control sample (from the spectral libraries) is presented in grey, with dwarf and giant stars represented by crosses and squares, respectively.

By applying the results obtained from SVM in Sect.\,\ref{sec_selec}, we classified the 508 miniJPAS stars as dwarfs or giants according to their locations in the colour-colour diagrams. A classification was given only if it was the same in all five diagrams. 
Thus, we classified 448 sources as dwarfs, 32 as giants, and the remaining 28 as inconclusive (those with a diverging class among the five diagrams). 
Since the training sample was defined to avoid transitional stages, with $3<\log{g}<4$\,dex, 
we expect this to be the case at least for some of these unclassified objects. 
A table with the 508 miniJPAS objects and their classifications, in addition to the parameters obtained from the SED fitting with \texttt{VOSA}, is available online 
at Vizier.

\begin{figure}[!t]
        \includegraphics[width=\columnwidth]{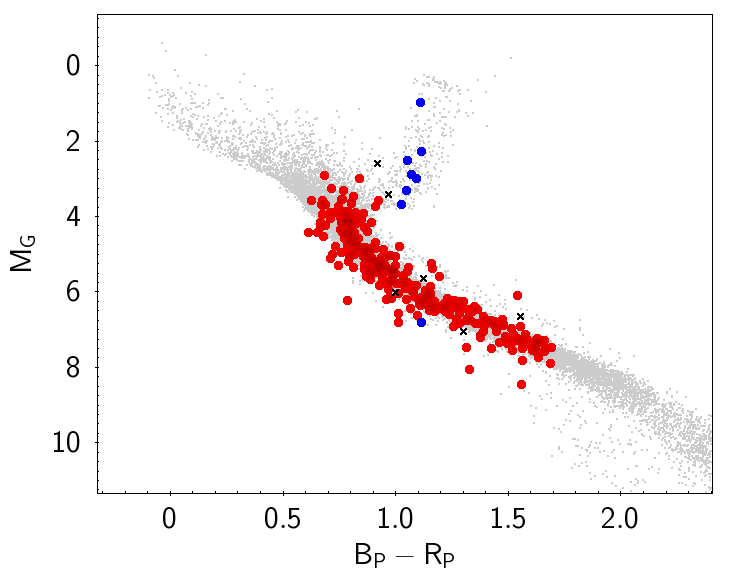}
    \caption{Colour-magnitude diagram showing FGK miniJPAS objects with good parallax, photometry, and RUWE measurements from {\sl Gaia}. Red and blue dots represent stars that were identified as dwarf and giant stars according to our method, respectively, and black crosses are the objects with an inconclusive classification. {\sl Gaia} nearby objects 
    are represented by small dots, for reference.} 
    \label{fig_CMD_FGK}
\end{figure}

To verify our results, we constructed the {\it Gaia} colour-absolute magnitude diagram (CMD) for those objects with known distances from {\it Gaia} DR3. 
Of our sample of 508 objects, only 291 stars have good parallax ($\pi$) measurements (those with positive parallaxes and a relative error of less than 20\%), good astrometric solutions ($\mathrm{RUWE} < 1.4$), 
and {\sl Gaia} $G$, $Bp$, $Rp$ photometry with relative flux errors below 10\%. These objects are shown in the CMD presented in Fig.\,\ref{fig_CMD_FGK}, in which stars defined as dwarfs and giants by our method are represented by red and blue dots, respectively. The black crosses are the objects with an inconclusive classification. A set with over 40\,000 {\sl Gaia} nearby stars taken from \citet{Torres22}, with the quality metric from \citet{Riello21} already applied, is also illustrated for reference (small grey dots). 

From their position on the CMD, we estimated a correct identification of dwarfs and giants with an accuracy of 0.996. This result shows a great improvement in accuracy when compared with previous results in the literature from broad-band photometry, such as those from \citet{Bilir_2006}, as is described in Sect.\,\ref{Class-appro}.

\subsection{A photometric scale for \teff}\label{sec61}

\begin{figure}
        \includegraphics[width=\columnwidth]{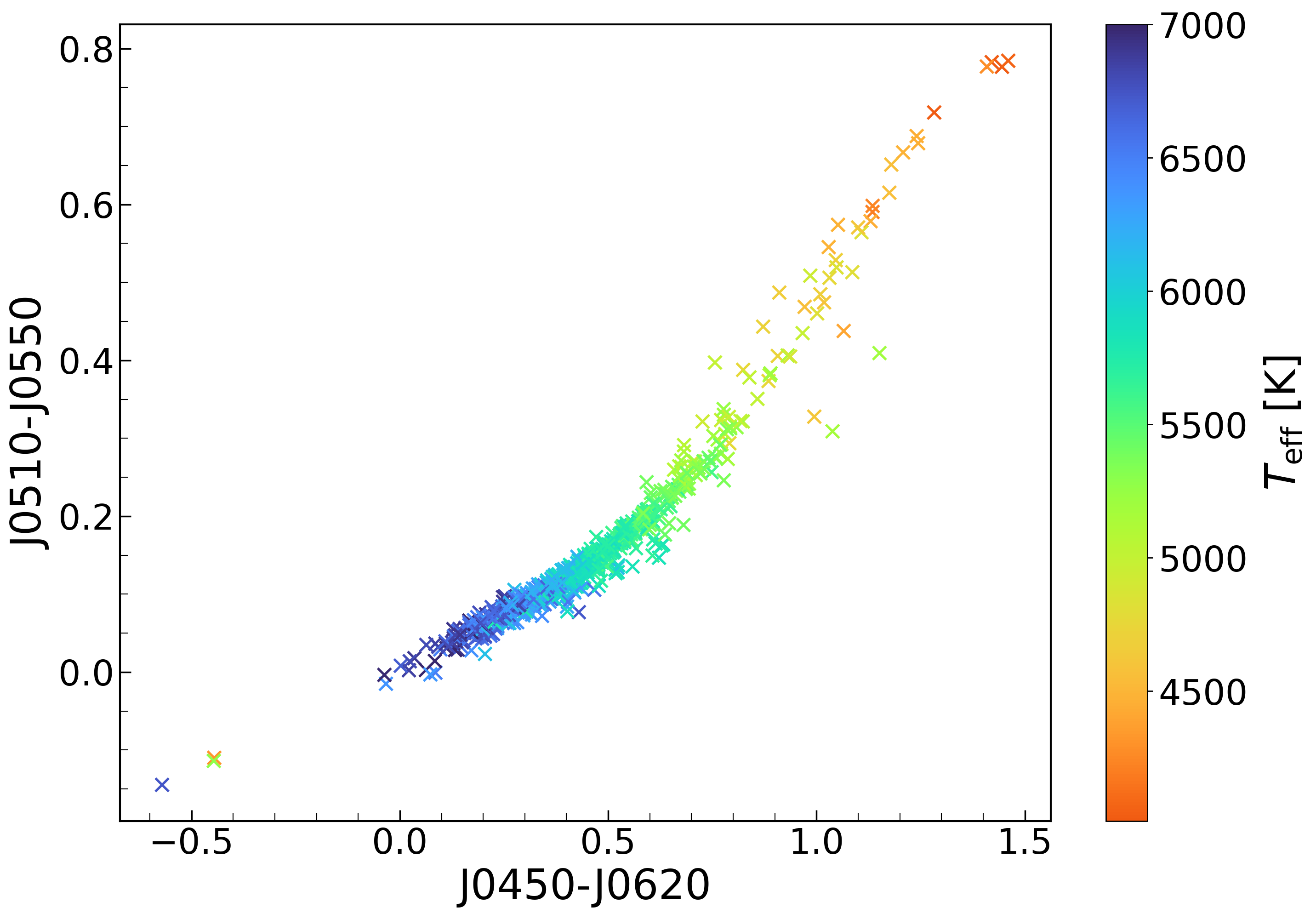}
    \caption{Colour-colour diagram showing our spectroscopic FGK dwarf sample (690 stars with $4\,000 \leq T_{\rm eff} \leq 7\,000$\,K and \logg\  $\ge 4$\,dex). The side bar is colour-coded with the stellar spectroscopic effective temperature obtained from the libraries.}
    \label{fig_CCD3_Teff}
\end{figure}

The reader may have noticed that the colour-colour diagram  ($J0510-J0550$ versus $J0450-J0620$) also presents a well-defined trend with the stellar \teff\  for dwarf stars, as is shown in Fig.\,\ref{fig_CCD3_Teff}. Therefore, we also explored the possibility of defining a mathematical colour-temperature relation that could be used to estimate \teff\ for FGK dwarf stars based on these two J-PAS colours. 
With that aim, we considered all of the FGK dwarfs from our libraries (690 stars with \logg\  $\ge 4$\,dex; Fig.\,\ref{fig_CCD3_Teff}) that have well-established \teff\  from spectroscopy.

\begin{table}
\centering
\caption{Table with polynomial multivariate regression coefficients.}
\begin{tabular}{lc}
\hline \hline
\teff\  range & 4\,000$-$7\,000\,K  \\
 \hline 
$x_1$ & $J0450-J0620$  \\
$x_2$ & $J0510-J0550$  \\
$\beta_0$ & 5916.1 $\pm$ 7.7  \\
$\beta_1$ & $-$8794 $\pm$ 1298 \\
$\beta_2$ & 1013 $\pm$ 779 \\ 
$\beta_3$ & 6976 $\pm$ 422 \\ 
$\beta_4$ & $-$5079 $\pm$ 1285 \\ 
$\beta_5$ & $-$2275 $\pm$ 820 \\ 
$\beta_6$ & $-$2600 $\pm$ 382 \\ 
$\epsilon$ & 204.6 \\
$R^2$ & 0.8742  \\
\hline \hline
\end{tabular}
\label{regressions}
\end{table}

To build the colour-temperature relation, we performed a polynomial multivariate regression of the two colours of the form
\begin{equation}
    y = \beta_0 + \beta_1 x_1^3 + \beta_2 x_1^2 + \beta_3 x_1+ \beta_4 x_2^3 + \beta_5 x_2^2 + \beta_6 x_2
    \label{eq_3}
,\end{equation}

\noindent
where $y$ represents the \teff, $x_1$ and $x_2$ are the colours $J0450-J0620$ and $J0510-J0550$, respectively, and $\beta_i$ are the regression coefficients. 
The model fitting was performed using the ordinary least squares (OLS) method. 
The results are shown in Table\,\ref{regressions}. The residual standard error, $\epsilon$, is $204.6$\,K. 

\begin{figure}
        \includegraphics[width=\columnwidth]{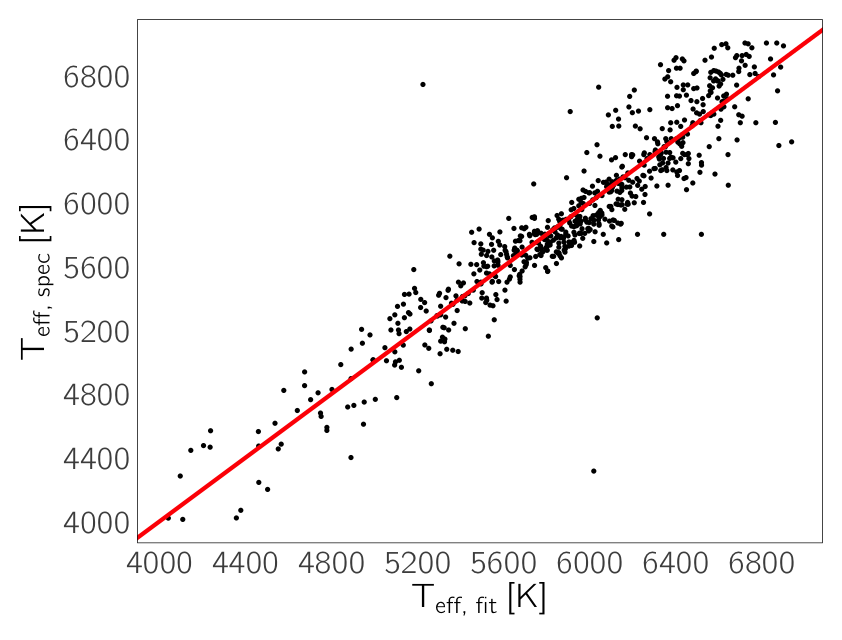}
        \includegraphics[width=\columnwidth]{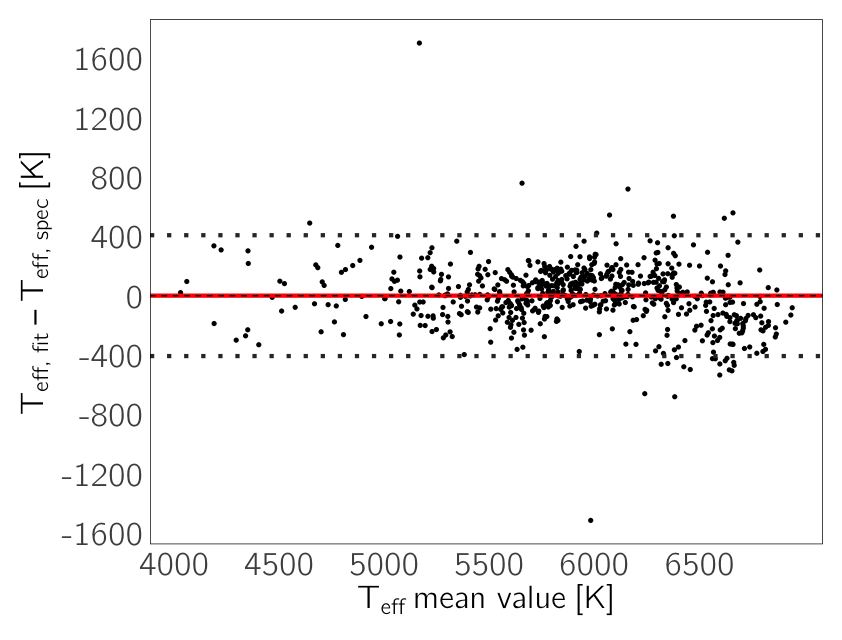}
    \caption{Comparison between photometric and spectroscopic temperatures. {\it Top panel:} The x axis shows the estimated photometric temperatures using the derived polynomial function and the y axis shows the spectroscopic values from libraries. 
    {\it Bottom panel:} The Bland-Altman plot comparing both values, in which the x axis shows the mean values between the two temperatures and the y axis shows the difference between them.}     \label{fig_BAplot_FGK}
\end{figure}

To verify the validity of the fitted polynomial, we used Eq.\,\ref{eq_3} to estimate the photometric temperatures of the objects from our sample and compared them to the spectroscopic values available from the libraries. This comparison is shown in Fig.\,\ref{fig_BAplot_FGK} (top panel), where the solid line represent the identity function. To assess the quality of the estimated values using the polynomial fit, we generated a Bland-Altman plot (Fig.\,\ref{fig_BAplot_FGK}; bottom panel), in which the horizontal axis represents the mean values between the two temperatures and the vertical axis represents the difference between them ($T_{\rm eff,fit}-T_{\rm eff,spec}$). The mean \teff\  difference is $\Delta T_{\rm eff}\sim0$\,K, represented by a horizontal line, which shows that both methods are in accordance, and the standard deviation, $\sigma$, is 203.8\,K. Dotted lines show the confidence limits of the measurements adopted as $2\sigma$ to define a confidence level of $\sim$95\%. Therefore, we can see that both the spectroscopic \teff\  measurements and the values estimated from J-PAS colours are in general in very good agreement.

\subsection{The case of M stars}

\begin{figure}
        \includegraphics[width=\columnwidth]{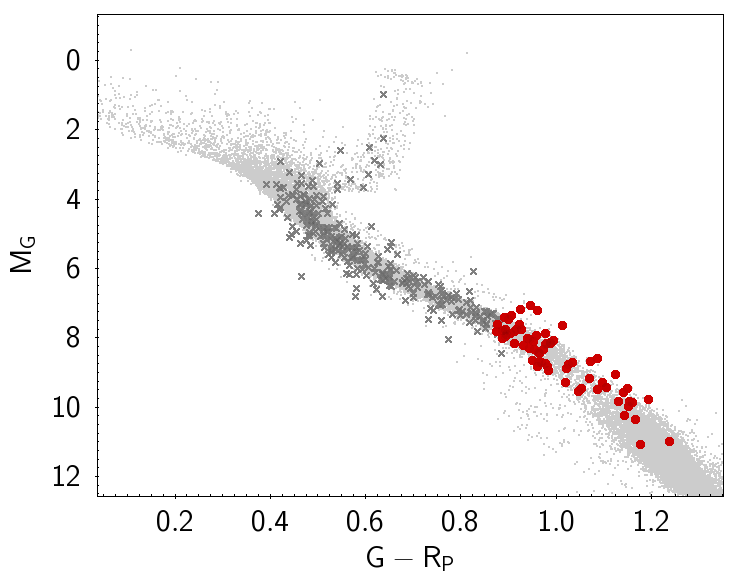}       
    \caption{{\sl Gaia} colour-magnitude diagram showing M stars from the miniJPAS sample (red dots) with good parallax, photometry, and RUWE measurements from {\sl Gaia} (see Sect.\,\ref{miniJPAS} for details). Other miniJPAS stars with high-quality data from {\sl Gaia} are shown as grey crosses. {\sl Gaia} nearby objects 
    (described in Sect.\,\ref{miniJPAS}) are represented by small dots, for reference.}
    \label{fig_CMD_Ms}
\end{figure}

\begin{figure*}
        \includegraphics[width=1\textwidth]{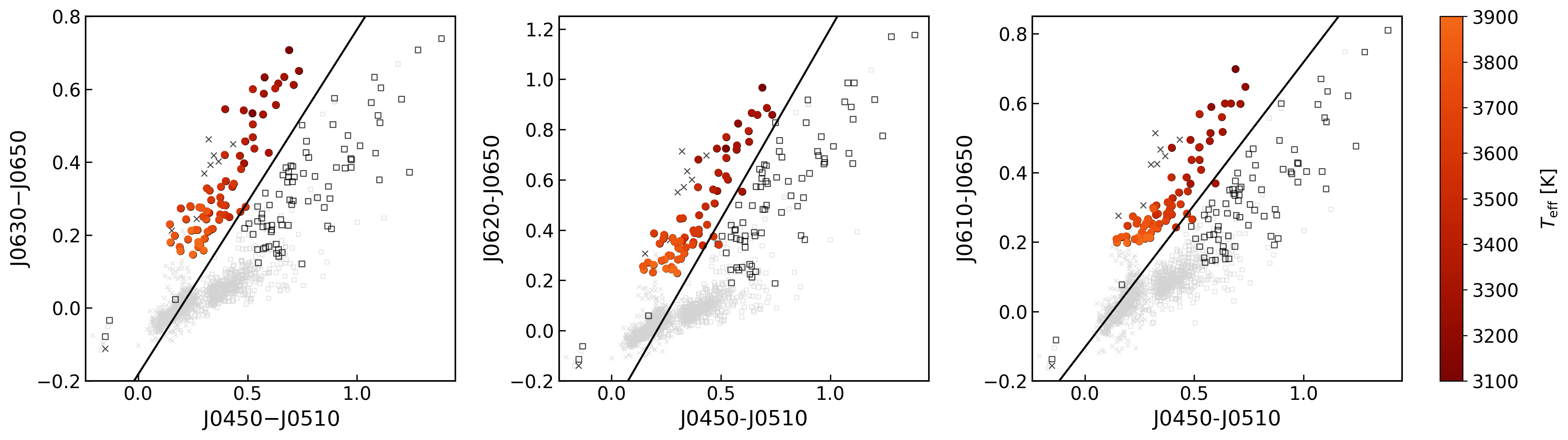}
                
    \caption{ Colour-colour diagrams for M stars ($T_{\rm eff} \leq 3\,900$\,K). The miniJPAS M-dwarf sample is colour-coded according to the derived effective temperature. The M stars from the libraries -- MILES, STELIB, and ELODIE -- are shown in black, and dwarfs and giants are represented by crosses and squares, respectively. The FGK sample from the libraries is presented in light grey, for reference.} 
    \label{fig_miniJPAS_CMD_Ms}
\end{figure*}

\begin{figure*}
        \includegraphics[width=0.64\columnwidth]{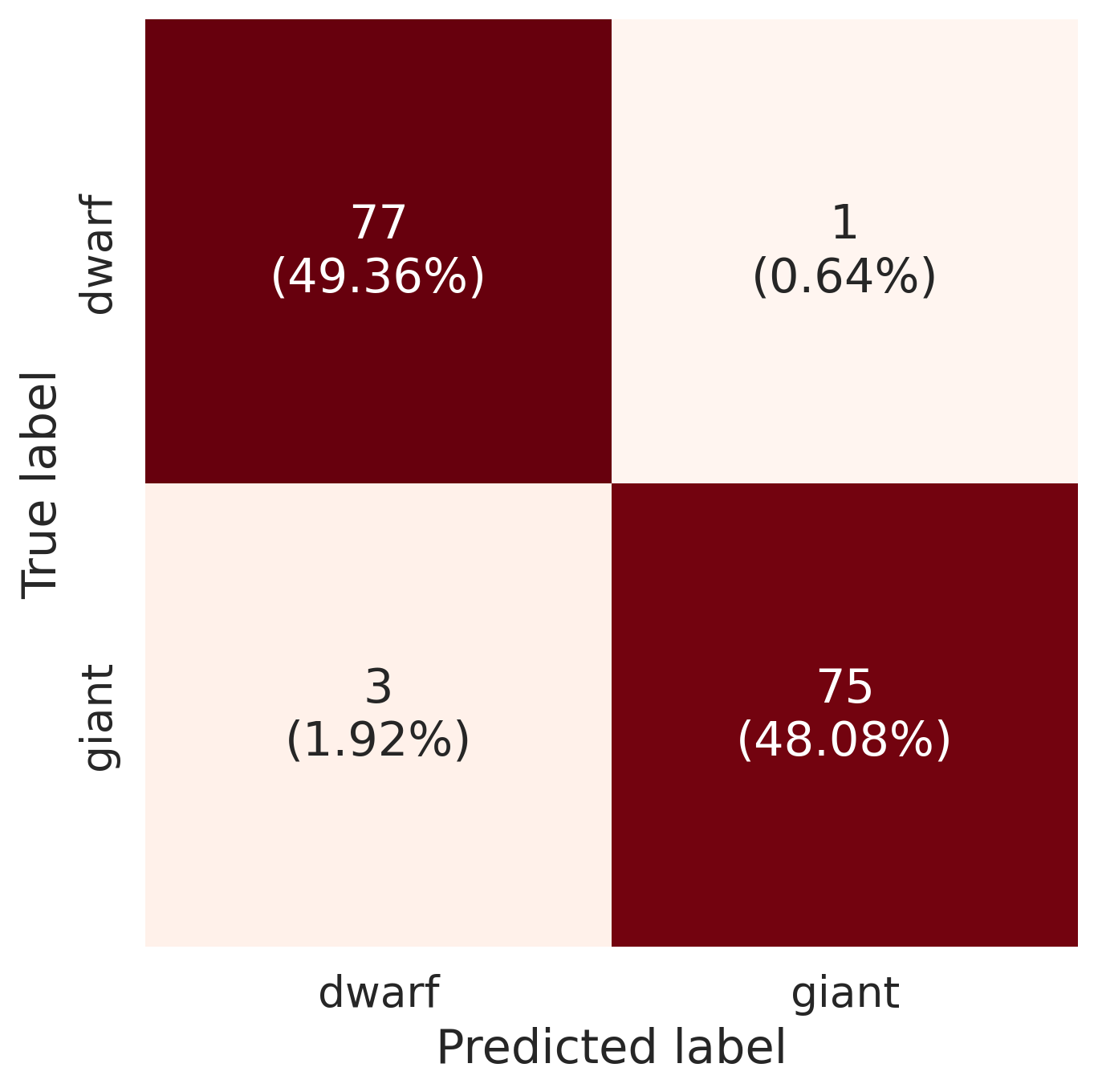}
        \includegraphics[width=0.64\columnwidth]{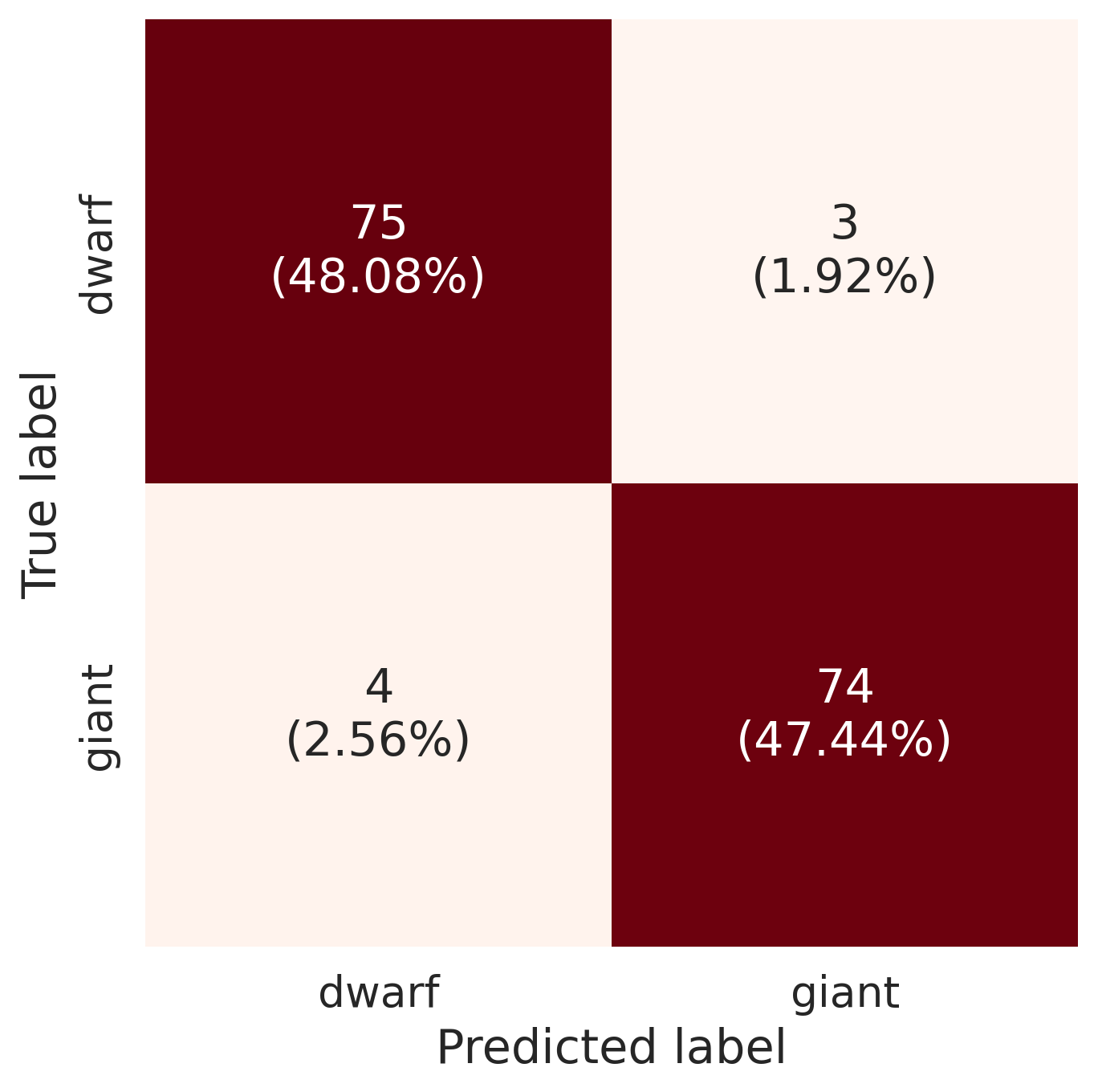}
        \includegraphics[width=0.75\columnwidth]{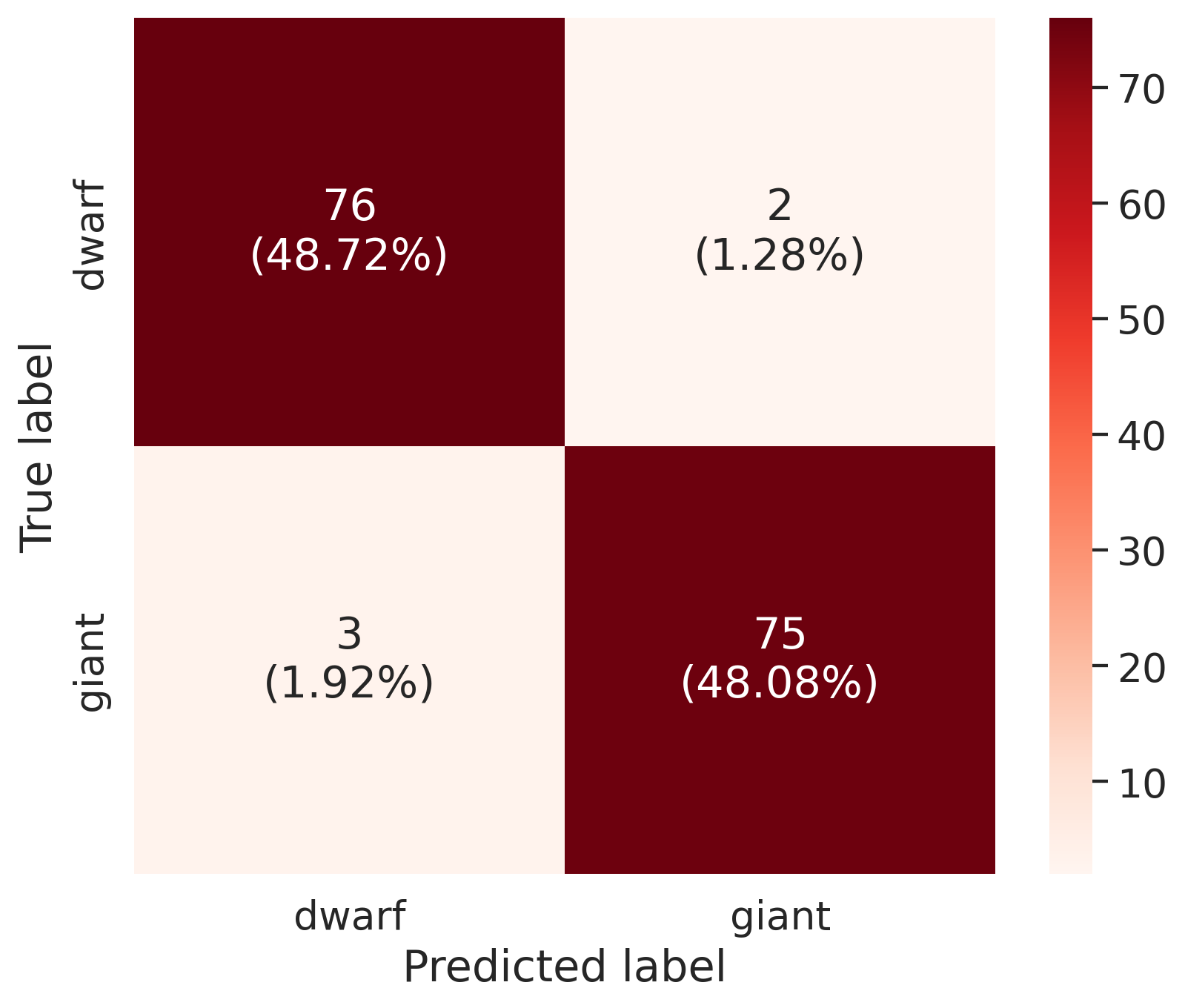}
    \caption{Confusion matrices from the dwarf-giant classification of M stars, obtained from the SVM solutions for the colour-colour diagrams 1, 2, and 4 presented in Table\,\ref{tab_GMMcolours} (from left to right, respectively). The accuracy of correctly identifying dwarfs and giants is 0.974, 0.955, and 0.968, respectively.}
    \label{fig_confusionMs}
\end{figure*}

As the miniJPAS sample has a range of temperatures that extends beyond the K type, we assessed the effectiveness of the colour-colour diagrams presented in Table\,\ref{tab_GMMcolours} at also discriminating between giant and dwarf M stars. To separate dwarf and giant M stars, we applied the same techniques described in Sect.\,\ref{sec_selec}. For that, we selected from the same libraries presented in Sect.\,\ref{libraries} a sample of 87 M stars, those with \teff\ $\le 3\,900$\,K, of which only nine are dwarf stars. In order to enlarge the number of M dwarfs, we selected sources from the miniJPAS survey for which the \teff\ obtained with \texttt{VOSA} (see Sect.\,\ref{miniJPAS}) was also lower than 3\,900\,K. To ensure that these objects are in fact dwarf stars, we plotted them in the {\sl Gaia} colour-magnitude diagram, as is shown in Fig.\,\ref{fig_CMD_Ms}, in which the selected mimiJPAS sources, shown in red, appear in the M dwarf region. The 69 miniJPAS stars depicted in this figure are those with the same quality criteria described in Sect.\,\ref{miniJPAS}, and they were added to the M dwarf sample.

Applying the SVM method to discriminate between dwarf and giant M stars, as was done in Sect.\,\ref{sec_selec}, it is interesting to note how three of the J-PAS colour pairs used to classify FGK stars can also be used for M stars. In this case, diagrams 1, 2, and 4 proved to be very useful, as shown in Fig.\,\ref{fig_miniJPAS_CMD_Ms}. The solutions obtained from the SVM are 
 represented by the following linear decision boundary functions:
\begin{subequations}
\begin{align}
    (J0630-J0650) =& -0.182 + 0.946 \cdot (J0450-J0510) \tag{\theequation\,a}\label{eq_CCDMa} \\
    (J0620-J0650) =& -0.314 +1.516 \cdot (J0450-J0510) \tag{\theequation\,b} \\
    (J0610-J0650) =& -0.104 +0.822 \cdot (J0450-J0510). \tag{\theequation\,c}\label{eq_CCDMc} 
\end{align}
\end{subequations}

\noindent
They are shown in Fig.\,\ref{fig_miniJPAS_CMD_Ms} as solid lines. 

From the SVM solutions obtained for M stars, we achieved an accuracy of 0.974, 0.955, and 0.968 for the three diagrams shown in Fig.\,\ref{fig_miniJPAS_CMD_Ms}, respectively, as is shown by their confusion matrices of Fig.\,\ref{fig_confusionMs}. Considering the three colour-colour diagrams, we estimated an average contamination of only $\sim$4.2\% of giants classified as dwarfs and $\sim$2.6\% of dwarfs classified as giants. It is worth emphasising that these results are based on a smaller sample than the previous one, with a total of 78 dwarfs and 78 giant stars, and also a more heterogeneous one, as it combines spectroscopic and photometric data. Nonetheless, although further validation is required, the results are very promising given its high accuracy. This validation will be possible in the near future with the upcoming J-PAS survey data releases.

\section{Conclusions}\label{conclusions}

This paper has two main goals. On the one hand, it aims to present the SVO FPS and 
the associated photometric tools that allow, in the framework of the VO, the combination and comparison of heterogeneous information coming from different photometric systems. 
On the other hand, the other natural objective is to show the power and capabilities of this service and tools by means of addressing one important issue in today's stellar astrophysics: the discrimination of dwarf and giant stars using photometry data.

We used 15 J-PAS narrow filters, especially selected to cover spectral features that are known to be sensitive to gravity, to develop a methodology of discriminating between dwarf and giant stars, limited to the F, G, and K spectral types. For this purpose, we selected a sample of known stars from the MILES, STELIB, and ELODIE stellar libraries, and used the SVO photometric tools to derive the J-PAS synthetic photometry from their spectra. 

We first selected the 15 J-PAS filters more suitable for our purposes, which resulted in 105 J-PAS colours with a larger probability of separating both types of stars. We then conducted a machine-learning approach to identify the colour-colour pairs adequate for the dwarf-giant segregation among the total of 5\,460 possible pairs generated. 
  
Then, using a Gaussian mixture model, we selected the most promising five colour-colour diagrams to discriminate between giant and dwarf stars. 
Finally, we used the support vector machine technique to define a criteria that maximises the separation of the two luminosity classes, which performs excellently, with an average accuracy of approximately 0.97 for FGK stars. 
We notice, though, that the rate of accurate classification depends on the type of star and the range of temperature, being better for dwarf stars in the higher temperature range (\teff\ $>$5\,500\,K) and better for giants in the lower temperature range, according to the adopted sample. The effectiveness of our methodology was demonstrated 
by comparing the results obtained from broad-band photometry to those obtained using J-PAS filters. 

Additionally, we defined a colour-temperature relation for the FGK dwarf stars that allows one to estimate the effective temperature of a star with an error of $\sim$200\,K from the photometric data of four J-PAS filters ($J0450$, $J0510$, $J0550$, and $J0620$). 
Finally, we extended our method of distinguishing between FGK giants and dwarfs to M-type stars with very promising results, also achieving an average accuracy of $\sim$0.97. 
Nevertheless, a validation with a larger sample of both dwarf and giant M stars is needed to draw definitive conclusions.

It has definitely been demonstrated that the publicly available services provided by the SVO and presented in this work are powerful tools for research in astrophysics. The capacity of these services to work with either observational or theoretical data enhances their usefulness and power. The science case presented in this work is just one drop in an ocean of possibilities, the only limit of which is the creativity of the researcher. These kinds of services are foreseen to play a fundamental role in the current era of huge 
photometric and astrometric surveys.

\begin{acknowledgements}

%
P.C. and A.A. acknowledge financial support from the Government of Comunidad Autónoma de Madrid (Spain) via postdoctoral grants `Atracción de Talento Investigador' 2019-T2/TIC-14760 and 2018-T2/TIC-11697, respectively. 
P.C. and F.J.E. acknowledge financial support from the Spanish Virtual Observatory project funded by the Spanish Ministry of Science and Innovation/State Agency of Research MCIN/AEI/10.13039/501100011033 through grant PID2020-112949GB-I00. 
J.A.'s contribution to this work is a product of his academic exercise as a professor at the Universidad Militar Nueva Granada, Bogotá, Colombia. 
M.C.G.O acknowledges financial support from the Agencia Estatal de Investigaci\'on (AEI/10.13039/501100011033) of the Ministerio de Ciencia e Innovaci\'on and the ERDF ``A way of making Europe'' through project PID2022-137241NB-C42. 
P.M.B. is funded by INTA through grant PRE-OVE.
A.B. acknowledges support from the Deutsche Forschungsgemeinschaft (Germany's Excellence Strategy – EXC 2094 – 39078331). 
M.C.C acknowledges financial support from the Universidad Complutense de Madrid (UCM) and the Agencia Estatal de Investigaci\'on (AEI/10.13039/501100011033) of the Ministerio de Ciencia e Innovaci\'on and the ERDF ``A way of making Europe'' through projects PID2019-109522GB-C5[4] and PID2022-137241NBC4[4]. 
R.M.O. is funded by INTA through grant PRE-OBSERVATORIO. 
This research has made use of the Spanish Virtual Observatory (\url{https://svo.cab.inta-csic.es}) project funded by the Spanish Ministry of Science and Innovation/State Agency of Research MCIN/AEI/10.13039/501100011033 through grant PID2020-112949GB-I00. 
%
This research has made use of the SVO Filter Profile Service (\url{http://svo2.cab.inta-csic.es/theory/fps/}). 
This publication makes use of VOSA, developed under the Spanish Virtual Observatory project. \texttt{VOSA} has been partially updated by using funding from the European Union's Horizon 2020 Research and Innovation Programme, under Grant Agreement nº 776403 (EXOPLANETS-A). 
This research has made use of TOPCAT \citep{Taylor05}. 
This research has made use of the SIMBAD and VizieR databases, operated by CDS at Strasbourg, France.
We made use of several Python packages, including \texttt{pandas}\footnote{\url{https://github.com/pandas-dev/pandas}}, \texttt{seaborn} \citep{seaborn}, \texttt{numpy} \citep{numpy}, \texttt{matplotlib}, and \texttt{umap-learn} \citep{mcinnes18}.
%
Based on observations made with the JST/T250 telescope and JPCam at the Observatorio Astrofísico de Javalambre (OAJ), in Teruel, owned, managed, and operated by the Centro de Estudios de Física del Cosmos de Aragón (CEFCA). We acknowledge the OAJ Data Processing and Archiving Unit (UPAD) for reducing and calibrating the OAJ data used in this work. Funding for the J-PAS Project has been provided by the Governments of Spain and Aragón through the Fondo de Inversión de Teruel, European FEDER funding and the Spanish Ministry of Science, Innovation and Universities, and by the Brazilian agencies FINEP, FAPESP, FAPERJ and by the National Observatory of Brazil. Additional funding was also provided by the Tartu Observatory and by the J-PAS Chinese Astronomical Consortium. Funding for OAJ, UPAD, and CEFCA has been provided by the Governments of Spain and Arag\'on through the Fondo de Inversiones de Teruel and their general budgets; the Aragonese Government through the Research Groups E96, E103, E16\_17R, E16\_20R and E16\_23R; the Spanish Ministry of Science and Innovation (MCIN/AEI/10.13039/501100011033 y FEDER, Una manera de hacer Europa) with grants PID2021-124918NB-C41, PID2021-124918NB-C42, PID2021-124918NA-C43, and PID2021-124918NB-C44; the Spanish Ministry of Science, Innovation and Universities (MCIU/AEI/FEDER, UE) with grant PGC2018-097585-B-C21; the Spanish Ministry of Economy and Competitiveness (MINECO) under AYA2015-66211-C2-1-P, AYA2015-66211-C2-2, AYA2012-30789, and ICTS-2009-14; and European FEDER funding (FCDD10-4E-867, FCDD13-4E-2685).

\end{acknowledgements}

%
%



\bibliographystyle{aa}
\bibliography{FPS_erratum}

\begin{thebibliography}{66}
\expandafter\ifx\csname natexlab\endcsname\relax\def\natexlab#1{#1}\fi

\bibitem[{{Adams} \& {Joy}(1917)}]{Adams1917}
{Adams}, W.~S. \& {Joy}, A.~H. 1917, \apj, 46, 313

\bibitem[{{Allard} {et~al.}(2011){Allard}, {Homeier}, \& {Freytag}}]{Allard11}
{Allard}, F., {Homeier}, D., \& {Freytag}, B. 2011, in Astronomical Society of the Pacific Conference Series, Vol. 448, 16th Cambridge Workshop on Cool Stars, Stellar Systems, and the Sun, ed. C.~{Johns-Krull}, M.~K. {Browning}, \& A.~A. {West}, 91

\bibitem[{{Bayo} {et~al.}(2008){Bayo}, {Rodrigo}, {Barrado Y Navascu{\'e}s}, {Solano}, {Guti{\'e}rrez}, {Morales-Calder{\'o}n}, \& {Allard}}]{Bayo08}
{Bayo}, A., {Rodrigo}, C., {Barrado Y Navascu{\'e}s}, D., {et~al.} 2008, \aap, 492, 277

\bibitem[{{Benitez} {et~al.}(2014){Benitez}, {Dupke}, {Moles}, {Sodre}, {Cenarro}, {Marin-Franch}, {Taylor}, {Cristobal}, {Fernandez-Soto}, {Mendes de Oliveira}, {Cepa-Nogue}, {Abramo}, {Alcaniz}, {Overzier}, {Hernandez-Monteagudo}, {Alfaro}, {Kanaan}, {Carvano}, {Reis}, {Martinez Gonzalez}, {Ascaso}, {Ballesteros}, {Xavier}, {Varela}, {Ederoclite}, {Vazquez Ramio}, {Broadhurst}, {Cypriano}, {Angulo}, {Diego}, {Zandivarez}, {Diaz}, {Melchior}, {Umetsu}, {Spinelli}, {Zitrin}, {Coe}, {Yepes}, {Vielva}, {Sahni}, {Marcos-Caballero}, {Kitaura}, {Maroto}, {Masip}, {Tsujikawa}, {Carneiro}, {Gonzalez Nuevo}, {Carvalho}, {Reboucas}, {Carvalho}, {Abdalla}, {Bernui}, {Pigozzo}, {Ferreira}, {Chandrachani Devi}, {Bengaly}, {Campista}, {Amorim}, {Asari}, {Bongiovanni}, {Bonoli}, {Bruzual}, {Cardiel}, {Cava}, {Cid Fernandes}, {Coelho}, {Cortesi}, {Delgado}, {Diaz Garcia}, {Espinosa}, {Galliano}, {Gonzalez-Serrano}, {Falcon-Barroso}, {Fritz}, {Fernandes}, {Gorgas}, {Hoyos}, {Jimenez-Teja}, {Lopez-Aguerri}, {Lopez-San Juan},
  {Mateus}, {Molino}, {Novais}, {OMill}, {Oteo}, {Perez-Gonzalez}, {Poggianti}, {Proctor}, {Ricciardelli}, {Sanchez-Blazquez}, {Storchi-Bergmann}, {Telles}, {Schoennell}, {Trujillo}, {Vazdekis}, {Viironen}, {Daflon}, {Aparicio-Villegas}, {Rocha}, {Ribeiro}, {Borges}, {Martins}, {Marcolino}, {Martinez-Delgado}, {Perez-Torres}, {Siffert}, {Calvao}, {Sako}, {Kessler}, {Alvarez-Candal}, {De Pra}, {Roig}, {Lazzaro}, {Gorosabel}, {Lopes de Oliveira}, {Lima-Neto}, {Irwin}, {Liu}, {Alvarez}, {Balmes}, {Chueca}, {Costa-Duarte}, {da Costa}, {Dantas}, {Diaz}, {Fabregat}, {Ferrari}, {Gavela}, {Gracia}, {Gruel}, {Gutierrez}, {Guzman}, {Hernandez-Fernandez}, {Herranz}, {Hurtado-Gil}, {Jablonsky}, {Laporte}, {Le Tiran}, {Licandro}, {Lima}, {Martin}, {Martinez}, {Montero}, {Penteado}, {Pereira}, {Peris}, {Quilis}, {Sanchez-Portal}, {Soja}, {Solano}, {Torra}, \& {Valdivielso}}]{Benitez14}
{Benitez}, N., {Dupke}, R., {Moles}, M., {et~al.} 2014, arXiv e-prints, arXiv:1403.5237

\bibitem[{{Bertin} \& {Arnouts}(1996)}]{Bertin96}
{Bertin}, E. \& {Arnouts}, S. 1996, \aaps, 117, 393

\bibitem[{{Bilir} {et~al.}(2006{\natexlab{a}}){Bilir}, {G{\"u}ver}, \& {Aslan}}]{Bilir_2006b}
{Bilir}, S., {G{\"u}ver}, T., \& {Aslan}, M. 2006{\natexlab{a}}, Astronomische Nachrichten, 327, 693

\bibitem[{{Bilir} {et~al.}(2006{\natexlab{b}}){Bilir}, {Karaali}, {G{\"u}ver}, {Karata{\c{s}}}, \& {Ak}}]{Bilir_2006}
{Bilir}, S., {Karaali}, S., {G{\"u}ver}, T., {Karata{\c{s}}}, Y., \& {Ak}, S.~G. 2006{\natexlab{b}}, Astronomische Nachrichten, 327, 72

\bibitem[{{Bohlin} {et~al.}(2014){Bohlin}, {Gordon}, \& {Tremblay}}]{Bohlin14}
{Bohlin}, R.~C., {Gordon}, K.~D., \& {Tremblay}, P.~E. 2014, \pasp, 126, 711

\bibitem[{{Bonoli} {et~al.}(2021){Bonoli}, {Mar{\'\i}n-Franch}, {Varela}, {V{\'a}zquez Rami{\'o}}, {Abramo}, {Cenarro}, {Dupke}, {V{\'\i}lchez}, {Crist{\'o}bal-Hornillos}, {Gonz{\'a}lez Delgado}, {Hern{\'a}ndez-Monteagudo}, {L{\'o}pez-Sanjuan}, {Muniesa}, {Civera}, {Ederoclite}, {Hern{\'a}n-Caballero}, {Marra}, {Baqui}, {Cortesi}, {Cypriano}, {Daflon}, {de Amorim}, {D{\'\i}az-Garc{\'\i}a}, {Diego}, {Mart{\'\i}nez-Solaeche}, {P{\'e}rez}, {Placco}, {Prada}, {Queiroz}, {Alcaniz}, {Alvarez-Candal}, {Cepa}, {Maroto}, {Roig}, {Siffert}, {Taylor}, {Benitez}, {Moles}, {Sodr{\'e}}, {Carneiro}, {Mendes de Oliveira}, {Abdalla}, {Angulo}, {Aparicio Resco}, {Balaguera-Antol{\'\i}nez}, {Ballesteros}, {Brito-Silva}, {Broadhurst}, {Carrasco}, {Castro}, {Cid Fernandes}, {Coelho}, {de Melo}, {Doubrawa}, {Fernandez-Soto}, {Ferrari}, {Finoguenov}, {Garc{\'\i}a-Benito}, {Iglesias-P{\'a}ramo}, {Jim{\'e}nez-Teja}, {Kitaura}, {Laur}, {Lopes}, {Lucatelli}, {Mart{\'\i}nez}, {Maturi}, {Overzier}, {Pigozzo}, {Quartin},
  {Rodr{\'\i}guez-Mart{\'\i}n}, {Salzano}, {Tamm}, {Tempel}, {Umetsu}, {Valdivielso}, {von Marttens}, {Zitrin}, {D{\'\i}az-Mart{\'\i}n}, {L{\'o}pez-Alegre}, {L{\'o}pez-Sainz}, {Yanes-D{\'\i}az}, {Rueda-Teruel}, {Rueda-Teruel}, {Abril Iba{\~n}ez}, {L Ant{\'o}n Bravo}, {Bello Ferrer}, {Bielsa}, {Casino}, {Castillo}, {Chueca}, {Cuesta}, {Garzar{\'a}n Calderaro}, {Iglesias-Marzoa}, {{\'I}niguez}, {Lamadrid Gutierrez}, {Lopez-Martinez}, {Lozano-P{\'e}rez}, {Ma{\'\i}cas Sacrist{\'a}n}, {Molina-Ib{\'a}{\~n}ez}, {Moreno-Signes}, {Rodr{\'\i}guez Llano}, {Royo Navarro}, {Tilve Rua}, {Andrade}, {Alfaro}, {Akras}, {Arnalte-Mur}, {Ascaso}, {Barbosa}, {Beltr{\'a}n Jim{\'e}nez}, {Benetti}, {Bengaly}, {Bernui}, {Blanco-Pillado}, {Borges Fernandes}, {Bregman}, {Bruzual}, {Calderone}, {Carvano}, {Casarini}, {Chaves-Montero}, {Chies-Santos}, {Coutinho de Carvalho}, {Dimauro}, {Duarte Puertas}, {Figueruelo}, {Gonz{\'a}lez-Serrano}, {Guerrero}, {Gurung-L{\'o}pez}, {Herranz}, {Huertas-Company}, {Irwin}, {Izquierdo-Villalba},
  {Kanaan}, {Kehrig}, {Kirkpatrick}, {Lim}, {Lopes}, {Lopes de Oliveira}, {Marcos-Caballero}, {Mart{\'\i}nez-Delgado}, {Mart{\'\i}nez-Gonz{\'a}lez}, {Mart{\'\i}nez-Somonte}, {Oliveira}, {Orsi}, {Penna-Lima}, {Reis}, {Spinoso}, {Tsujikawa}, {Vielva}, {Vitorelli}, {Xia}, {Yuan}, {Arroyo-Polonio}, {Dantas}, {Galarza}, {Gon{\c{c}}alves}, {Gon{\c{c}}alves}, {Gonzalez}, {Gonzalez}, {Greisel}, {Jim{\'e}nez-Esteban}, {Landim}, {Lazzaro}, {Magris}, {Monteiro-Oliveira}, {Pereira}, {Rebou{\c{c}}as}, {Rodriguez-Espinosa}, {Santos da Costa}, \& {Telles}}]{Bonoli21}
{Bonoli}, S., {Mar{\'\i}n-Franch}, A., {Varela}, J., {et~al.} 2021, \aap, 653, A31

\bibitem[{{Casey} {et~al.}(2018){Casey}, {Kennedy}, {Hartle}, \& {Schlaufman}}]{Casey18}
{Casey}, A.~R., {Kennedy}, G.~M., {Hartle}, T.~R., \& {Schlaufman}, K.~C. 2018, \mnras, 478, 2812

\bibitem[{{Cayrel de Strobel} {et~al.}(2001){Cayrel de Strobel}, {Soubiran}, \& {Ralite}}]{Cayrel2001}
{Cayrel de Strobel}, G., {Soubiran}, C., \& {Ralite}, N. 2001, A\&A, 373, 159

\bibitem[{{Cenarro} {et~al.}(2007){Cenarro}, {Peletier}, {S{\'a}nchez-Bl{\'a}zquez}, {Selam}, {Toloba}, {Cardiel}, {Falc{\'o}n-Barroso}, {Gorgas}, {Jim{\'e}nez-Vicente}, \& {Vazdekis}}]{Cenarro07}
{Cenarro}, A.~J., {Peletier}, R.~F., {S{\'a}nchez-Bl{\'a}zquez}, P., {et~al.} 2007, \mnras, 374, 664

\bibitem[{{Chambers} {et~al.}(2016){Chambers}, {Magnier}, {Metcalfe}, {Flewelling}, {Huber}, {Waters}, {Denneau}, {Draper}, {Farrow}, {Finkbeiner}, {Holmberg}, {Koppenhoefer}, {Price}, {Rest}, {Saglia}, {Schlafly}, {Smartt}, {Sweeney}, {Wainscoat}, {Burgett}, {Chastel}, {Grav}, {Heasley}, {Hodapp}, {Jedicke}, {Kaiser}, {Kudritzki}, {Luppino}, {Lupton}, {Monet}, {Morgan}, {Onaka}, {Shiao}, {Stubbs}, {Tonry}, {White}, {Ba{\~n}ados}, {Bell}, {Bender}, {Bernard}, {Boegner}, {Boffi}, {Botticella}, {Calamida}, {Casertano}, {Chen}, {Chen}, {Cole}, {Deacon}, {Frenk}, {Fitzsimmons}, {Gezari}, {Gibbs}, {Goessl}, {Goggia}, {Gourgue}, {Goldman}, {Grant}, {Grebel}, {Hambly}, {Hasinger}, {Heavens}, {Heckman}, {Henderson}, {Henning}, {Holman}, {Hopp}, {Ip}, {Isani}, {Jackson}, {Keyes}, {Koekemoer}, {Kotak}, {Le}, {Liska}, {Long}, {Lucey}, {Liu}, {Martin}, {Masci}, {McLean}, {Mindel}, {Misra}, {Morganson}, {Murphy}, {Obaika}, {Narayan}, {Nieto-Santisteban}, {Norberg}, {Peacock}, {Pier}, {Postman}, {Primak}, {Rae}, {Rai},
  {Riess}, {Riffeser}, {Rix}, {R{\"o}ser}, {Russel}, {Rutz}, {Schilbach}, {Schultz}, {Scolnic}, {Strolger}, {Szalay}, {Seitz}, {Small}, {Smith}, {Soderblom}, {Taylor}, {Thomson}, {Taylor}, {Thakar}, {Thiel}, {Thilker}, {Unger}, {Urata}, {Valenti}, {Wagner}, {Walder}, {Walter}, {Watters}, {Werner}, {Wood-Vasey}, \& {Wyse}}]{Chambers16}
{Chambers}, K.~C., {Magnier}, E.~A., {Metcalfe}, N., {et~al.} 2016, arXiv e-prints, arXiv:1612.05560

\bibitem[{{Cortes} \& {Vapnik}(1995)}]{Cortes95}
{Cortes}, C. \& {Vapnik}, V. 1995, Machine Learning, 20, 273

\bibitem[{{Epchtein} {et~al.}(1999){Epchtein}, {Deul}, {Derriere}, {Borsenberger}, {Egret}, {Simon}, {Alard}, {Bal{\'a}zs}, {de Batz}, {Cioni}, {Copet}, {Dennefeld}, {Forveille}, {Fouqu{\'e}}, {Garz{\'o}n}, {Habing}, {Holl}, {Hron}, {Kimeswenger}, {Lacombe}, {Le Bertre}, {Loup}, {Mamon}, {Omont}, {Paturel}, {Persi}, {Robin}, {Rouan}, {Tiph{\`e}ne}, {Vauglin}, \& {Wagner}}]{Epchteun_1999}
{Epchtein}, N., {Deul}, E., {Derriere}, S., {et~al.} 1999, \aap, 349, 236

\bibitem[{{Fiorucci} \& {Munari}(2003)}]{Fiorucci03}
{Fiorucci}, M. \& {Munari}, U. 2003, \aap, 401, 781

\bibitem[{{Fitzpatrick}(1999)}]{Fitzpatrick99}
{Fitzpatrick}, E.~L. 1999, \pasp, 111, 63

\bibitem[{{Gaia Collaboration} {et~al.}(2016){Gaia Collaboration}, {Prusti}, {de Bruijne}, {Brown}, {Vallenari}, {Babusiaux}, {Bailer-Jones}, {Bastian}, {Biermann}, {Evans}, {Eyer}, {Jansen}, {Jordi}, {Klioner}, {Lammers}, {Lindegren}, {Luri}, {Mignard}, {Milligan}, {Panem}, {Poinsignon}, {Pourbaix}, {Randich}, {Sarri}, {Sartoretti}, {Siddiqui}, {Soubiran}, {Valette}, {van Leeuwen}, {Walton}, {Aerts}, {Arenou}, {Cropper}, {Drimmel}, {H{\o}g}, {Katz}, {Lattanzi}, {O'Mullane}, {Grebel}, {Holland}, {Huc}, {Passot}, {Bramante}, {Cacciari}, {Casta{\~n}eda}, {Chaoul}, {Cheek}, {De Angeli}, {Fabricius}, {Guerra}, {Hern{\'a}ndez}, {Jean-Antoine-Piccolo}, {Masana}, {Messineo}, {Mowlavi}, {Nienartowicz}, {Ord{\'o}{\~n}ez-Blanco}, {Panuzzo}, {Portell}, {Richards}, {Riello}, {Seabroke}, {Tanga}, {Th{\'e}venin}, {Torra}, {Els}, {Gracia-Abril}, {Comoretto}, {Garcia-Reinaldos}, {Lock}, {Mercier}, {Altmann}, {Andrae}, {Astraatmadja}, {Bellas-Velidis}, {Benson}, {Berthier}, {Blomme}, {Busso}, {Carry}, {Cellino}, {Clementini},
  {Cowell}, {Creevey}, {Cuypers}, {Davidson}, {De Ridder}, {de Torres}, {Delchambre}, {Dell'Oro}, {Ducourant}, {Fr{\'e}mat}, {Garc{\'\i}a-Torres}, {Gosset}, {Halbwachs}, {Hambly}, {Harrison}, {Hauser}, {Hestroffer}, {Hodgkin}, {Huckle}, {Hutton}, {Jasniewicz}, {Jordan}, {Kontizas}, {Korn}, {Lanzafame}, {Manteiga}, {Moitinho}, {Muinonen}, {Osinde}, {Pancino}, {Pauwels}, {Petit}, {Recio-Blanco}, {Robin}, {Sarro}, {Siopis}, {Smith}, {Smith}, {Sozzetti}, {Thuillot}, {van Reeven}, {Viala}, {Abbas}, {Abreu Aramburu}, {Accart}, {Aguado}, {Allan}, {Allasia}, {Altavilla}, {{\'A}lvarez}, {Alves}, {Anderson}, {Andrei}, {Anglada Varela}, {Antiche}, {Antoja}, {Ant{\'o}n}, {Arcay}, {Atzei}, {Ayache}, {Bach}, {Baker}, {Balaguer-N{\'u}{\~n}ez}, {Barache}, {Barata}, {Barbier}, {Barblan}, {Baroni}, {Barrado y Navascu{\'e}s}, {Barros}, {Barstow}, {Becciani}, {Bellazzini}, {Bellei}, {Bello Garc{\'\i}a}, {Belokurov}, {Bendjoya}, {Berihuete}, {Bianchi}, {Bienaym{\'e}}, {Billebaud}, {Blagorodnova}, {Blanco-Cuaresma}, {Boch},
  {Bombrun}, {Borrachero}, {Bouquillon}, {Bourda}, {Bouy}, {Bragaglia}, {Breddels}, {Brouillet}, {Br{\"u}semeister}, {Bucciarelli}, {Budnik}, {Burgess}, {Burgon}, {Burlacu}, {Busonero}, {Buzzi}, {Caffau}, {Cambras}, {Campbell}, {Cancelliere}, {Cantat-Gaudin}, {Carlucci}, {Carrasco}, {Castellani}, {Charlot}, {Charnas}, {Charvet}, {Chassat}, {Chiavassa}, {Clotet}, {Cocozza}, {Collins}, {Collins}, {Costigan}, {Crifo}, {Cross}, {Crosta}, {Crowley}, {Dafonte}, {Damerdji}, {Dapergolas}, {David}, {David}, {De Cat}, {de Felice}, {de Laverny}, {De Luise}, {De March}, {de Martino}, {de Souza}, {Debosscher}, {del Pozo}, {Delbo}, {Delgado}, {Delgado}, {di Marco}, {Di Matteo}, {Diakite}, {Distefano}, {Dolding}, {Dos Anjos}, {Drazinos}, {Dur{\'a}n}, {Dzigan}, {Ecale}, {Edvardsson}, {Enke}, {Erdmann}, {Escolar}, {Espina}, {Evans}, {Eynard Bontemps}, {Fabre}, {Fabrizio}, {Faigler}, {Falc{\~a}o}, {Farr{\`a}s Casas}, {Faye}, {Federici}, {Fedorets}, {Fern{\'a}ndez-Hern{\'a}ndez}, {Fernique}, {Fienga}, {Figueras}, {Filippi},
  {Findeisen}, {Fonti}, {Fouesneau}, {Fraile}, {Fraser}, {Fuchs}, {Furnell}, {Gai}, {Galleti}, {Galluccio}, {Garabato}, {Garc{\'\i}a-Sedano}, {Gar{\'e}}, {Garofalo}, {Garralda}, {Gavras}, {Gerssen}, {Geyer}, {Gilmore}, {Girona}, {Giuffrida}, {Gomes}, {Gonz{\'a}lez-Marcos}, {Gonz{\'a}lez-N{\'u}{\~n}ez}, {Gonz{\'a}lez-Vidal}, {Granvik}, {Guerrier}, {Guillout}, {Guiraud}, {G{\'u}rpide}, {Guti{\'e}rrez-S{\'a}nchez}, {Guy}, {Haigron}, {Hatzidimitriou}, {Haywood}, {Heiter}, {Helmi}, {Hobbs}, {Hofmann}, {Holl}, {Holland}, {Hunt}, {Hypki}, {Icardi}, {Irwin}, {Jevardat de Fombelle}, {Jofr{\'e}}, {Jonker}, {Jorissen}, {Julbe}, {Karampelas}, {Kochoska}, {Kohley}, {Kolenberg}, {Kontizas}, {Koposov}, {Kordopatis}, {Koubsky}, {Kowalczyk}, {Krone-Martins}, {Kudryashova}, {Kull}, {Bachchan}, {Lacoste-Seris}, {Lanza}, {Lavigne}, {Le Poncin-Lafitte}, {Lebreton}, {Lebzelter}, {Leccia}, {Leclerc}, {Lecoeur-Taibi}, {Lemaitre}, {Lenhardt}, {Leroux}, {Liao}, {Licata}, {Lindstr{\o}m}, {Lister}, {Livanou}, {Lobel}, {L{\"o}ffler},
  {L{\'o}pez}, {Lopez-Lozano}, {Lorenz}, {Loureiro}, {MacDonald}, {Magalh{\~a}es Fernandes}, {Managau}, {Mann}, {Mantelet}, {Marchal}, {Marchant}, {Marconi}, {Marie}, {Marinoni}, {Marrese}, {Marschalk{\'o}}, {Marshall}, {Mart{\'\i}n-Fleitas}, {Martino}, {Mary}, {Matijevi{\v{c}}}, {Mazeh}, {McMillan}, {Messina}, {Mestre}, {Michalik}, {Millar}, {Miranda}, {Molina}, {Molinaro}, {Molinaro}, {Moln{\'a}r}, {Moniez}, {Montegriffo}, {Monteiro}, {Mor}, {Mora}, {Morbidelli}, {Morel}, {Morgenthaler}, {Morley}, {Morris}, {Mulone}, {Muraveva}, {Musella}, {Narbonne}, {Nelemans}, {Nicastro}, {Noval}, {Ord{\'e}novic}, {Ordieres-Mer{\'e}}, {Osborne}, {Pagani}, {Pagano}, {Pailler}, {Palacin}, {Palaversa}, {Parsons}, {Paulsen}, {Pecoraro}, {Pedrosa}, {Pentik{\"a}inen}, {Pereira}, {Pichon}, {Piersimoni}, {Pineau}, {Plachy}, {Plum}, {Poujoulet}, {Pr{\v{s}}a}, {Pulone}, {Ragaini}, {Rago}, {Rambaux}, {Ramos-Lerate}, {Ranalli}, {Rauw}, {Read}, {Regibo}, {Renk}, {Reyl{\'e}}, {Ribeiro}, {Rimoldini}, {Ripepi}, {Riva}, {Rixon},
  {Roelens}, {Romero-G{\'o}mez}, {Rowell}, {Royer}, {Rudolph}, {Ruiz-Dern}, {Sadowski}, {Sagrist{\`a} Sell{\'e}s}, {Sahlmann}, {Salgado}, {Salguero}, {Sarasso}, {Savietto}, {Schnorhk}, {Schultheis}, {Sciacca}, {Segol}, {Segovia}, {Segransan}, {Serpell}, {Shih}, {Smareglia}, {Smart}, {Smith}, {Solano}, {Solitro}, {Sordo}, {Soria Nieto}, {Souchay}, {Spagna}, {Spoto}, {Stampa}, {Steele}, {Steidelm{\"u}ller}, {Stephenson}, {Stoev}, {Suess}, {S{\"u}veges}, {Surdej}, {Szabados}, {Szegedi-Elek}, {Tapiador}, {Taris}, {Tauran}, {Taylor}, {Teixeira}, {Terrett}, {Tingley}, {Trager}, {Turon}, {Ulla}, {Utrilla}, {Valentini}, {van Elteren}, {Van Hemelryck}, {van Leeuwen}, {Varadi}, {Vecchiato}, {Veljanoski}, {Via}, {Vicente}, {Vogt}, {Voss}, {Votruba}, {Voutsinas}, {Walmsley}, {Weiler}, {Weingrill}, {Werner}, {Wevers}, {Whitehead}, {Wyrzykowski}, {Yoldas}, {{\v{Z}}erjal}, {Zucker}, {Zurbach}, {Zwitter}, {Alecu}, {Allen}, {Allende Prieto}, {Amorim}, {Anglada-Escud{\'e}}, {Arsenijevic}, {Azaz}, {Balm}, {Beck}, {Bernstein},
  {Bigot}, {Bijaoui}, {Blasco}, {Bonfigli}, {Bono}, {Boudreault}, {Bressan}, {Brown}, {Brunet}, {Bunclark}, {Buonanno}, {Butkevich}, {Carret}, {Carrion}, {Chemin}, {Ch{\'e}reau}, {Corcione}, {Darmigny}, {de Boer}, {de Teodoro}, {de Zeeuw}, {Delle Luche}, {Domingues}, {Dubath}, {Fodor}, {Fr{\'e}zouls}, {Fries}, {Fustes}, {Fyfe}, {Gallardo}, {Gallegos}, {Gardiol}, {Gebran}, {Gomboc}, {G{\'o}mez}, {Grux}, {Gueguen}, {Heyrovsky}, {Hoar}, {Iannicola}, {Isasi Parache}, {Janotto}, {Joliet}, {Jonckheere}, {Keil}, {Kim}, {Klagyivik}, {Klar}, {Knude}, {Kochukhov}, {Kolka}, {Kos}, {Kutka}, {Lainey}, {LeBouquin}, {Liu}, {Loreggia}, {Makarov}, {Marseille}, {Martayan}, {Martinez-Rubi}, {Massart}, {Meynadier}, {Mignot}, {Munari}, {Nguyen}, {Nordlander}, {Ocvirk}, {O'Flaherty}, {Olias Sanz}, {Ortiz}, {Osorio}, {Oszkiewicz}, {Ouzounis}, {Palmer}, {Park}, {Pasquato}, {Peltzer}, {Peralta}, {P{\'e}turaud}, {Pieniluoma}, {Pigozzi}, {Poels}, {Prat}, {Prod'homme}, {Raison}, {Rebordao}, {Risquez}, {Rocca-Volmerange}, {Rosen},
  {Ruiz-Fuertes}, {Russo}, {Sembay}, {Serraller Vizcaino}, {Short}, {Siebert}, {Silva}, {Sinachopoulos}, {Slezak}, {Soffel}, {Sosnowska}, {Strai{\v{z}}ys}, {ter Linden}, {Terrell}, {Theil}, {Tiede}, {Troisi}, {Tsalmantza}, {Tur}, {Vaccari}, {Vachier}, {Valles}, {Van Hamme}, {Veltz}, {Virtanen}, {Wallut}, {Wichmann}, {Wilkinson}, {Ziaeepour}, \& {Zschocke}}]{Gaia16}
{Gaia Collaboration}, {Prusti}, T., {de Bruijne}, J.~H.~J., {et~al.} 2016, \aap, 595, A1

\bibitem[{{Gaia Collaboration} {et~al.}(2022){Gaia Collaboration}, {Vallenari}, {Brown}, {Prusti}, {de Bruijne}, {Arenou}, {Babusiaux}, {Biermann}, {Creevey}, {Ducourant}, {Evans}, {Eyer}, {Guerra}, {Hutton}, {Jordi}, {Klioner}, {Lammers}, {Lindegren}, {Luri}, {Mignard}, {Panem}, {Pourbaix}, {Randich}, {Sartoretti}, {Soubiran}, {Tanga}, {Walton}, {Bailer-Jones}, {Bastian}, {Drimmel}, {Jansen}, {Katz}, {Lattanzi}, {van Leeuwen}, {Bakker}, {Cacciari}, {Casta{\~n}eda}, {De Angeli}, {Fabricius}, {Fouesneau}, {Fr{\'e}mat}, {Galluccio}, {Guerrier}, {Heiter}, {Masana}, {Messineo}, {Mowlavi}, {Nicolas}, {Nienartowicz}, {Pailler}, {Panuzzo}, {Riclet}, {Roux}, {Seabroke}, {Sordo{\o}rcit}, {Th{\'e}venin}, {Gracia-Abril}, {Portell}, {Teyssier}, {Altmann}, {Andrae}, {Audard}, {Bellas-Velidis}, {Benson}, {Berthier}, {Blomme}, {Burgess}, {Busonero}, {Busso}, {C{\'a}novas}, {Carry}, {Cellino}, {Cheek}, {Clementini}, {Damerdji}, {Davidson}, {de Teodoro}, {Nu{\~n}ez Campos}, {Delchambre}, {Dell'Oro}, {Esquej},
  {Fern{\'a}ndez-Hern{\'a}ndez}, {Fraile}, {Garabato}, {Garc{\'\i}a-Lario}, {Gosset}, {Haigron}, {Halbwachs}, {Hambly}, {Harrison}, {Hern{\'a}ndez}, {Hestroffer}, {Hodgkin}, {Holl}, {Jan{\ss}en}, {Jevardat de Fombelle}, {Jordan}, {Krone-Martins}, {Lanzafame}, {L{\"o}ffler}, {Marchal}, {Marrese}, {Moitinho}, {Muinonen}, {Osborne}, {Pancino}, {Pauwels}, {Recio-Blanco}, {Reyl{\'e}}, {Riello}, {Rimoldini}, {Roegiers}, {Rybizki}, {Sarro}, {Siopis}, {Smith}, {Sozzetti}, {Utrilla}, {van Leeuwen}, {Abbas}, {{\'A}brah{\'a}m}, {Abreu Aramburu}, {Aerts}, {Aguado}, {Ajaj}, {Aldea-Montero}, {Altavilla}, {{\'A}lvarez}, {Alves}, {Anders}, {Anderson}, {Anglada Varela}, {Antoja}, {Baines}, {Baker}, {Balaguer-N{\'u}{\~n}ez}, {Balbinot}, {Balog}, {Barache}, {Barbato}, {Barros}, {Barstow}, {Bartolom{\'e}}, {Bassilana}, {Bauchet}, {Becciani}, {Bellazzini}, {Berihuete}, {Bernet}, {Bertone}, {Bianchi}, {Binnenfeld}, {Blanco-Cuaresma}, {Blazere}, {Boch}, {Bombrun}, {Bossini}, {Bouquillon}, {Bragaglia}, {Bramante}, {Breedt},
  {Bressan}, {Brouillet}, {Brugaletta}, {Bucciarelli}, {Burlacu}, {Butkevich}, {Buzzi}, {Caffau}, {Cancelliere}, {Cantat-Gaudin}, {Carballo}, {Carlucci}, {Carnerero}, {Carrasco}, {Casamiquela}, {Castellani}, {Castro-Ginard}, {Chaoul}, {Charlot}, {Chemin}, {Chiaramida}, {Chiavassa}, {Chornay}, {Comoretto}, {Contursi}, {Cooper}, {Cornez}, {Cowell}, {Crifo}, {Cropper}, {Crosta}, {Crowley}, {Dafonte}, {Dapergolas}, {David}, {David}, {de Laverny}, {De Luise}, {De March}, {De Ridder}, {de Souza}, {de Torres}, {del Peloso}, {del Pozo}, {Delbo}, {Delgado}, {Delisle}, {Demouchy}, {Dharmawardena}, {Di Matteo}, {Diakite}, {Diener}, {Distefano}, {Dolding}, {Edvardsson}, {Enke}, {Fabre}, {Fabrizio}, {Faigler}, {Fedorets}, {Fernique}, {Fienga}, {Figueras}, {Fournier}, {Fouron}, {Fragkoudi}, {Gai}, {Garcia-Gutierrez}, {Garcia-Reinaldos}, {Garc{\'\i}a-Torres}, {Garofalo}, {Gavel}, {Gavras}, {Gerlach}, {Geyer}, {Giacobbe}, {Gilmore}, {Girona}, {Giuffrida}, {Gomel}, {Gomez}, {Gonz{\'a}lez-N{\'u}{\~n}ez},
  {Gonz{\'a}lez-Santamar{\'\i}a}, {Gonz{\'a}lez-Vidal}, {Granvik}, {Guillout}, {Guiraud}, {Guti{\'e}rrez-S{\'a}nchez}, {Guy}, {Hatzidimitriou}, {Hauser}, {Haywood}, {Helmer}, {Helmi}, {Sarmiento}, {Hidalgo}, {Hilger}, {H{\l}adczuk}, {Hobbs}, {Holland}, {Huckle}, {Jardine}, {Jasniewicz}, {Jean-Antoine Piccolo}, {Jim{\'e}nez-Arranz}, {Jorissen}, {Juaristi Campillo}, {Julbe}, {Karbevska}, {Kervella}, {Khanna}, {Kontizas}, {Kordopatis}, {Korn}, {K{\'o}sp{\'a}l}, {Kostrzewa-Rutkowska}, {Kruszy{\'n}ska}, {Kun}, {Laizeau}, {Lambert}, {Lanza}, {Lasne}, {Le Campion}, {Lebreton}, {Lebzelter}, {Leccia}, {Leclerc}, {Lecoeur-Taibi}, {Liao}, {Licata}, {Lindstr{\o}m}, {Lister}, {Livanou}, {Lobel}, {Lorca}, {Loup}, {Madrero Pardo}, {Magdaleno Romeo}, {Managau}, {Mann}, {Manteiga}, {Marchant}, {Marconi}, {Marcos}, {Marcos Santos}, {Mar{\'\i}n Pina}, {Marinoni}, {Marocco}, {Marshall}, {Polo}, {Mart{\'\i}n-Fleitas}, {Marton}, {Mary}, {Masip}, {Massari}, {Mastrobuono-Battisti}, {Mazeh}, {McMillan}, {Messina}, {Michalik},
  {Millar}, {Mints}, {Molina}, {Molinaro}, {Moln{\'a}r}, {Monari}, {Mongui{\'o}}, {Montegriffo}, {Montero}, {Mor}, {Mora}, {Morbidelli}, {Morel}, {Morris}, {Muraveva}, {Murphy}, {Musella}, {Nagy}, {Noval}, {Oca{\~n}a}, {Ogden}, {Ordenovic}, {Osinde}, {Pagani}, {Pagano}, {Palaversa}, {Palicio}, {Pallas-Quintela}, {Panahi}, {Payne-Wardenaar}, {Pe{\~n}alosa Esteller}, {Penttil{\"a}}, {Pichon}, {Piersimoni}, {Pineau}, {Plachy}, {Plum}, {Poggio}, {Pr{\v{s}}a}, {Pulone}, {Racero}, {Ragaini}, {Rainer}, {Raiteri}, {Rambaux}, {Ramos}, {Ramos-Lerate}, {Re Fiorentin}, {Regibo}, {Richards}, {Rios Diaz}, {Ripepi}, {Riva}, {Rix}, {Rixon}, {Robichon}, {Robin}, {Robin}, {Roelens}, {Rogues}, {Rohrbasser}, {Romero-G{\'o}mez}, {Rowell}, {Royer}, {Ruz Mieres}, {Rybicki}, {Sadowski}, {S{\'a}ez N{\'u}{\~n}ez}, {Sagrist{\`a} Sell{\'e}s}, {Sahlmann}, {Salguero}, {Samaras}, {Sanchez Gimenez}, {Sanna}, {Santove{\~n}a}, {Sarasso}, {Schultheis}, {Sciacca}, {Segol}, {Segovia}, {S{\'e}gransan}, {Semeux}, {Shahaf}, {Siddiqui}, {Siebert},
  {Siltala}, {Silvelo}, {Slezak}, {Slezak}, {Smart}, {Snaith}, {Solano}, {Solitro}, {Souami}, {Souchay}, {Spagna}, {Spina}, {Spoto}, {Steele}, {Steidelm{\"u}ller}, {Stephenson}, {S{\"u}veges}, {Surdej}, {Szabados}, {Szegedi-Elek}, {Taris}, {Taylo}, {Teixeira}, {Tolomei}, {Tonello}, {Torra}, {Torra}, {Torralba Elipe}, {Trabucchi}, {Tsounis}, {Turon}, {Ulla}, {Unger}, {Vaillant}, {van Dillen}, {van Reeven}, {Vanel}, {Vecchiato}, {Viala}, {Vicente}, {Voutsinas}, {Weiler}, {Wevers}, {Wyrzykowski}, {Yoldas}, {Yvard}, {Zhao}, {Zorec}, {Zucker}, \& {Zwitter}}]{Gaia22}
{Gaia Collaboration}, {Vallenari}, A., {Brown}, A.~G.~A., {et~al.} 2022, arXiv e-prints, arXiv:2208.00211

\bibitem[{{Geisler}(1984)}]{Geisler_1984}
{Geisler}, D. 1984, \pasp, 96, 723

\bibitem[{{Girardi} {et~al.}(2004){Girardi}, {Grebel}, {Odenkirchen}, \& {Chiosi}}]{Girardi04}
{Girardi}, L., {Grebel}, E.~K., {Odenkirchen}, M., \& {Chiosi}, C. 2004, \aap, 422, 205

\bibitem[{{Gray} \& {Corbally}(2009)}]{Gray09}
{Gray}, R.~O. \& {Corbally}, Christopher, J. 2009, {Stellar Spectral Classification}

\bibitem[{{Harris} {et~al.}(2020){Harris}, {Millman}, {van der Walt}, {Gommers}, {Virtanen}, {Cournapeau}, {Wieser}, {Taylor}, {Berg}, {Smith}, {Kern}, {Picus}, {Hoyer}, {van Kerkwijk}, {Brett}, {Haldane}, {del R{\'\i}o}, {Wiebe}, {Peterson}, {G{\'e}rard-Marchant}, {Sheppard}, {Reddy}, {Weckesser}, {Abbasi}, {Gohlke}, \& {Oliphant}}]{numpy}
{Harris}, C.~R., {Millman}, K.~J., {van der Walt}, S.~J., {et~al.} 2020, \nat, 585, 357

\bibitem[{{Huang} {et~al.}(2019){Huang}, {Chen}, {Yuan}, {Zhang}, {Xiang}, {Wang}, {Wang}, {Wolf}, {Liu}, \& {Liu}}]{Huang2019}
{Huang}, Y., {Chen}, B.~Q., {Yuan}, H.~B., {et~al.} 2019, \apjs, 243, 7

\bibitem[{{Indebetouw} {et~al.}(2005){Indebetouw}, {Mathis}, {Babler}, {Meade}, {Watson}, {Whitney}, {Wolff}, {Wolfire}, {Cohen}, {Bania}, {Benjamin}, {Clemens}, {Dickey}, {Jackson}, {Kobulnicky}, {Marston}, {Mercer}, {Stauffer}, {Stolovy}, \& {Churchwell}}]{Indebetouw05}
{Indebetouw}, R., {Mathis}, J.~S., {Babler}, B.~L., {et~al.} 2005, \apj, 619, 931

\bibitem[{{Jordi} {et~al.}(2006){Jordi}, {H{\o}g}, {Brown}, {Lindegren}, {Bailer-Jones}, {Carrasco}, {Knude}, {Strai{\v{z}}ys}, {de Bruijne}, {Claeskens}, {Drimmel}, {Figueras}, {Grenon}, {Kolka}, {Perryman}, {Tautvai{\v{s}}ien{\.{e}}}, {Vansevi{\v{c}}ius}, {Willemsen}, {Brid{\v{z}}ius}, {Evans}, {Fabricius}, {Fiorucci}, {Heiter}, {Kaempf}, {Kazlauskas}, {Ku{\v{c}}inskas}, {Malyuto}, {Munari}, {Reyl{\'e}}, {Torra}, {Vallenari}, {Zdanavi{\v{c}}ius}, {Korakitis}, {Malkov}, \& {Smette}}]{Jordi06}
{Jordi}, C., {H{\o}g}, E., {Brown}, A.~G.~A., {et~al.} 2006, \mnras, 367, 290

\bibitem[{{Klement} {et~al.}(2011){Klement}, {Bailer-Jones}, {Fuchs}, {Rix}, \& {Smith}}]{Klement2011}
{Klement}, R.~J., {Bailer-Jones}, C.~A.~L., {Fuchs}, B., {Rix}, H.~W., \& {Smith}, K.~W. 2011, \apj, 726, 103

\bibitem[{{Koornneef} {et~al.}(1986){Koornneef}, {Bohlin}, {Buser}, {Horne}, \& {Turnshek}}]{Koornneef86}
{Koornneef}, J., {Bohlin}, R., {Buser}, R., {Horne}, K., \& {Turnshek}, D. 1986, Highlights of Astronomy, 7, 833

\bibitem[{{Lawrence} {et~al.}(2007){Lawrence}, {Warren}, {Almaini}, {Edge}, {Hambly}, {Jameson}, {Lucas}, {Casali}, {Adamson}, {Dye}, {Emerson}, {Foucaud}, {Hewett}, {Hirst}, {Hodgkin}, {Irwin}, {Lodieu}, {McMahon}, {Simpson}, {Smail}, {Mortlock}, \& {Folger}}]{Lawrence07}
{Lawrence}, A., {Warren}, S.~J., {Almaini}, O., {et~al.} 2007, \mnras, 379, 1599

\bibitem[{{Le Borgne} {et~al.}(2003){Le Borgne}, {Bruzual}, {Pell{\'o}}, {Lan{\c{c}}on}, {Rocca-Volmerange}, {Sanahuja}, {Schaerer}, {Soubiran}, \& {V{\'\i}lchez-G{\'o}mez}}]{LeBorgne03}
{Le Borgne}, J.~F., {Bruzual}, G., {Pell{\'o}}, R., {et~al.} 2003, \aap, 402, 433

\bibitem[{Lindblad(1919)}]{Lindblad1919}
Lindblad, B. 1919, AJ, 49, 289

\bibitem[{{L{\'o}pez-Sanjuan} {et~al.}(2019){L{\'o}pez-Sanjuan}, {V{\'a}zquez Rami{\'o}}, {Varela}, {Spinoso}, {Angulo}, {Muniesa}, {Viironen}, {Crist{\'o}bal-Hornillos}, {Cenarro}, {Ederoclite}, {Mar{\'\i}n-Franch}, {Moles}, {Ascaso}, {Bonoli}, {Chies-Santos}, {Coelho}, {Costa-Duarte}, {Cortesi}, {D{\'\i}az-Garc{\'\i}a}, {Dupke}, {Galbany}, {Hern{\'a}ndez-Monteagudo}, {Logro{\~n}o-Garc{\'\i}a}, {Molino}, {Orsi}, {Placco}, {Sampedro}, {San Roman}, {Vilella-Rojo}, {Whitten}, {Mendes de Oliveira}, \& {Sodr{\'e}}}]{LopezSanjuan19}
{L{\'o}pez-Sanjuan}, C., {V{\'a}zquez Rami{\'o}}, H., {Varela}, J., {et~al.} 2019, \aap, 622, A177

\bibitem[{{Majewski} {et~al.}(2003){Majewski}, {Skrutskie}, {Weinberg}, \& {Ostheimer}}]{Majewski_2003}
{Majewski}, S.~R., {Skrutskie}, M.~F., {Weinberg}, M.~D., \& {Ostheimer}, J.~C. 2003, \apj, 599, 1082

\bibitem[{{McClure} \& {van den Bergh}(1968)}]{McClure_1668}
{McClure}, R.~D. \& {van den Bergh}, S. 1968, \aj, 73, 313

\bibitem[{{McInnes} {et~al.}(2018){McInnes}, {Healy}, \& {Melville}}]{mcinnes18}
{McInnes}, L., {Healy}, J., \& {Melville}, J. 2018, arXiv e-prints, arXiv:1802.03426

\bibitem[{{Monet} {et~al.}(2003){Monet}, {Levine}, {Canzian}, {Ables}, {Bird}, {Dahn}, {Guetter}, {Harris}, {Henden}, {Leggett}, {Levison}, {Luginbuhl}, {Martini}, {Monet}, {Munn}, {Pier}, {Rhodes}, {Riepe}, {Sell}, {Stone}, {Vrba}, {Walker}, {Westerhout}, {Brucato}, {Reid}, {Schoening}, {Hartley}, {Read}, \& {Tritton}}]{Monet_2003}
{Monet}, D.~G., {Levine}, S.~E., {Canzian}, B., {et~al.} 2003, \aj, 125, 984

\bibitem[{{Moore} {et~al.}(1966){Moore}, {Minnaert}, \& {Houtgast}}]{Moore66}
{Moore}, C.~E., {Minnaert}, M.~G.~J., \& {Houtgast}, J. 1966, {The solar spectrum 2935 A to 8770 A}

\bibitem[{{Moro} \& {Munari}(2000)}]{Moro00}
{Moro}, D. \& {Munari}, U. 2000, \aaps, 147, 361

\bibitem[{{Neff}(1966)}]{Neff1966}
{Neff}, J.~S. 1966, \aj, 71, 202

\bibitem[{{Oke}(1974)}]{Oke74}
{Oke}, J.~B. 1974, \apjs, 27, 21

\bibitem[{{Pecaut} \& {Mamajek}(2013)}]{Pecaut13}
{Pecaut}, M.~J. \& {Mamajek}, E.~E. 2013, \apjs, 208, 9

\bibitem[{{Prugniel} {et~al.}(2007){Prugniel}, {Soubiran}, {Koleva}, \& {Le Borgne}}]{Prugniel07}
{Prugniel}, P., {Soubiran}, C., {Koleva}, M., \& {Le Borgne}, D. 2007, arXiv e-prints, astro

\bibitem[{{Riello} {et~al.}(2021){Riello}, {De Angeli}, {Evans}, {Montegriffo}, {Carrasco}, {Busso}, {Palaversa}, {Burgess}, {Diener}, {Davidson}, {Rowell}, {Fabricius}, {Jordi}, {Bellazzini}, {Pancino}, {Harrison}, {Cacciari}, {van Leeuwen}, {Hambly}, {Hodgkin}, {Osborne}, {Altavilla}, {Barstow}, {Brown}, {Castellani}, {Cowell}, {De Luise}, {Gilmore}, {Giuffrida}, {Hidalgo}, {Holland}, {Marinoni}, {Pagani}, {Piersimoni}, {Pulone}, {Ragaini}, {Rainer}, {Richards}, {Sanna}, {Walton}, {Weiler}, \& {Yoldas}}]{Riello21}
{Riello}, M., {De Angeli}, F., {Evans}, D.~W., {et~al.} 2021, \aap, 649, A3

\bibitem[{{Salgado} {et~al.}(2013){Salgado}, {Osuna}, {Rodrigo}, {Allen}, {Louys}, {McDowell}, {Baines}, {Maiz Apellaniz}, {Hatziminaoglou}, {Derriere}, \& {Lemson}}]{Salgado13}
{Salgado}, J., {Osuna}, P., {Rodrigo}, C., {et~al.} 2013, {IVOA Photometry Data Model Version 1.0}, IVOA Recommendation 05 October 2013

\bibitem[{{S{\'a}nchez-Bl{\'a}zquez} {et~al.}(2006){S{\'a}nchez-Bl{\'a}zquez}, {Peletier}, {Jim{\'e}nez-Vicente}, {Cardiel}, {Cenarro}, {Falc{\'o}n-Barroso}, {Gorgas}, {Selam}, \& {Vazdekis}}]{Sanchez06}
{S{\'a}nchez-Bl{\'a}zquez}, P., {Peletier}, R.~F., {Jim{\'e}nez-Vicente}, J., {et~al.} 2006, \mnras, 371, 703

\bibitem[{Scrucca(2016)}]{Scrucca16}
Scrucca, L. 2016, Computational Statistics \& Data Analysis, 93, 5

\bibitem[{{Skrutskie} {et~al.}(2006){Skrutskie}, {Cutri}, {Stiening}, {Weinberg}, {Schneider}, {Carpenter}, {Beichman}, {Capps}, {Chester}, {Elias}, {Huchra}, {Liebert}, {Lonsdale}, {Monet}, {Price}, {Seitzer}, {Jarrett}, {Kirkpatrick}, {Gizis}, {Howard}, {Evans}, {Fowler}, {Fullmer}, {Hurt}, {Light}, {Kopan}, {Marsh}, {McCallon}, {Tam}, {Van Dyk}, \& {Wheelock}}]{Skrutskie06}
{Skrutskie}, M.~F., {Cutri}, R.~M., {Stiening}, R., {et~al.} 2006, \aj, 131, 1163

\bibitem[{{Sousa} {et~al.}(2014){Sousa}, {Santos}, {Adibekyan}, {Delgado-Mena}, {Tabernero}, {Gonz{\'a}lez Hern{\'a}ndez}, {Montes}, {Smiljanic}, {Korn}, {Bergemann}, {Soubiran}, \& {Mikolaitis}}]{Sousa14}
{Sousa}, S.~G., {Santos}, N.~C., {Adibekyan}, V., {et~al.} 2014, \aap, 561, A21

\bibitem[{{Sousa} {et~al.}(2008){Sousa}, {Santos}, {Mayor}, {Udry}, {Casagrande}, {Israelian}, {Pepe}, {Queloz}, \& {Monteiro}}]{Sousa08}
{Sousa}, S.~G., {Santos}, N.~C., {Mayor}, M., {et~al.} 2008, \aap, 487, 373

\bibitem[{{Steinmetz} {et~al.}(2006){Steinmetz}, {Zwitter}, {Siebert}, {Watson}, {Freeman}, {Munari}, {Campbell}, {Williams}, {Seabroke}, {Wyse}, {Parker}, {Bienaym{\'e}}, {Roeser}, {Gibson}, {Gilmore}, {Grebel}, {Helmi}, {Navarro}, {Burton}, {Cass}, {Dawe}, {Fiegert}, {Hartley}, {Russell}, {Saunders}, {Enke}, {Bailin}, {Binney}, {Bland-Hawthorn}, {Boeche}, {Dehnen}, {Eisenstein}, {Evans}, {Fiorucci}, {Fulbright}, {Gerhard}, {Jauregi}, {Kelz}, {Mijovi{\'c}}, {Minchev}, {Parmentier}, {Pe{\~n}arrubia}, {Quillen}, {Read}, {Ruchti}, {Scholz}, {Siviero}, {Smith}, {Sordo}, {Veltz}, {Vidrih}, {von Berlepsch}, {Boyle}, \& {Schilbach}}]{Steinmetz_2006}
{Steinmetz}, M., {Zwitter}, T., {Siebert}, A., {et~al.} 2006, \aj, 132, 1645

\bibitem[{{Stoughton} {et~al.}(2002){Stoughton}, {Lupton}, {Bernardi}, {Blanton}, {Burles}, {Castander}, {Connolly}, {Eisenstein}, {Frieman}, {Hennessy}, {Hindsley}, {Ivezi{\'c}}, {Kent}, {Kunszt}, {Lee}, {Meiksin}, {Munn}, {Newberg}, {Nichol}, {Nicinski}, {Pier}, {Richards}, {Richmond}, {Schlegel}, {Smith}, {Strauss}, {SubbaRao}, {Szalay}, {Thakar}, {Tucker}, {Vanden Berk}, {Yanny}, {Adelman}, {Anderson}, {Anderson}, {Annis}, {Bahcall}, {Bakken}, {Bartelmann}, {Bastian}, {Bauer}, {Berman}, {B{\"o}hringer}, {Boroski}, {Bracker}, {Briegel}, {Briggs}, {Brinkmann}, {Brunner}, {Carey}, {Carr}, {Chen}, {Christian}, {Colestock}, {Crocker}, {Csabai}, {Czarapata}, {Dalcanton}, {Davidsen}, {Davis}, {Dehnen}, {Dodelson}, {Doi}, {Dombeck}, {Donahue}, {Ellman}, {Elms}, {Evans}, {Eyer}, {Fan}, {Federwitz}, {Friedman}, {Fukugita}, {Gal}, {Gillespie}, {Glazebrook}, {Gray}, {Grebel}, {Greenawalt}, {Greene}, {Gunn}, {de Haas}, {Haiman}, {Haldeman}, {Hall}, {Hamabe}, {Hansen}, {Harris}, {Harris}, {Harvanek}, {Hawley}, {Hayes},
  {Heckman}, {Helmi}, {Henden}, {Hogan}, {Hogg}, {Holmgren}, {Holtzman}, {Huang}, {Hull}, {Ichikawa}, {Ichikawa}, {Johnston}, {Kauffmann}, {Kim}, {Kimball}, {Kinney}, {Klaene}, {Kleinman}, {Klypin}, {Knapp}, {Korienek}, {Krolik}, {Kron}, {Krzesi{\'n}ski}, {Lamb}, {Leger}, {Limmongkol}, {Lindenmeyer}, {Long}, {Loomis}, {Loveday}, {MacKinnon}, {Mannery}, {Mantsch}, {Margon}, {McGehee}, {McKay}, {McLean}, {Menou}, {Merelli}, {Mo}, {Monet}, {Nakamura}, {Narayanan}, {Nash}, {Neilsen}, {Newman}, {Nitta}, {Odenkirchen}, {Okada}, {Okamura}, {Ostriker}, {Owen}, {Pauls}, {Peoples}, {Peterson}, {Petravick}, {Pope}, {Pordes}, {Postman}, {Prosapio}, {Quinn}, {Rechenmacher}, {Rivetta}, {Rix}, {Rockosi}, {Rosner}, {Ruthmansdorfer}, {Sandford}, {Schneider}, {Scranton}, {Sekiguchi}, {Sergey}, {Sheth}, {Shimasaku}, {Smee}, {Snedden}, {Stebbins}, {Stubbs}, {Szapudi}, {Szkody}, {Szokoly}, {Tabachnik}, {Tsvetanov}, {Uomoto}, {Vogeley}, {Voges}, {Waddell}, {Walterbos}, {Wang}, {Watanabe}, {Weinberg}, {White}, {White}, {Wilhite},
  {Wolfe}, {Yasuda}, {York}, {Zehavi}, \& {Zheng}}]{Stoughton2002}
{Stoughton}, C., {Lupton}, R.~H., {Bernardi}, M., {et~al.} 2002, \aj, 123, 485

\bibitem[{{Sutherland} {et~al.}(2015){Sutherland}, {Emerson}, {Dalton}, {Atad-Ettedgui}, {Beard}, {Bennett}, {Bezawada}, {Born}, {Caldwell}, {Clark}, {Craig}, {Henry}, {Jeffers}, {Little}, {McPherson}, {Murray}, {Stewart}, {Stobie}, {Terrett}, {Ward}, {Whalley}, \& {Woodhouse}}]{Sutherland15}
{Sutherland}, W., {Emerson}, J., {Dalton}, G., {et~al.} 2015, \aap, 575, A25

\bibitem[{{Taylor}(2005)}]{Taylor05}
{Taylor}, M.~B. 2005, in Astronomical Society of the Pacific Conference Series, Vol. 347, Astronomical Data Analysis Software and Systems XIV, ed. P.~{Shopbell}, M.~{Britton}, \& R.~{Ebert}, 29

\bibitem[{{Thomas} {et~al.}(2019){Thomas}, {Annau}, {McConnachie}, {Fabbro}, {Teimoorinia}, {C{\^o}t{\'e}}, {Cuillandre}, {Gwyn}, {Ibata}, {Starkenburg}, {Carlberg}, {Famaey}, {Fantin}, {Ferrarese}, {H{\'e}nault-Brunet}, {Jensen}, {Lan{\c{c}}on}, {Lewis}, {Martin}, {Navarro}, {Reyl{\'e}}, \& {S{\'a}nchez-Janssen}}]{Thomas2019}
{Thomas}, G.~F., {Annau}, N., {McConnachie}, A., {et~al.} 2019, \apj, 886, 10

\bibitem[{{Thomas} {et~al.}(2018){Thomas}, {McConnachie}, {Ibata}, {C{\^o}t{\'e}}, {Martin}, {Starkenburg}, {Carlberg}, {Chapman}, {Fabbro}, {Famaey}, {Fantin}, {Gwyn}, {H{\'e}nault-Brunet}, {Malhan}, {Navarro}, {Robin}, \& {Scott}}]{Thomas2018}
{Thomas}, G.~F., {McConnachie}, A.~W., {Ibata}, R.~A., {et~al.} 2018, \mnras, 481, 5223

\bibitem[{{Torres} {et~al.}(2022){Torres}, {Canals}, {Jim{\'e}nez-Esteban}, {Rebassa-Mansergas}, \& {Solano}}]{Torres22}
{Torres}, S., {Canals}, P., {Jim{\'e}nez-Esteban}, F.~M., {Rebassa-Mansergas}, A., \& {Solano}, E. 2022, \mnras, 511, 5462

\bibitem[{{Wang} {et~al.}(2020){Wang}, {L{\'o}pez-Corredoira}, {Huang}, {Chang}, {Zhang}, {Carlin}, {Chen}, {Chrob{\'a}kov{\'a}}, \& {Chen}}]{Wang2020}
{Wang}, H.~F., {L{\'o}pez-Corredoira}, M., {Huang}, Y., {et~al.} 2020, \apj, 897, 119

\bibitem[{{Waskom}(2021)}]{seaborn}
{Waskom}, M. 2021, The Journal of Open Source Software, 6, 3021

\bibitem[{{Wenger} {et~al.}(2000){Wenger}, {Ochsenbein}, {Egret}, {Dubois}, {Bonnarel}, {Borde}, {Genova}, {Jasniewicz}, {Lalo{\"e}}, {Lesteven}, \& {Monier}}]{Wenger00}
{Wenger}, M., {Ochsenbein}, F., {Egret}, D., {et~al.} 2000, \aaps, 143, 9

\bibitem[{{Wolf} {et~al.}(2018){Wolf}, {Onken}, {Luvaul}, {Schmidt}, {Bessell}, {Chang}, {Da Costa}, {Mackey}, {Martin-Jones}, {Murphy}, {Preston}, {Scalzo}, {Shao}, {Smillie}, {Tisserand}, {White}, \& {Yuan}}]{Wolf2018}
{Wolf}, C., {Onken}, C.~A., {Luvaul}, L.~C., {et~al.} 2018, \pasa, 35, e010

\bibitem[{{Wolthoff} {et~al.}(2022){Wolthoff}, {Reffert}, {Quirrenbach}, {Jones}, {Wittenmyer}, \& {Jenkins}}]{Wolthoff2022}
{Wolthoff}, V., {Reffert}, S., {Quirrenbach}, A., {et~al.} 2022, \aap, 661, A63

\bibitem[{{Wright} {et~al.}(2010){Wright}, {Eisenhardt}, {Mainzer}, {Ressler}, {Cutri}, {Jarrett}, {Kirkpatrick}, {Padgett}, {McMillan}, {Skrutskie}, {Stanford}, {Cohen}, {Walker}, {Mather}, {Leisawitz}, {Gautier}, {McLean}, {Benford}, {Lonsdale}, {Blain}, {Mendez}, {Irace}, {Duval}, {Liu}, {Royer}, {Heinrichsen}, {Howard}, {Shannon}, {Kendall}, {Walsh}, {Larsen}, {Cardon}, {Schick}, {Schwalm}, {Abid}, {Fabinsky}, {Naes}, \& {Tsai}}]{Wright10}
{Wright}, E.~L., {Eisenhardt}, P. R.~M., {Mainzer}, A.~K., {et~al.} 2010, \aj, 140, 1868

\bibitem[{{York} {et~al.}(2000){York}, {Adelman}, {Anderson}, {Anderson}, {Annis}, {Bahcall}, {Bakken}, {Barkhouser}, {Bastian}, {Berman}, {Boroski}, {Bracker}, {Briegel}, {Briggs}, {Brinkmann}, {Brunner}, {Burles}, {Carey}, {Carr}, {Castander}, {Chen}, {Colestock}, {Connolly}, {Crocker}, {Csabai}, {Czarapata}, {Davis}, {Doi}, {Dombeck}, {Eisenstein}, {Ellman}, {Elms}, {Evans}, {Fan}, {Federwitz}, {Fiscelli}, {Friedman}, {Frieman}, {Fukugita}, {Gillespie}, {Gunn}, {Gurbani}, {de Haas}, {Haldeman}, {Harris}, {Hayes}, {Heckman}, {Hennessy}, {Hindsley}, {Holm}, {Holmgren}, {Huang}, {Hull}, {Husby}, {Ichikawa}, {Ichikawa}, {Ivezi{\'c}}, {Kent}, {Kim}, {Kinney}, {Klaene}, {Kleinman}, {Kleinman}, {Knapp}, {Korienek}, {Kron}, {Kunszt}, {Lamb}, {Lee}, {Leger}, {Limmongkol}, {Lindenmeyer}, {Long}, {Loomis}, {Loveday}, {Lucinio}, {Lupton}, {MacKinnon}, {Mannery}, {Mantsch}, {Margon}, {McGehee}, {McKay}, {Meiksin}, {Merelli}, {Monet}, {Munn}, {Narayanan}, {Nash}, {Neilsen}, {Neswold}, {Newberg}, {Nichol}, {Nicinski},
  {Nonino}, {Okada}, {Okamura}, {Ostriker}, {Owen}, {Pauls}, {Peoples}, {Peterson}, {Petravick}, {Pier}, {Pope}, {Pordes}, {Prosapio}, {Rechenmacher}, {Quinn}, {Richards}, {Richmond}, {Rivetta}, {Rockosi}, {Ruthmansdorfer}, {Sandford}, {Schlegel}, {Schneider}, {Sekiguchi}, {Sergey}, {Shimasaku}, {Siegmund}, {Smee}, {Smith}, {Snedden}, {Stone}, {Stoughton}, {Strauss}, {Stubbs}, {SubbaRao}, {Szalay}, {Szapudi}, {Szokoly}, {Thakar}, {Tremonti}, {Tucker}, {Uomoto}, {Vanden Berk}, {Vogeley}, {Waddell}, {Wang}, {Watanabe}, {Weinberg}, {Yanny}, {Yasuda}, \& {SDSS Collaboration}}]{York00}
{York}, D.~G., {Adelman}, J., {Anderson}, John~E., J., {et~al.} 2000, \aj, 120, 1579

\bibitem[{{Yuan} {et~al.}(2023{\natexlab{a}}){Yuan}, {Liu}, {Yang}, {Bu}, {Yi}, {Kong}, {Wu}, \& {Zhang}}]{Yuan2023a}
{Yuan}, H., {Liu}, M., {Yang}, Z., {et~al.} 2023{\natexlab{a}}, \aj, 166, 244

\bibitem[{{Yuan} {et~al.}(2023{\natexlab{b}}){Yuan}, {Yang}, {Cruz}, {Jim{\'e}nez-Esteban}, {Daflon}, {Placco}, {Akras}, {Alfaro}, {Galarza}, {Gon{\c{c}}alves}, {Duan}, {Liu}, {Laur}, {Solano}, {Borges Fernandes}, {Cenarro}, {Mar{\'\i}n-Franch}, {Varela}, {Ederoclite}, {L{\'o}pez-Sanjuan}, {Abramo}, {Alcaniz}, {Ben{\'\i}tez}, {Bonoli}, {Crist{\'o}bal-Hornillos}, {Dupke}, {Hern{\'a}n-Caballero}, {Mendes de Oliveira}, {Moles}, {Sodr{\'e}}, {V{\'a}zquez Rami{\'o}}, \& {Taylor}}]{Yuan23b}
{Yuan}, H.~B., {Yang}, L., {Cruz}, P., {et~al.} 2023{\natexlab{b}}, \mnras, 518, 2018

\bibitem[{{Zhang} {et~al.}(2021){Zhang}, {Yuan}, {Liu}, {Xiang}, {Huang}, \& {Chen}}]{Zhang2021}
{Zhang}, R.-Y., {Yuan}, H.-B., {Liu}, X.-W., {et~al.} 2021, Research in Astronomy and Astrophysics, 21, 319

\end{thebibliography}


\begin{appendix} 

\section{Description of the information provided by the Filter Profile Service}\label{appenA}

The information provided for each filter by the FPS is structured into four main blocks. 

\subsection{Description}
\begin{itemize}
    \item Filter identifier:
Filter identifiers are uniquely assigned to each filter. They follow a predictable syntax defined as a Category/Subcategory.Filter. In many cases, a ‘category’ is a physical facility (for instance, a telescope or an observatory), a ‘subcategory’ is the name of the instrument, and a ‘filter’ is the name of the photometric band (e.g. Paranal/NACO.J). For surveys, we often use the survey name both as a category
and subcategory (e.g. 2MASS/2MASS.J). This nomenclature also applies to filter collections of a given observatory (e.g. CAHA/CAHA.353\_41). Finally, a ‘generic’ category was created to include standard filters not associated with any particular observatory (e.g. Generic/Johnson.R). These rules are not rigid, as it is difficult to define a unique best way to label filters. In all cases, we tried to find intuitive names to ease the filter identification. Moreover, a short readable description of the filter and the photometric system is also provided.

\item Detector type associated with the photometric system. This property defines how the transmission curve must be used to compute the average 
flux associated with a given band. Two types of detectors are considered: energy counters and photon counters, for which the averaged 
flux is respectively calculated as follows: 
    \begin{equation}
       f(\lambda_{\rm eff}) = \frac{\int T(\lambda)\, f(\lambda) \, d\lambda }%
                {\int T(\lambda)\,d\lambda}
    \end{equation}
    and
    \begin{equation}
       f(\lambda_{\rm eff}) = \frac{\int T(\lambda)\, f(\lambda)\, \lambda \, d\lambda }%
                {\int T(\lambda)\, \lambda\,d\lambda}
    ,\end{equation}
 where T($\lambda$) is the filter transmission and f($\lambda$) is the flux. 

\item Band name:
This is an standard representation of the spectral band associated with this filter when appropriate (e.g. B, J, V,...). It is sometimes useful for human interpretation but it is not very useful for discovery purposes as it is empty for many filters that are not associated with any standard band.

\item Observational facility:
In most cases, this is the name of the observatory where the filter is used. In some other cases it corresponds to the name of a survey or a space mission (e.g. Paranal, 2MASS, or Herschel). 

\item Instrument: 
This corresponds to the name of the instrument associated with the filter (e.g. MIPS, VISIR, or BUSCA). 

\item Comments:
This includes a text description of some matters relating to the filter that could be of interest to the user.

 \end{itemize}

\subsection{Transmission curve}
\label{app.trans}

\begin{figure}
\centering
    \includegraphics[width=0.95\columnwidth]{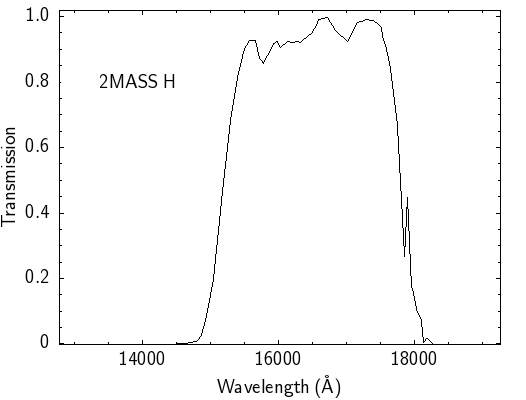}
    \caption{2MASS H transmission curve.}
    \label{Paperfigs/2massh_tr}
\end{figure}

The transmission curve is a 2D set of data, the wavelength in \AA\ and the transmission normalized to the unit, which describes the transmission properties of the filter over the wavelength range defined by the filter bandpass. In addition to the graphical representation (Fig.\,\ref{Paperfigs/2massh_tr}), the system also offers the possibility of downloading the transmission curve in two formats (aSc\,II and VOTable). The VOTable also provides the origin of the information and the components (filter, instrument, atmosphere) considered in the transmission curve.  

\subsection{Mathematical properties}\label{app.math}

\begin{figure*}
    \includegraphics[width=0.69\columnwidth]{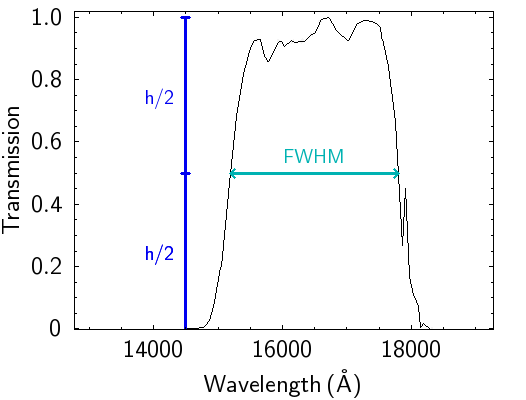}
    \includegraphics[width=0.69\columnwidth]{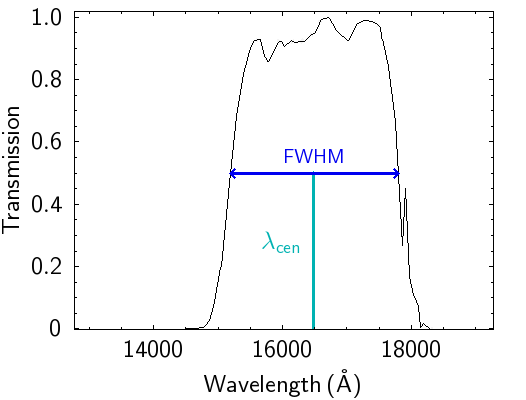}
    \includegraphics[width=0.69\columnwidth]{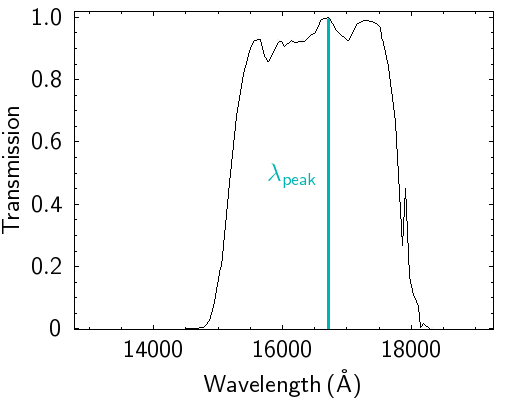}
    \includegraphics[width=0.69\columnwidth]{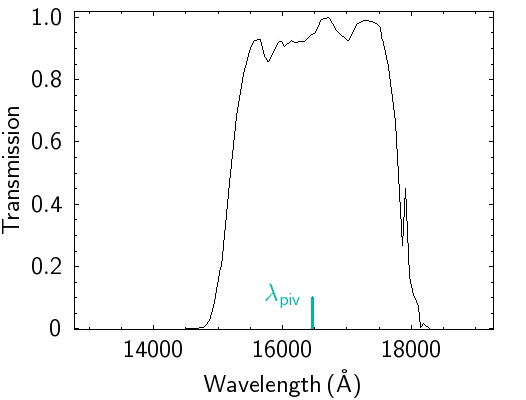}
    \includegraphics[width=0.69\columnwidth]{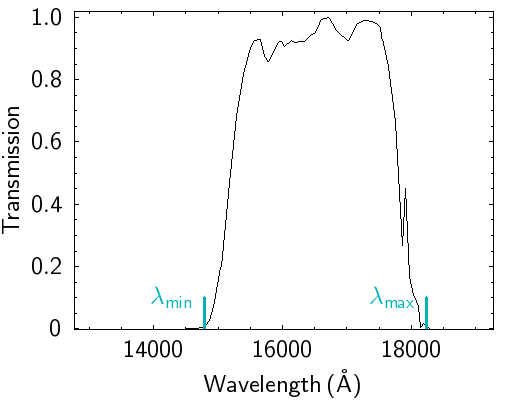}
    \includegraphics[width=0.69\columnwidth]{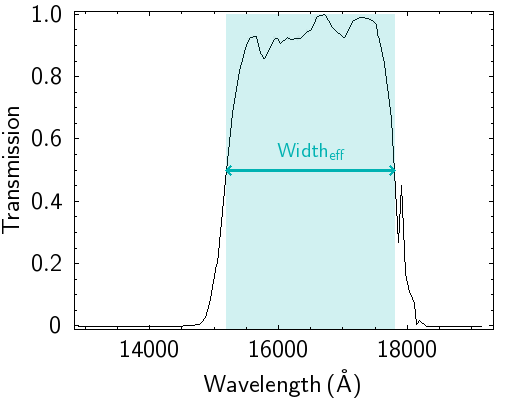}
    \caption{Definitions of several quantities illustrated using the 2MASS H filter: the FWHM ({\sl upper left panel}); the central wavelength ({\sl upper central panel}); the peak wavelength ({\sl upper right panel}); the pivot wavelength ({\sl lower left panel}); the minimum and maximum wavelengths ({\sl lower central panel}); and the effective width ({\sl lower right panel}). For more details, see Sect.\,\ref{app.math}.}
    \label{fig_filterProp}
\end{figure*}

A total of 13 properties (nine different wavelengths, the effective width, the FWHM, the extinction ratio, and the solar flux) are provided. The IVOA Photometry Data Model \citep{Salgado13} only requires one to define a value for a ‘reference wavelength’, another for a ‘width’ of the filter, and two wavelengths identifying where the transmission starts and ends. But, in practice, different authors have their own preferences about how to define these quantities in the most useful way. That is why we have included up to nine different definitions of wavelengths that can characterize the filter. If any of these properties are computed by the owner or developer of the photometric system, they are used by default. Otherwise, they are calculated following these equations:

\begin{itemize}

\item Mean wavelength ($\lambda_{\rm mean}$). It is calculated by the service as
 \begin{equation}
       \lambda_{\rm mean} = \frac{\int \lambda \,T(\lambda) \, d\lambda }%
                {\int T(\lambda)\,d\lambda},
 \end{equation}
 \noindent
 where T($\lambda$) is the filter transmission.

\item  Full width at half maximum (FWHM). This is defined as the difference between the two wavelengths for which the filter transmission is half maximum (Fig.\,\ref{fig_filterProp}, upper left panel).

\item Central wavelength ($\lambda_{\rm cen}$). This is defined as the wavelength at the middle position between the two wavelengths used to compute the FWHM (Fig.\,\ref{fig_filterProp}, upper central panel).

\item Effective wavelength ($\lambda_{\rm eff}$). It is obtained by:
 \begin{equation}
       \lambda_{\rm eff} = \frac{\int \lambda \,T(\lambda) \, {\rm Vg}(\lambda) \, d\lambda }%
                {\int T(\lambda)\,{\rm Vg}(\lambda) \, d\lambda},
 \end{equation}
 \noindent
 where Vg($\lambda$) is the Vega spectrum. For this calculation, and all other calculations that involve the Vega spectrum, the system uses the alpha\_lyr\_stis\_010.fits spectrum from the CALSPEC\footnote{\url{https://www.stsci.edu/hst/instrumentation/reference-data-for-calibration-and-tools/astronomical-catalogs/calspec}} database \citep{Bohlin14}.

\item Peak wavelength ($\lambda_{\rm peak}$). This is defined as the  wavelength value with the largest transmission (Fig.\,\ref{fig_filterProp}, upper right panel).

\item Pivot wavelength ($\lambda_{\rm piv}$; Fig.\,\ref{fig_filterProp}, lower left panel). This is defined as
 \begin{equation}
       \lambda_{piv} = \sqrt{\frac{\int \lambda \,T(\lambda) \, d\lambda }%
                {\int T(\lambda)\, d\lambda\, /  \lambda^{2}}}.
 \end{equation}
 \noindent
This is the reference wavelength for the FPS. 

\item Photon distribution wavelength ($\lambda_{phot}$). This is defined as
 \begin{equation}
       \lambda_{phot} = \frac{\int \lambda^{2} \,T(\lambda) \, {\rm Vg}(\lambda) \, d\lambda }%
                {\int \lambda\,T(\lambda)\,{\rm Vg}(\lambda) \, d\lambda}.
 \end{equation}
    
\item Minimum or maximum wavelength ($\lambda_{\rm min}$, $\lambda_{\rm max}$). This is defined as the first or last wavelength value with a transmission at least 1\% of the maximum transmission (Fig.\,\ref{fig_filterProp}, lower central panel).

\item Effective width ($W_{\rm eff}$). This is defined as the width of a rectangle with a height equal to the maximum transmission and with the same area that the one covered by the filter transmission curve (Fig.\,\ref{fig_filterProp}, lower right panel), where
 \begin{equation}
       W_{\rm eff} =  \frac{\int T(\lambda) \, d\lambda }%
                {{\rm max}(T(\lambda))}.
 \end{equation}

\item Ratio between extinction at the effective wavelength and at the V band ($A_{\rm \lambda}/A_{\rm V}$). It is calculated using the extinction law of \citet{Fitzpatrick99}, improved by \citet{Indebetouw05} in the infrared (Fig.\,\ref{extlaw}). 

\begin{figure}
\centering
    \includegraphics[width=0.95\columnwidth]{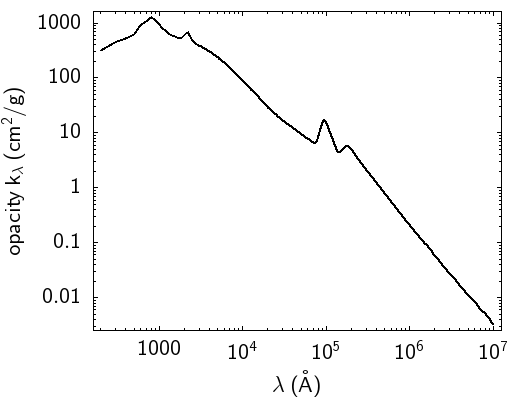}
    \caption{Opacity as a function of wavelength used for calculating the extinction ($A_\lambda = A_V \cdot k_\lambda/k_V$, where $k_V=211.4$).}
    \label{extlaw}
\end{figure}

\item Solar flux ($F_{\rm sun}$). This is defined as
 \begin{equation}
       F_{\rm sun} = \frac{\int \ T(\lambda) \, {\rm Sun}(\lambda) \,d\lambda }%
                {\int T(\lambda)\, d\lambda},
 \end{equation}
 \noindent
 where $T(\lambda)$ is the filter transmission, and Sun($\lambda$) is the solar spectrum. 
 The FPS uses a solar spectrum, which is a concatenation of two files -- the sun\_reference\_stis\_002.fits and sun\_mod\_001.fits (for larger wavelengths) -- available at the CALSPEC database.

\end{itemize}

\subsection{Calibration properties}

Zero points are used to transform magnitudes into fluxes and vice versa. This transformation is made in different ways depending on the magnitude system (Vega, AB, ST) and zero-point type (Pogson, Asinh, Linear). In fact, the same filter could be used in different surveys, catalogues, or observations with a different calibration (and thus different zero points).

\subsubsection{Magnitude systems}

The different magnitude systems are defined in terms of which spectrum is used as a reference. 

The zero point in the Vega system is calculated by the service as
 \begin{equation}
       F_{0,\lambda}{\rm (Vega)} = \frac{\int T(\lambda) \, Vg(\lambda)\,d\lambda}%
                {\int T(\lambda)\,d(\lambda)}
 ,\end{equation}

\noindent
where T($\lambda$) is the filter transmission and Vg($\lambda$) is the Vega spectrum. 

To calculate the zero point in the AB system \citep{Oke74}, we made use of a reference spectrum of constant flux density per unit frequency equal to 3631 Jy. Thus, unless otherwise specified by the filter provider, this implies that F$_{0,\nu}$(AB)\,=\,3631\,Jy. 

To calculate the zero point in the ST system \citep{Koornneef86}, we used a reference spectrum of constant flux density per unit wavelength equal to 3.631e-9\,\fluxunit, which implies that F$_{0,\lambda}$(ST)\,=\,3.631e-9 \,\fluxunit.

In all cases, the transformation between \fluxunit\ and Jy is given by the following relationship:
 \begin{equation}
    F_{0,\nu} = (2.9979246)^{-1} \times 10^{5} \times \lambda_{\rm ref}^{2} \times F_{0,\lambda} ,
 \end{equation}

\noindent
where F$_{0,\nu}$ is given in Jy and F$_{0,\lambda}$ in \fluxunit.

\subsubsection{Zero-point type}

The zero-point type specifies how fluxes are calculated from magnitudes or vice versa. This parameter can take three values (Pogson, Asinh, or Linear) with the corresponding transformation equations: 
\begin{itemize}

\item Pogson
    \begin{equation}
        {\rm F} = {\rm F}_{0} \times 10^{-({\rm  mag}-{\rm mag}_{0})/2.5} \\
    \end{equation}
    or
    \begin{equation}
        {\rm  mag} = {\rm  mag}_{0} -2.5 \log ({\rm F}/{\rm F}_{0})
    \end{equation}

\item Asinh
    \begin{equation}
        {\rm F} = {\rm F}_{0} \times 10^{-({\rm  mag}-{\rm mag}_{0})/2.5} \times [1-b^{2} \times 10^{2 ({\rm  mag}-{\rm mag}_{0})/2.5}]
    \end{equation}
    or
    \begin{equation}
    {\rm  mag} = {\rm  mag}_{0} - (2.5/\ln10)[{\rm asinh}({\rm F}/2b{\rm F}_{0})+ \ln(b)]    
    \end{equation}

\item Linear
    \begin{equation}
        {\rm F} = {\rm F}_{0} \times {\rm  mag}/{\rm  mag}_{0}
        \end{equation}
        or
        \begin{equation}
{\rm  mag} = {\rm  mag}_{0} \times {\rm F}/{\rm F}_{0}.
    \end{equation}

\end{itemize}

\noindent
In the above-mentioned equations, F$_{0}$ is the zero-point value, mag$_{0}$ the reference magnitude, and $b$ the softening parameter. More details about Asinh magnitudes can be found in \cite{Girardi04} and the SDSS documentation.\footnote{\url{http://classic.sdss.org/dr7/algorithms/photometry.html}}

\section{The synthetic photometry services}
\label{appenB}

\FloatBarrier

\begin{figure*} 

\includegraphics[width=\textwidth]{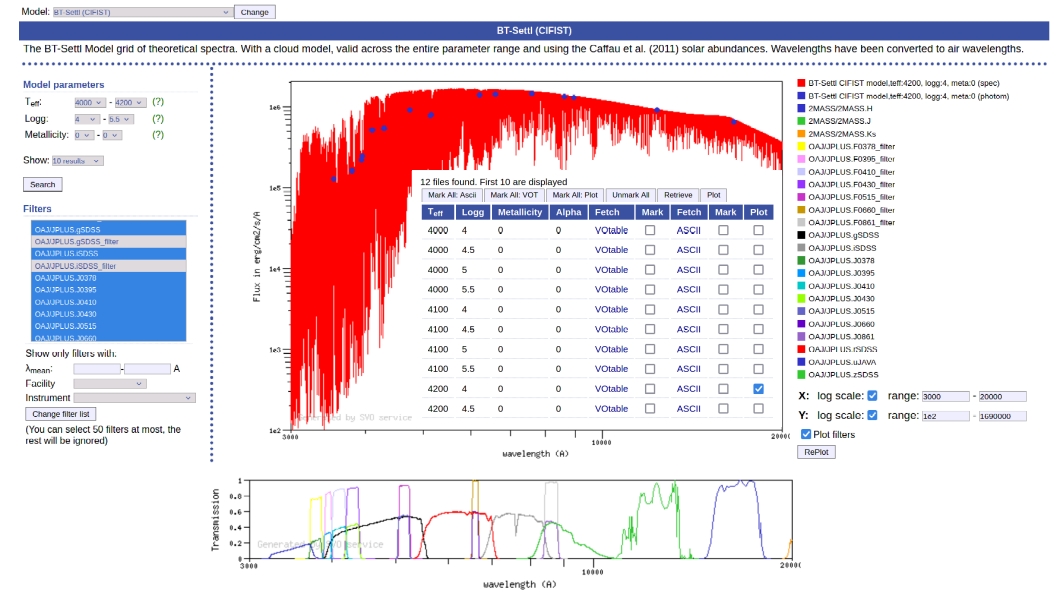}
    \caption{Example of an output from the SVO Synthetic Photometry Server. Once the spectral collection (the CFIST version of the BT-Settl models), the range of physical parameters (4000\,K < \teff < 4\,250\,K; 4 < \logg < 5.5\,dex), and the collection of filters (2MASS and J-PLUS) have been selected, the system returns, 
    for each combination of a filter and the physical parameters, 
    the corresponding spectrum (in red), the filter transmission curves (in different colours at the bottom), and the computed synthetic photometry in absolute flux units (\fluxunit; blue dots overplotted on the red spectrum). }
    \label{syntphot}
\end{figure*}

\begin{figure*}
    \includegraphics[width=\textwidth]{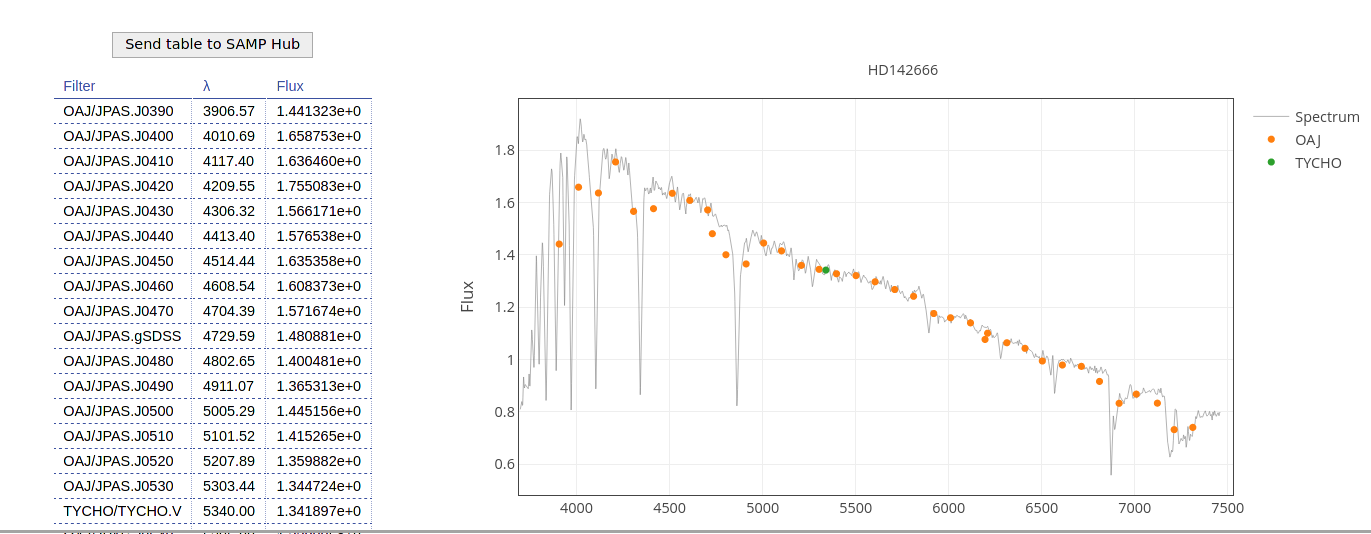}
    \caption{Output from {\it Specphot} for the input spectrum of HD\,142666. The synthetic photometry for J-PAS filters are shown in orange and for TYCHO V-band in green. A subset of the filters are listed on the left.}
    \label{fig_specphot}
\end{figure*}

One of the main utilities of the FPS is to calculate synthetic photometry from either observational or theoretical spectra, defined as the averaged flux of a spectrum  weighted by a photometric filter, using the following expression:

    \begin{equation}
    \label{synphotint}
       F(\lambda_{\rm ref}) = \frac{\int T(\lambda) \ {\rm F}(\lambda) \ d\lambda}%
                {\int T(\lambda) \ d\lambda},
    \end{equation}

\noindent
where F($\lambda_{\rm ref}$) is the synthetic flux at the reference wavelength and T($\lambda$) and F($\lambda$) are the filter transmission and spectra flux density, respectively.



Synthetic photometry for $\sim$ 70 collections of both theoretical and observational spectra in all the filters included in the FPS and within the wavelength coverage of the collection are available at the SVO Theoretical Model Services.\footnote{\url{http://svo2.cab.inta-csic.es/theory/main/}} This encompasses 290\,000 spectra and $3 \times 10^9$ synthetic photometry values. All this information is available through the SVO Synthetic Photometry Server web interface.\footnote{\url{http://svo2.cab.inta-csic.es/theory//newov2/syph.php}} 

Using the interface, the user can select a collection of spectra, a range of physical parameters, and a set of filters. 
For each combination of a filter and the physical parameters, 
the system returns the associated synthetic photometry, which can be downloaded in either an aSc\,II or VOTable format (See Fig.\,\ref{syntphot}).

For those cases, like the one described in this paper, in which the handling of a large number of colours is required, the previous service is nicely complemented with a second facility, \texttt{COLCA},\footnote{\url{http://svo2.cab.inta-csic.es/theory/newov2/wcolors.php}} also developed by the SVO, to directly obtain photometric colours given a collection of theoretical models or an observational templates, a range of physical parameters (\teff, \logg, [M/H],...), and a collection of filters. \texttt{COLCA} allows the user to calculate up to a maximum of 20 colours at a time for a particular collection of theoretical models. Here, it is important to note that, if a theoretical spectrum does not cover the entire wavelength range of selected filters, the value for the corresponding colour will be empty.

Finally, in addition to the generation of synthetic photometry from theoretical spectra, there are other cases (e.g. filling in the gaps in photometric SEDs using spectra, obtaining photometric information for bright sources that can appear saturated in the photometric catalogues...) in which it may be useful to convert spectroscopic information into photometry. 

{\it Specphot}\footnote{\url{http://svo2.cab.inta-csic.es/theory/specphot/}} is the tool that makes this type of operations in the Virtual Observatory framework. It is a web application that allows one to upload user spectra, grouped as collections, and then calculate the synthetic photometry from these spectra for the filters available at the SVO FPS.

As an example, Fig.~\ref{fig_specphot} illustrates the output obtained from {\it Specphot} for the spectrum of HD\,142666 taken from the BESS database.\footnote{\url{http://basebe.obspm.fr/basebe/}} 
Overplotted on the input spectrum, there are two collections of synthetic photometry values obtained using J-PAS filters (orange dots) and TYCHO\footnote{\url{http://svo2.cab.inta-csic.es/theory/fps/index.php?mode=browse&gname=TYCHO}} V band (green dot).
Moreover, if information on coordinates (right ascension and declination) is provided, the computed photometry can be used by other SVO tools like VOSA. Detailed information on the {\it Specphot} capabilities can be found in the Help\footnote{\url{http://svo2.cab.inta-csic.es/theory/specphot/index.php?action=help&&idx=1}} section.

\end{appendix}
\end{document}